\documentclass[a4paper]{article}
\pdfoutput=1

\usepackage{jcappub} 

\usepackage[T1]{fontenc} 
\usepackage[utf8]{inputenc}
\usepackage{pythonhighlight} 
\usepackage{textcomp,gensymb}
\usepackage{autobreak}

\usepackage{autobreak}
\usepackage{physics}
\usepackage{feynmp-auto}
\usepackage{slashed}
\usepackage{lipsum}
\usepackage{adjustbox}

\usepackage{ulem}
\usepackage{amsmath}
\usepackage{amssymb}
\usepackage{bm}
\usepackage{bbold}
\usepackage{multirow}
\usepackage{makecell}

\newcommand\Tstrut{\rule{0pt}{2.6ex}}         

\usepackage{graphicx}
\usepackage{hyperref}
\usepackage[american]{babel}
\usepackage{enumerate}
\usepackage{mathtools}
\usepackage{support-caption}
\usepackage{subcaption}
\usepackage{soul}
\setstcolor{red}
\usepackage{makebox}
\usepackage{comment}
\usepackage{booktabs}

\newcommand{\U}[1]{\mathrm{U}(1)_{\mathrm{#1}}}			

\renewcommand{\[}{\left[}

\usepackage{pifont}

\newcommand\varpm{\mathbin{\vcenter{\hbox{%
  \oalign{\hfil$\scriptstyle\hspace{-0.2ex}+\hspace{-0.2ex}$\hfil\cr
          \noalign{\kern-.5ex}
          $\scriptscriptstyle({-})$\cr}%
}}}}

\newcommand\varmp{\mathbin{\vcenter{\hbox{%
  \oalign{\hfil$\scriptstyle\hspace{-0.2ex}-\hspace{-0.2ex}$\hfil\cr
          \noalign{\kern-.5ex}
          $\scriptscriptstyle({+})$\cr}%
}}}}

\DeclareMathSymbol{\comma}{\mathpunct}{letters}{"3B} 

\definecolor{ForestGreen}{rgb}{0.13, 0.55, 0.13}

\usepackage{xspace}

\usepackage[capitalise]{cleveref}

\crefname{section}{Section}{Sections}
\crefname{table}{Table}{Tables}
\crefname{figure}{Fig.}{Figs.}
\crefname{equation}{Eq.}{Eqs.}
\crefname{appendix}{Appendix}{Appendices}

\graphicspath{{figures/}}

\allowdisplaybreaks

\title{Primordial black holes and magnetic fields in conformal neutrino mass models}

\author[a]{Shyam Balaji,}
\author[b,e,g]{Jo\~ao~Gon\c{c}alves,}
\author[c]{Danny Marfatia,}
\author[d,e,g]{Ant\'onio~P.~Morais}
\author[f]{and Roman Pasechnik}

\affiliation[a]{Physics Department, King's College London, Strand, London, WC2R 2LS, United Kingdom}
\affiliation[b]{Departamento de F\'{i}sica da Universidade de Aveiro, Campus de Santiago, 3810-183 Aveiro, Portugal.}
\affiliation[c]{Department of Physics and Astronomy, University of Hawaii at Manoa, Honolulu, HI 96822, USA}
\affiliation[d]{Departamento de F\'{i}sica, Escola de Ci\^{e}ncias, Universidade do Minho, 4710-057 Braga, Portugal}
\affiliation[e]{Laborat\'{o}rio de Instrumenta\c{c}\~{a}o e F\'{i}sica Experimental de Part\'{i}culas (LIP), Universidade do Minho, 4710-057 Braga, Portugal}
\affiliation[f]{Department of Physics, Lund University, 221 00 Lund, Sweden}
\affiliation[g]{Centre for Research and Development in Mathematics and Applications (CIDMA), Campus de Santiago, 3810-183 Aveiro, Portugal}

\emailAdd{shyam.balaji@kcl.ac.uk}
\emailAdd{jpedropino@ua.pt}
\emailAdd{dmarf8@hawaii.edu}
\emailAdd{amorais@fisica.uminho.pt}
\emailAdd{roman.pasechnik@fysik.lu.se}

\abstract{Sufficiently strong and long-lasting first-order phase transitions can produce primordial black holes (PBHs) that contribute substantially to the dark matter abundance of the Universe, and can produce large-scale primordial magnetic fields.  We study these mechanisms in a generic class of conformal $\mathrm{U(1)}^\prime$ models that also explain active neutrino oscillation data via the type-I seesaw mechanism. 
We find that phase transitions that occur at seesaw scales between $10^4$~GeV and $10^{11}$~GeV produce gravitational wave signals (from the dynamics of the phase transition and from the decay of cosmic string loops) at LISA/ET that can be correlated with microlensing signals of PBHs at the Roman Space Telescope, while scales near $10^{11}$~GeV can be correlated with Hawking evaporation signals at future gamma-ray telescopes. LISA can probe the entire range of PBH masses between $1\times 10^{-16}M_\odot$ and $8\times 10^{-11}M_\odot$ if PBHs fully account for the dark matter abundance. For $\mathrm{Z^\prime}$ masses between 5~TeV and 100~TeV, and $\sim 3$~TeV right-handed neutrinos, helical magnetic fields can be produced with magnitudes $\sim 10^{-16}$--$10^{-13}$~G and coherence lengths $\sim 10^{-4}$--$10^{-2}$~Mpc, above current blazar lower bounds. 
}
\begin{document}

\maketitle
\flushbottom

\section{Introduction}\label{sec:intro}

A mechanism for the formation of primordial black holes (PBHs) as a candidate for dark matter (DM) involves overdensities in the primordial plasma that collapse gravitationally. This could happen if a sufficiently strong first-order phase transition (FOPT) occurred between the end of inflation and the beginning of big bang nucleosynthesis. This idea has been explored over the past four decades, although less frequently than the inflation scenario~\cite{Hawking:1982ga,Kodama:1982sf,Moss:1994iq}. Among the FOPT mechanisms that trigger PBH formation, the creation of large false vacuum remnants from stochastically late-nucleated patches during supercooled FOPTs is particularly intriguing~\cite{Liu:2021svg,Hashino:2021qoq,Lewicki:2023ioy,Gouttenoire:2023naa,Salvio:2023ynn,Baldes:2023rqv,Balaji:2024rvo}. This process is inevitable if the FOPT is sufficiently strong, leading to Hubble-sized overdensities in the plasma after reheating and closely mimicking the inflation mechanism. 

A paradigmatic example of a strongly supercooled FOPT arises in models with classical scale invariance. In this framework, a potential barrier between the true and false vacua emerges solely due to thermal effects and can persist for an extended period as the Universe cools down. As a result, the nucleation of true vacuum bubbles is delayed to temperatures far below the critical temperature. The formation of PBHs can be triggered in FOPT scenarios with strong and prolonged supercooling. Moreover, the FOPT generates primordial gravitational waves (GWs) that contribute to the stochastic GW background (SGWB), thus providing a correlated signature with the formation of PBHs.

Primordial magnetogenesis has been extensively studied in the context of the electroweak (EW) phase transition~\cite{Vachaspati:1991nm, Ellis:2019tjf} and the QCD phase transition~\cite{Sigl:1996dm, Tevzadze:2012kk}. The idea of magnetogenesis during a first-order EW phase transition, first proposed in Ref.~\cite{Vachaspati:1991nm}, suggests that magnetic fields are generated through EW sphaleron decays~\cite{Vachaspati:2001nb, Copi:2008he, Chu:2011tx}. As the true vacuum bubbles grow, collide, and merge, they drive the primordial plasma into high Reynolds number motion, leading to magnetohydrodynamic (MHD) turbulence in the magnetic fields~\cite{Witten:1984rs, Hogan:1986qda, Kamionkowski:1993fg, Brandenburg:1996fc, Christensson:2000sp, Kahniashvili:2010gp, Brandenburg:2017neh}. These processes are particularly relevant for explanations of the origin of coherent intergalactic magnetic fields (IGMFs), which are indirectly supported by blazar observations~\cite{Fermi-LAT:2018jdy,MAGIC:2022piy,HESS:2023zwb,Neronov:2010gir}. While astrophysical mechanisms, such as the Biermann battery effect~\cite{PhysRev.82.863} combined with dynamo amplification~\cite{AlvesBatista:2021sln}, can produce IGMFs with long correlation lengths, cosmological scenarios involving FOPTs offer a compelling alternative, naturally accommodating magnetic fields with extremely large coherence lengths. Thus, early universe processes like FOPTs remain an attractive explanation for the observed IGMFs.

In this work, we focus on a class of generic $\mathrm{U(1)}^\prime$ extensions of the Standard Model (SM) in which the $\mathrm{U(1)}^\prime$ gauge group is a linear combination of the SM $\mathrm{U(1)_Y}$ and $\mathrm{U(1)_{B-L}}$ gauge groups, and that exhibit classical scale invariance in both the visible and dark sectors. These models are designed to explain neutrino masses and mixing via a type-I seesaw mechanism with three generations of right-handed neutrinos. A subset of the model parameter space in Ref.~\cite{Goncalves:2024lrk} allows FOPTs and cosmic strings that generate strong GW signals while also providing the necessary conditions for the formation of PBHs and primordial magnetic fields.

This article is organized as follows. In \cref{sec:model} we introduce a generic class of classically scale-invariant $\mathrm{U(1)}^\prime$ models of neutrino mass at both zero and finite temperatures. In \cref{sec:FOPT_GWs} we provide a brief overview of the formalism of supercooled FOPTs and the corresponding SGWB production. In Sections~\ref{sec:PBHformation} and~\ref{sec:pmf} we discuss PBH formation and the process of primordial magnetogenesis during these FOPTs, respectively.  Our results are presented in \cref{sec:results}, and we conclude in~\cref{sec:summary}.

\section{Generic scale-invariant \texorpdfstring{$\bm{\mathrm{U(1)^\prime}}$}{U(1)p} models}\label{sec:model}

 Imposing scale invariance on the classical action forbids tree-level dimensionful parameters in both the fermionic and bosonic sectors, thereby reducing the number of free parameters. Analogous to standard Majoron models~\cite{Chikashige:1980qk,Chikashige:1980ui,Gelmini:1980re}, neutrino masses originate from the vacuum expectation value (vev) of a complex scalar singlet $\sigma$, which we refer to as the Majoron due to its role in generating Majorana masses.
\begin{table}[t]
	\centering
	\begin{tabular}{c|c|c|c|c}
		\textbf{Field} & $\mathbf{U(1)^\prime}$ & $\mathbf{SU(3)_\text{C}}$ & $\mathbf{SU(2)_\text{L}}$ & $\mathbf{U(1)_\text{Y}}$  \\ \hline\Tstrut
		$L$ & $ -x_\mathcal{H} - \frac{1}{2} x_\sigma$ & $\bm{1}$ & $\bm{2}$ & $-1/2$ \\[0.5em]
		$\mathcal{H}$ & $x_\mathcal{H}$ & $\bm{1}$ & $\bm{2}$ & $1/2$ \\[0.5em]
        $N $ & $-\frac{1}{2} x_\sigma$ & $\bm{1}$ & $\bm{1}$ & $0$ \\[0.5em]
        $\sigma$ & $x_\sigma$ & $\bm{1}$ & $\bm{1}$ & $0$ \\ [0.5em]
        \hline
	\end{tabular}
	\caption{\small Anomaly-free $\mathrm{U(1)}^\prime$ charges of the SM lepton ($L$) and Higgs ($\mathcal{H}$) doublets and the dark sector fields expressed in terms of the $\mathcal{H}$ charge, $x_\mathcal{H}$, and the Majoron charge, $x_\sigma$~\cite{Oda:2015gna}.}
 \label{tab:charges}
\end{table}

\cref{tab:charges} summarizes the model's field content and quantum numbers. The second column lists the $\U{}^\prime$ charges that respect the anomaly cancellation conditions. 
We provide a brief description of the model, and 
refer the reader to Ref.~\cite{Goncalves:2024lrk} for details.

\subsection*{Yukawa sector}

Due to the $\mathrm{U(1)^\prime}$ symmetry, explicit mass terms for the right-handed neutrinos are forbidden. Instead, their scale is dynamically generated through radiative symmetry breaking once $\sigma$ acquires a vev. The type I seesaw mechanism arises from the Lagrangian,
\begin{equation}
    \mathcal{L}_\nu = y_\nu^{ij} \overline{N}_{i} \mathcal{H} L_j + y_\sigma^{ij} \Bar{N}_{i}^c N_{j} \sigma
    + \mathrm{h.c.}\,, \qquad i,j = 1,2,3 \,.
    \label{eq:Lnu}
\end{equation}
Once both $\mathcal{H}$ and $\sigma$ acquire vevs,
$v \simeq 246~\mathrm{GeV}$ and $v_\sigma$, respectively, 
the Lagrangian parameters can be cast in terms of physical parameters -- neutrino mass squared differences and the 
Pontercorvo-Maki-Nakagawa-Sakata mixing matrix -- through well-known master formulas~\cite{Cordero-Carrion:2019qtu}. The latter enable us to utilize as inputs the oscillation parameters determined from the global fit by the NuFIT collaboration~\cite{Esteban:2024eli} assuming a normal ordering of the neutrino masses. We also impose the cosmological bound on the sum of active neutrino masses, $\sum m_\nu < 0.12~\mathrm{eV}$~\cite{Planck:2018vyg}.

\subsection*{Scalar sector}

In the classically conformal scalar sector, the tree-level potential is given by 
\begin{equation}\label{eq:tree_potential}
V_{0}(\mathcal{H},\sigma) = \lambda_h(\mathcal{H}^{\dagger}\mathcal{H})^2 + \lambda_{\sigma}(\sigma^{\dagger}\sigma)^2 + \lambda_{\sigma h}(\mathcal{H}^{\dagger}\mathcal{H}) (\sigma^{\dagger}\sigma)\,.
\end{equation}
Both the EW and $\mathrm{U(1)}^\prime$ symmetries are radiatively broken at one-loop level giving rise to two physical scalar fields: the SM-like Higgs boson, $h_1$, and its CP-even partner, $h_2$. We consider the mass hierarchy, $M_{h_2} > M_{h_1}$, which induces strongly supercooled FOPTs that yield an observable SGWB at frequencies above a mHz~\cite{Goncalves:2024lrk}.\footnote{In the opposite case with $M_{h_1} > M_{h_2}$, a quadratic term for $\mathcal{H}$ must be incorporated in \cref{eq:tree_potential} to ensure consistent EW symmetry breaking. Then, strongly supercooled FOPTs produce GWs at nHz frequencies for $M_{h_2} \sim \mathcal{O}(\mathrm{MeV})$~\cite{Goncalves:2025uwh}.}

The scalar sector also incorporates the Coleman-Weinberg (CW) potential~\cite{Coleman:1973jx}, which in the $\overline{\mathrm{MS}}$ renormalization scheme is at one-loop accuracy,
\begin{equation}\label{eq:CW_potential}
    V_{\mathrm{CW}}(\phi_h, \phi_\sigma) = \frac{1}{64\pi^2} \sum_a n_a M_a^4(\phi_h, \phi_\sigma) \qty(\ln \frac{M_a^2(\phi_h, \phi_\sigma)}{Q^2} - c_a) \,, 
\end{equation}
where $\phi_h$ and $\phi_\sigma$ are the classical field configurations of $\mathcal{H}$ and $\sigma$, respectively, $M_a(\phi_h,\phi_\sigma)$ represents the field-dependent mass for a vector, scalar or femionic particle $a$, and $Q$ is the renormalization scale. The constants $c_a$ take values $3/2$ for fermions and scalars, and $5/6$ for vectors, and $n_a$ accounts for the number of degrees of freedom of each particle $a$. 

The symmetry breaking patterns are analyzed by minimizing the total effective potential, $V_{\rm eff} = V_0 + V_{\mathrm{CW}}$. The inclusion of the CW potential impacts not only the vacuum structure but also the mass spectrum of the theory. Consequently, the masses are also computed at one-loop accuracy. A detailed description of the one-loop minimization procedure and the calculation of the one-loop mass spectrum is provided in Ref.~\cite{Goncalves:2024lrk}. This approach fixes the values of $\lambda_h$, $\lambda_\sigma$, $\lambda_{\sigma h}$ and $v_\sigma$, while we fix the mass of the SM-like Higgs boson to its experimentally measured value $M_{h_1} = 125.11~\mathrm{GeV}$~\cite{ATLAS:2023oaq}. This leaves the mass of the heavy CP-even Higgs boson, $M_{h_2}$, as the only free parameter in the scalar sector.

\subsection*{Gauge sector}

The spontaneous breaking of the $\mathrm{U(1)}^\prime$ symmetry gives rise to a heavy vector boson, $\mathrm{Z^\prime}$. It mixes with the SM $\mathrm{Z^0}$ boson through both kinetic mixing and gauge-field mass mixing parametrized by $g_{12}$ and $g_L x_\mathcal{H}$, respectively, where $g_L$ is the $\U{}^\prime$ gauge coupling. In the $v_\sigma \gg v$ limit, the field-dependent masses of the $\mathrm{Z^\prime}$ and $\mathrm{Z^0}$ are given by
\begin{equation}\label{eq:Mzprime_Mz_masses}
M^2_{\mathrm{Z^0}} = \frac{v^2}{16}(g_1^2 + g_2^2)\qty(4 - \frac{(g_{12} + 2g_L x_\mathcal{H})\phi_h^2}{g_L^2 x_\sigma^2 \phi_\sigma^2} )\,, \quad
M^2_{\mathrm{Z^\prime}} = \frac{1}{4}\qty(g_{12} + 2g_L x_\mathcal{H})^2\phi_h^2 + g_L^2 x_\sigma^2 \phi_\sigma^2\,,
\end{equation}
where $g_1$ and $g_2$ are the $\mathrm{U(1)_Y}$ and $\mathrm{SU(2)_L}$ gauge couplings, respectively. Given that kinetic mixing is not relevant for the dynamics of supercooled phase transitions, we fix $g_{12} = 0$ at the EW scale, although renormalization group (RG) evolution generates a non-zero value at higher energies, which can help stabilize the Higgs vacuum~\cite{Marzo:2018nov}. As a result, the only free parameters in the gauge sector are $g_L$, $x_\mathcal{H}$ and $x_\sigma$. 
To ensure consistency with experimental limits, we require $M_{\mathrm{Z^\prime}} > 5~\mathrm{TeV}$~\cite{ATLAS:2019erb,ATLAS:2017eiz,ATLAS:2018tfk,ATLAS:2020lks}.

\subsection*{Thermal potential and RG improvement}

Variations in the RG scale are known to significantly affect the SGWB~\cite{Gould:2021oba,Croon:2020cgk}. To minimize these uncertainties, we employ an RG-improved potential, where both the couplings and fields are adjusted according to their RG equations: 
\begin{equation}\label{eq:RG_transformations}
\begin{aligned}
&\lambda_i \rightarrow \lambda_i(t)\,, \\
&\phi_\sigma^2 \rightarrow \frac{\phi_\sigma^2}{2} \exp{\int_0^t dt' [3x_\sigma^2 g_L^2(t^\prime) - 2\mathrm{Tr}(\bm{y_\sigma}(t^\prime) \bm{y_\sigma}^*(t^\prime))]  }\,.
\end{aligned}
\end{equation}
Here $\lambda_i = \lambda_\sigma\,,\lambda_h\,,\lambda_{\sigma h}\,,y_t\,,\bm{y_\sigma}\,,\bm{y_\nu}\,,g_L\,,g_{12}$, and $t \equiv \mathrm{ln}(Q/Q_{\mathrm{ref}})$, with the reference scale $Q_{\mathrm{ref}}$ set to $M_{\mathrm{Z^0}} = 91~\mathrm{GeV}$. The couplings $\lambda_i(t)$ evolve according to their respective beta-functions, which are detailed in Ref.~\cite{Goncalves:2024lrk}. An analogous rescaling applies to $\phi_h$. However, since $v_\sigma \gg v$, only $\phi_\sigma$ is relevant for the FOPT. The field-dependent RG scale is chosen to be $Q = \mathrm{max}[M_{\mathrm{Z^\prime}}(\phi_\sigma), \pi T]$, where $M_{\mathrm{Z^\prime}}(\phi_\sigma)$ is the field-dependent mass of the $\mathrm{Z^\prime}$ boson in \cref{eq:Mzprime_Mz_masses}. 

At high temperatures, the effective potential receives thermal and Daisy corrections, $V_T$ and $V_\mathrm{Daisy}$, respectively, which are essential to accurately describe thermally induced phase transitions. $V_T$ is derived from one-loop thermal bosonic and fermionic functions and is expressed as a sum over all particle species in the plasma. $V_\mathrm{Daisy}$ arises from resumming the leading infrared-divergent diagrams that dominate at high temperatures and are important for perturbative stability. In our case, the Daisy contribution is small because for supercooled FOPTs the phase transition temperature is below the $\mathrm{Z^\prime}$ and $h_2$ masses. Consequently, the full thermal effective potential is given by $V_\mathrm{eff} = V_0 + V_\mathrm{CW} + V_T + V_\mathrm{Daisy}$; see Ref.~\cite{Goncalves:2024lrk} for analytic expressions. Note that for both the zero-temperature and finite-temperature components of the potential, the couplings and $\phi_\sigma$ are RG-evolved in accordance with \cref{eq:RG_transformations}.

\section{Primordial gravitational waves}\label{sec:FOPT_GWs}

\subsection{First-order phase transitions}
 As true vacuum bubbles expand and occupy $34\%$ of the Universe's volume, they become causally connected and prevent a return to the symmetric phase. This defines the percolation temperature, $T_p$, at which all thermodynamic parameters of the phase transition are evaluated and which marks the epoch of SGWB generation (for further details, see {\it e.g.} Ref.~\cite{Goncalves:2024lrk}). For strong supercooling the temperature drops significantly before bubble nucleation occurs, and the bubbles runaway with a wall velocity approaching $v_w = 1$. To ensure that percolation completes, we require that the false vacuum volume is decreasing at $T_p$. 

The strength of the phase transition, $\alpha$, characterizes both the dynamics of the phase transition and its associated GW signal. It is defined as the ratio of the latent heat released during the transition to the total radiation energy density of the Universe at $T_p$
\begin{equation}\label{eq:alpha_param}
\alpha = \frac{\Delta V_\mathrm{eff}}{\rho_R}\Biggr|_{\substack{T = T_p}} - \frac{T}{\rho_R} \frac{\partial \Delta V_\mathrm{eff}}{\partial T} \Biggr|_{\substack{T = T_p}}\,,
\end{equation}
where $\Delta V_\mathrm{eff}(T)$ is the potential energy difference between the false and true vacuum at temperature $T$: $\Delta V_\mathrm{eff}(T) \equiv V_\mathrm{eff}(0,T) - V_\mathrm{eff}(v_\mathrm{True}(T),T)$. Here, $v_\mathrm{True}(T)$ is the vev of the true vacuum, and $\rho_R(T) = g_*(T)(\pi^2/30)T^4$ is the radiation energy density, with $g_*(T)$ the total number of relativistic degrees of freedom in the dark and visible sectors. Another important characteristic of the phase transition, its inverse time-scale, is given by
\begin{equation} \label{eq:beta}   
\frac{\beta}{H(T_p)} = T_p\frac{d (S_3/T)}{dT} \Biggr|_{\substack{T = T_p}}\,,
\end{equation}
where $H(T_p)$ is the Hubble parameter at $T_p$, and $S_3$ is the three-dimensional Euclidean action. In supercooled FOPTs a substantial amount of energy is released into the plasma, which is reheated to a temperature $T_\mathrm{RH}$. In scenarios where the Universe enters a period of radiation domination immediately after percolation ends, the reheating temperature is
\begin{equation}\label{eq:treh}
    T_\mathrm{RH} = T_p (1 + \alpha)^{1/4} \,.
\end{equation} 
It is important to note that while thermodynamic parameters are evaluated at $T_p$, the SGWB must be calculated at $T_\mathrm{RH}$.


The thermodynamic parameters discussed above are used in spectral templates to estimate the SGWB. We use the latest templates provided by the LISA Cosmology Working Group~\cite{Caprini:2024hue}, and only include contributions from bubble collisions and sound waves in the plasma, with the efficiency factors of
Ref.~\cite{Ellis:2020nnr}. For more details, see Ref.~\cite{Goncalves:2024lrk}. 

\subsection{Cosmic strings}

Cosmic strings form once the phase transition spontaneously breaks the $\mathrm{U(1)^\prime}$ symmetry~\cite{Kibble:1976sj}. Since 
$\mathrm{U(1)^\prime}$ is gauged, the strings decay predominantly via gravitational radiation, providing an additional component to the SGWB. Their dynamics is governed by a single parameter, the string tension $\mu$, which is set by the $\mathrm{U(1)^\prime}$-breaking  scale~\cite{Blanco-Pillado:2024aca}:
\begin{equation}\label{eq:GMU}
	G\mu \approx 10^{-6}\left( \frac{v_\sigma}{10^{16}~\mathrm{GeV}} \right)^2\,.
\end{equation}
Here $G = 1/M_p^2$ is the gravitational constant with the Planck mass $M_p = 1.22 \times 10^{19}~\mathrm{GeV}$.

The SGWB spectrum from cosmic strings is sourced by long strings and closed loops, with the latter typically providing the dominant, although more model-dependent, signal~\cite{CamargoNevesdaCunha:2022mvg}. For oscillating loops, the spectral energy density is given by a weighted sum over harmonic modes $k$, 
\begin{equation}\label{eq:Om_GW_CS}
	\Omega_{\mathrm{GW}}(f) = \frac{8\pi}{3H_0^2} (G\mu)^2 f \sum_{k=1}^\infty P_k\, C_k(f)\,,
\end{equation}
where $H_0 = 67.85~\mathrm{km}~\mathrm{s}^{-1}~\mathrm{Mpc^{-1}}$ is the Hubble constant, $P_k$ is the normalized GW power (in units of $G\mu^2$) emitted in the $k$-th harmonic, and $C_k(f)$ encodes the spectral shape. The GW power for harmonic mode $k$ follows a power law~\cite{Blanco-Pillado:2024aca},
\begin{equation}\label{eq:power_k}
	P_k = \frac{\Gamma\, k^{-q}}{\zeta(q)}\,,
\end{equation}
where $\Gamma \approx 50$ is the total emission power in units of $G\mu^2$, and $\zeta(q)$ is the Riemann zeta function. The spectral index $q$ depends on the GW emission process: $q = 5/3$ for kinks, $q = 4/3$ for cusps, and $q = 2$ for kink-kink collisions. We focus on GW emission from cusps because it is dominant. The weight function $C_k(f)$ is given by
\begin{equation}\label{eq:CK_weight}
	C_k(f) = \frac{2k}{f^2} \int_0^\infty \frac{dz}{H(z)(1+z)^6} \, n\left( \frac{2k}{(1+z)f}, t(z) \right),
\end{equation}
where $n(l, t)$ is the loop number density distribution at cosmic time $t(z)$ and redshift $z$. We assume a standard Friedmann-Robertson-Walker cosmology with
\begin{equation}\label{eq:Hubble}
	H(z) = H_0 \sqrt{1 - \Omega_M - \Omega_R + \Omega_M (1 + z)^3 + \Omega_R \mathcal{C}(z)(1 + z)^4}\,,
\end{equation}
where $\Omega_M = 0.3081$ and $\Omega_R = 1.291\times 10^{-5}$~\cite{Planck:2018vyg}. The correction factor $\mathcal{C}(z)$ accounts for the redshift evolution of the relativistic degrees of freedom.


We employ the analytical approximations derived in Ref.~\cite{Marfatia:2023fvh}. In the nHz range, relevant for Pulsar Timing Array (PTA) observations, the energy density is given by
\begin{equation} \label{eq:Omega_hsq_PTA_approx}
	h^2 \Omega_{\rm GW}(f)\big|_{\rm PTA} \approx 
    4.2 \times 10^{-9} \left( \frac{f}{f_{\text{yr}}} \right)^{3/2} \frac{\Gamma}{50} \left( \frac{G\mu}{10^{-11}} \right)^2\times
	\sum_{k=1}^{\infty} \frac{k^{-17/6}}{(1 + 2.075\, u_k)^{1.945} \, \zeta(17/6)}\,,
\end{equation}
where $f_{\mathrm{yr}} \simeq 32~\mathrm{nHz}$, and
\begin{equation}\label{eq:uk_fun}
	u_k = \frac{2.89}{2k} \frac{f}{f_{\rm yr}} \, \frac{\Gamma}{50}  \frac{G\mu}{10^{-11}}\,.
\end{equation}
For higher frequencies ($f \gtrsim \mathrm{mHz}$), relevant for space- and ground-based laser interferometers, the spectrum flattens and is approximately
\begin{equation}\label{eq:Omega_hsq_laser}
	h^2 \Omega_{\rm GW}(f)\big|_{\rm laser} \simeq 4.78 \times 10^{-5} \, \mathcal{C} \sqrt{G\mu}\,.
\end{equation}
Setting $\mathcal{C} = 0.8$ is a valid approximation in the frequency range $[10^{-3}, 10]~\mathrm{Hz}$~\cite{Marfatia:2023fvh}. {

\section{Primordial black holes}\label{sec:PBHformation}

During a FOPT, the Universe transitions from the false vacuum to the true vacuum through the process of bubble nucleation. Supercooling occurs when the Universe remains trapped in the false vacuum for an extended period, delaying bubble nucleation to temperatures far below the critical temperature $T_c$. In this regime, bubble expansion is primarily driven by the vacuum energy $\Delta V_\mathrm{eff}$ rather than by radiation. This triggers a period of thermal inflation that persists until the phase transition is complete. Upon completion, the vacuum energy is converted into radiation, which reheats the plasma and marks the onset of the radiation domination era.

Bubble nucleation is an inherently stochastic process, so that different regions of the Universe undergo nucleation at different times. For an average Hubble patch, nucleation occurs at a cosmic time $\tau_{\rm nuc}$. However, late-nucleating patches, labeled by $i$, nucleate at a time $\tau^i_{\rm nuc} > \tau_{\rm nuc}$ and remain vacuum-dominated for an extended period. Consequently, the false vacuum energy density of the late-nucleating patches remains approximately constant, while the energy density of the true-vacuum background, which is dominated by radiation, rapidly decreases as the Universe cools. As a result, late-nucleating patches become overdense relative to the background.

If the overdensity, $\delta(\tau, \tau^i_{\rm nuc})$, in a patch $i$ exceeds a critical threshold $\delta_c$ \cite{Gouttenoire:2023naa}, 
\begin{equation}
	\delta\left(\tau ; \tau^i_{\rm nuc}\right) \equiv \frac{\rho_\mathrm{tot}^{\mathrm{late}}\left(\tau ; \tau^i_{\rm nuc}\right) - \rho_\mathrm{tot}^{\mathrm{bkg}}(\tau)}{\rho_{R}^{\mathrm{bkg}}(\tau)} > \delta_{\mathrm{c}} \,,
\end{equation}
then the patch can collapse into a PBH. Here, $\rho_\mathrm{tot}^{\mathrm{late}}\left(\tau ; \tau^i_{\rm nuc}\right)$ is the total energy density at cosmic time $\tau$ for a patch $i$ that nucleated at time $\tau_{\rm nuc, i}$, and $\rho_\mathrm{tot}^{\mathrm{bkg}}(\tau)$ and $\rho_{R}^{\mathrm{bkg}}(\tau)$ are the total and radiation energy densities of the background true-vacuum regions, respectively. The overdensity in late-nucleating patches increases over time and reaches its peak shortly after the patch percolates. This maximum arises because the energy density of the surrounding true-vacuum regions begins to dilute slightly earlier, while the late-nucleating patches maintain a nearly constant energy density dominated by vacuum energy until they start to percolate and their vacuum energy is gradually converted into radiation. 

It is possible to derive an approximate analytical expression for the probability that a given patch $i$ collapses into a PBH. Assuming strongly supercooled FOPTs with $\alpha > 100$, this probability depends solely on the inverse time duration of the FOPT $\beta/H(T_p)$ and on the critical threshold $\delta_c$ as~\cite{Gouttenoire:2023naa}
\begin{equation}\label{eq:prob_PBH}
    \mathcal{P}_\mathrm{coll} \approx \exp{-a_1 \qty(\frac{\beta}{H(T_p)})^{a_2} (1 + \delta_c)^{a_3 [\beta/H(T_p)]}} \,,
\end{equation}
where $a_1=0.56468$, $a_2=1.266$ and $a_3 = 0.6639$ are dimensionless fitting parameters. Simulations suggest that 
$\delta_c$ ranges between 0.4 and 0.66 if the origin of the overdensities is similar to that of inflation~\cite{Carr:1975qj, Musco:2020jjb,Escriva:2019phb, Stamou:2023vxu}. 

The fraction of dark matter in the form of PBHs, normalized to the total DM relic abundance today, is given by~\cite{Gouttenoire:2023naa}\footnote{A similar PBH production mechanism was presented in Ref.~\cite{Lewicki:2023ioy}. The main  difference is that Ref.~\cite{Gouttenoire:2023naa} assumes that collapsing patches can contain true vacuum bubbles, while Ref.~\cite{Lewicki:2023ioy} assumes they can not. This affects the threshold value of $\beta/H(T_p)$ required to fully account for DM. In Ref.~\cite{Lewicki:2023ioy}, $f_\mathrm{PBH} = 1$ for $\beta/H(T_p) \gtrsim 3.8$, whereas in Ref.~\cite{Gouttenoire:2023naa} $\beta/H(T_p) \gtrsim 8$.}
\begin{equation}\label{eq:abundance_PBH}
    f_\textrm{PBH} \approx \frac{\mathcal{P}_\mathrm{coll}}{2.2\times 10^{-8}}\, \frac{T_\mathrm{RH}}{0.14~\mathrm{GeV}}\,,
\end{equation}
and the PBH mass is~\cite{Gouttenoire:2023naa}
\begin{equation}\label{eq:pbhmasses}
    M_{\rm PBH} = M_\odot\left(\frac{20}{g_{*}(T_\textrm{RH})}\right)^{1/2}\left(\frac{0.14\,\textrm{GeV}}{T_\textrm{RH}}\right)^2\,.
\end{equation}
Given the strong sensitivity of $f_\mathrm{PBH}$ on $\delta_c$, we treat $\delta_c$ as a free parameter in the range $[0.40,0.66]$. 

A recent study~\cite{Franciolini:2025ztf} revises $\delta_c$ upward by an order of magnitude, which leads to drastically reduced PBH abundances. To achieve $f_\mathrm{PBH} \sim 1$ requires $\beta/H(T_p) \lesssim 2.5$. However, such low values of $\beta/H(T_p)$ are not obtainable in our model, as the FOPT does not percolate and cannot complete. It is worth noting that additional contributions not included in Ref.~\cite{Franciolini:2025ztf}, particularly the energy density in bubble walls, may play an important role in the formation of PBHs in supercooled FOPTs~\cite{Flores:2024lng, Hashino:2025fse, Lewicki:2023ioy}. However, it remains to be demonstrated whether the bubble wall contribution validates the conclusions of Ref.~\cite{Gouttenoire:2023naa} and negates the conclusions of Ref.~\cite{Franciolini:2025ztf}. While this issue is being resolved, we optimistically take 
$\delta_c \in [0.40,0.66]$.

\section{Primordial magnetic fields}
\label{sec:pmf}

The initial conditions of the magnetic fields generated during a FOPT play a crucial role in determining the evolution of MHD turbulence. In particular, the initial magnetic helicity has a significant impact on the MHD decay process~\cite{Banerjee:2004df}.  MHD decay refers to the dissipation of magnetic energy and the relaxation of turbulent motions in a conducting plasma, governed by MHD equations. Magnetic helicity density characterizes the twist and linkage of magnetic field lines, defined as $H^M \equiv \langle \mathbf{A} \cdot \mathbf{B} \rangle$, where $\mathbf{B} = \nabla \times \mathbf{A}$, with $\mathbf{A}$ and $\mathbf{B}$ denoting the vector potential and magnetic field, respectively. For helical fields, this decay is typically slower because helicity is approximately conserved. Consequently, a maximally helical field increases its correlation length as magnetic energy dissipates, driving an inverse cascade of energy from small to large scales. This process generates coherent magnetic structures significantly larger than the initial energy injection scale. Such dynamics may have been critical for the persistence and evolution of primordial magnetic fields, enabling their survival and large-scale coherence in the present universe. In contrast, non-helical fields decay more rapidly, with energy dissipating at small scales through viscous and resistive effects. 

A maximally helical magnetic field satisfies $|H^M| = (\pi/k)\langle|\mathbf{B}|^2\rangle$ at each scale $k$, where $\langle|\mathbf{B}|^2\rangle/2$ is the magnetic energy density. For such a field, the decay of MHD turbulence during the radiation-dominated epoch follows a power-law dependence on conformal time $\eta$~\cite{Brandenburg:2017rnt}:
\begin{equation}
B \sim \eta^{-1/3}\,, \quad \lambda \sim \eta^{2/3}\,,
\end{equation}
where $B$ is the magnetic field strength and $\lambda$ is the correlation length of the magnetic field. A non-helical magnetic field has $H^M= 0$ at all scales, implying that its Fourier modes satisfy $\mathbf{B}_\mathbf{k} \cdot \mathbf{A}_\mathbf{k}^* = 0$ for all wavenumbers $\mathbf{k}$. Then, inverse transfer can still occur due to kinetic helicity in the plasma~\cite{Brandenburg:2017rnt}. Kinetic helicity $H^M_\mathrm{kin} \equiv \langle \mathbf{v} \cdot \boldsymbol{\omega} \rangle$ measures the correlation between a fluid's velocity $\mathbf{v}$ and vorticity $\boldsymbol{\omega} \equiv \nabla \times \mathbf{v}$.  
In plasmas, a non-zero $H_\mathrm{kin}^M$ can induce an inverse transfer of magnetic energy, even for non-helical magnetic fields ($H^M = 0$), by exciting helical plasma motions that reorganize the magnetic field structure at larger scales~\cite{Brandenburg:2017rnt}. However, the strength of this effect remains debated~\cite{Armua:2022rvx}. In this case, $B$ and $\lambda$ scale as
\begin{equation}
B \sim \eta^{-1/2}\,, \quad \lambda \sim \eta^{1/2} \,.
\end{equation}
These scaling laws are applicable during the radiation-dominated epoch, during which the scale factor evolves as $a \sim \eta$. After recombination, the magnetic field strength decays as $B \sim a^{-2}$ due to the expansion of the Universe. To encompass the helical and non-helical cases, we introduce the generalized parameters,
\begin{equation}
p_b = \frac{2}{b+3}(b+1)\,, \quad q_b = \frac{2}{b+3}\,, 
\end{equation}
yielding the power-law scalings
\begin{equation}
B \sim \eta^{- p_b / 2}\,, \quad \lambda \sim \eta^{q_b}\,,
\end{equation}
where $b=0$ and $b=1$ correspond to the helical and non-helical cases, respectively. These two scenarios serve as benchmarks that set upper and lower bounds on the amplitude for the magnetic fields. The magnetic energy density at the time of percolation is estimated as~\cite{Ellis:2019tjf, RoperPol:2023bqa}
\begin{equation}
\rho_{B,*} = 0.1\frac{ \kappa_{\rm col} \alpha}{1+ \alpha} \rho_{*}\,,
\end{equation}
where $\rho_{*} \approx \rho_{\rm vac}$ is the total energy density at percolation. This expression assumes a $10\%$ efficiency in converting plasma motion into magnetic fields~\cite{Kahniashvili:2009qi, Durrer:2013pga, Brandenburg:2017neh}.

The present-day magnetic field spectrum is given by~\cite{Ellis:2019tjf}
\begin{align}\label{eq:mag_fields}
&B_0(\lambda) = \left(\frac{a_{\rm RH}}{a_{\rm rec}}\right)^{p_b/2} \left(\frac{a_{\rm RH}}{a_0}\right)^2 \sqrt{\frac{17}{10}\,\kappa_h\,\rho_{B,*}} 
\begin{cases}
\left(\frac{\lambda}{\lambda_0}\right)^{-5/2},  & \lambda \geq \lambda_0 \\
\left(\frac{\lambda}{\lambda_0}\right)^{1/3}, & \lambda < \lambda_0\,,
\end{cases}
\end{align}
where the coherence scale of the magnetic field today is~\cite{Ellis:2019tjf}
\begin{equation}\label{eq:coherence_today}
\lambda_0 = \frac{a_0}{a_{\rm RH}} \left(\frac{a_{\rm rec}}{a_{\rm RH}}\right)^{q_b} \lambda_{*}\,,
\end{equation}
and the initial correlation length $\lambda_{*}$ is determined by the bubble size at percolation~\cite{Caprini:2019egz}:
\begin{equation}
\lambda_{*} = \frac{(8 \pi)^{1/3}}{H(T_p)} \left( \frac{\beta}{H(T_p)} \right)^{-1}\,,
\end{equation}
assuming a wall velocity $v_w=1$ for strongly supercooled transitions. The redshift factors are given by
\begin{align}
\frac{a_{\rm RH}}{a_0} &= 8 \times 10^{-14} \left( \frac{100}{g_{*}(T_{\rm RH})} \right)^{1/3} \frac{\textrm{GeV}}{T_{\rm RH}}\,, \\
\frac{a_{\rm RH}}{a_{\rm rec}} &= 8 \times 10^{-11} \left( \frac{100}{g_{*}(T_{\rm RH})} \right)^{1/3} \frac{\textrm{GeV}}{T_{\rm RH}} \,.
\label{eq:redshift_magfields}
\end{align}
After the phase transition ends, a fraction of the bubble wall energy of the dark scalar is transferred to the Higgs field, due to the Higgs portal coupling $\lambda_{\sigma h}$. To account for this we introduce an efficiency factor,
\begin{equation}\label{eq:kappa_h}
    \kappa_h = 1 - \frac{\mathcal{V}_\mathrm{eff}(v_\sigma(T_p),v)}{\mathcal{V}_\mathrm{eff}(v_\sigma(T_p),0)}\,,
\end{equation}
where $\mathcal{V}_\mathrm{eff}$ is the 2-field effective potential, where the first argument corresponds to the $\sigma$ direction, and the second argument corresponds to the Higgs direction. The vev is evaluated at $T_p$. 

\section{Numerical results}\label{sec:results}
\begin{table}[t]
	\centering
    \captionsetup{justification=raggedright}
    \begin{tabular}{c|c|c|c|c|c}
		\toprule
		$M_{h_2}$ (GeV) & $g_{L}$ & $x_\mathcal{H}$ & $x_\sigma$ & $(\bm{y_\sigma})_{ii}$ & $\delta_c$\\
		\midrule
		$  \left[ 150 , 10^{18} \right]$ & $ \left[0.20, 1\right]$ & $\left[-2, 2\right]$ & $\left[0, 5\right]$ & $\left[10^{-10}, 1\right]$ & $[0.40,0.66]$
		\\
		\bottomrule
	\end{tabular}%
	\caption{\small Ranges of input parameters, defined at $Q = M_{\mathrm{Z}^0}$, in our numerical scan. All parameters are sampled linearly except for $M_{h_2}$ and the Majorana neutrino Yukawa couplings $(\bm{y_\sigma})_{ii}$, which are sampled logarithmically. The gauge charges $x_\mathcal{H}$ and $x_\sigma$ take rational values only.}
	\label{tab:num_ranges_scan}
\end{table}
We perform a numerical scan over the model parameters, defined at the electroweak scale $Q = M_{\mathrm{Z}^0}$, in the ranges specified in \cref{tab:num_ranges_scan}. The calculation of the GW signal follows the procedure of Ref.~\cite{Goncalves:2024lrk}. To satisfy constraints on the extra effective number of neutrino species, $\Delta N_\mathrm{eff}$, we require the dark sector to remain in thermal equilibrium with the SM one after reheating. We focus on scenarios where percolation occurs after the QCD phase transition ($T_p > 0.17~\mathrm{GeV}$), and only consider the parameter space that yields an SGWB, which is observable at current and planned GW experiments, i.e.~satisfying $h^2 \Omega_\mathrm{GW} > 10^{-17}$. Cosmic microwave background data impose an upper bound on the string tension $G\mu \lesssim 10^{-7}$ \cite{Planck:2015fie}, while the 15-year NANOGrav dataset excludes 
$G\mu \gtrsim 1.32\times 10^{-10}$ for GWs emitted from cusps, including the contribution from supermassive black hole binaries~\cite{NANOGrav:2023hvm}.

\subsection{Phenomenology of the conformal \texorpdfstring{$\mathrm{U(1)_{B-L}}$ model}{U(1)BL} }

We first study the $\mathrm{U(1)_{B-L}}$ model, which corresponds to the charges $x_\mathcal{H} = 0$ and $x_\sigma = 2$.

\subsubsection{Phase transition thermodynamics and PBH formation}

In \cref{fig:PHBs_proj_plots_Thermo}, we present predictions for the PBH mass and relic abundance in terms of the thermodynamic parameters of the phase transition $\alpha$, $\beta/H(T_p)$,  $T_p$, and $T_\mathrm{RH}$.
\begin{figure*}[t]
	\centering
	\subfloat{\includegraphics[width=0.5\textwidth]{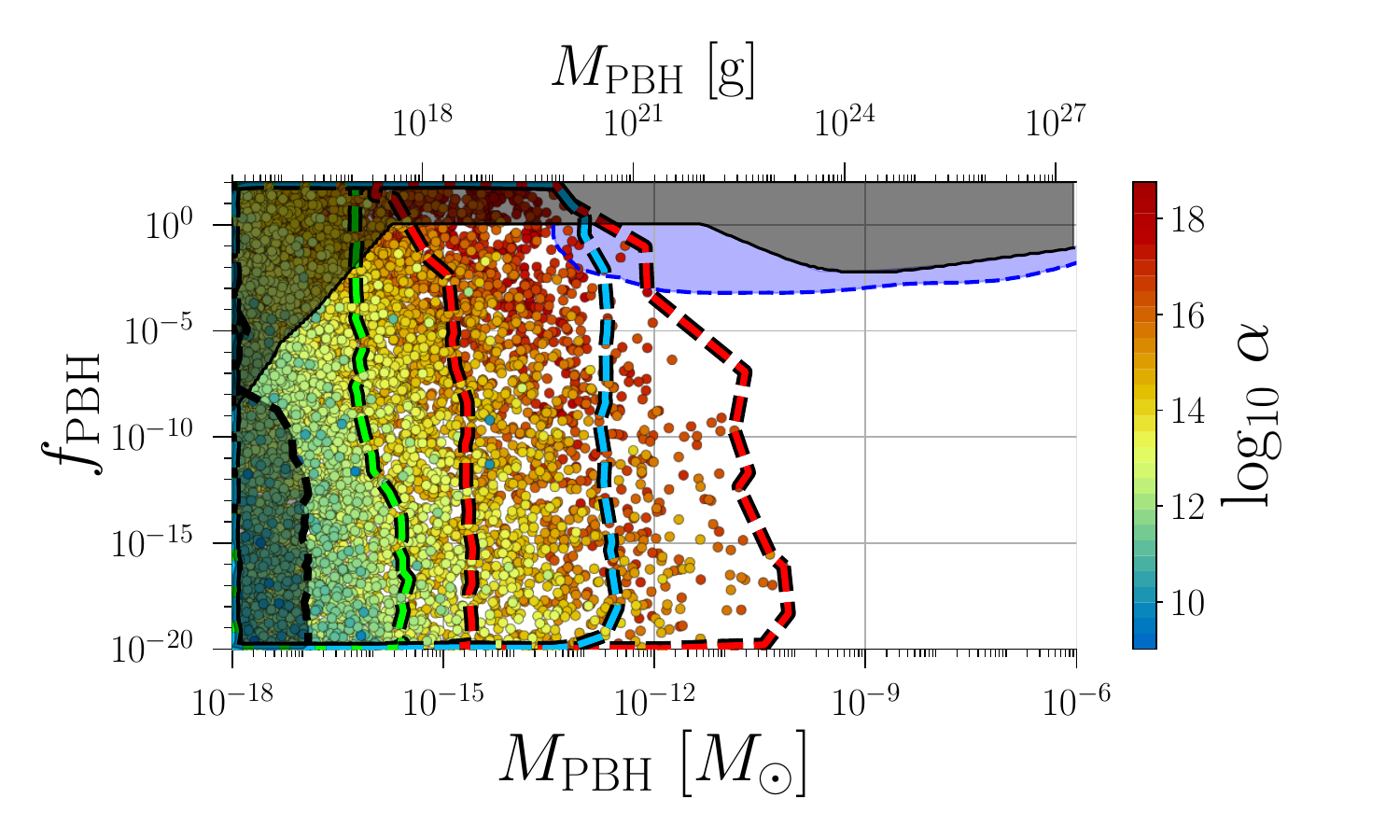}}
    \subfloat{\includegraphics[width=0.5\textwidth]{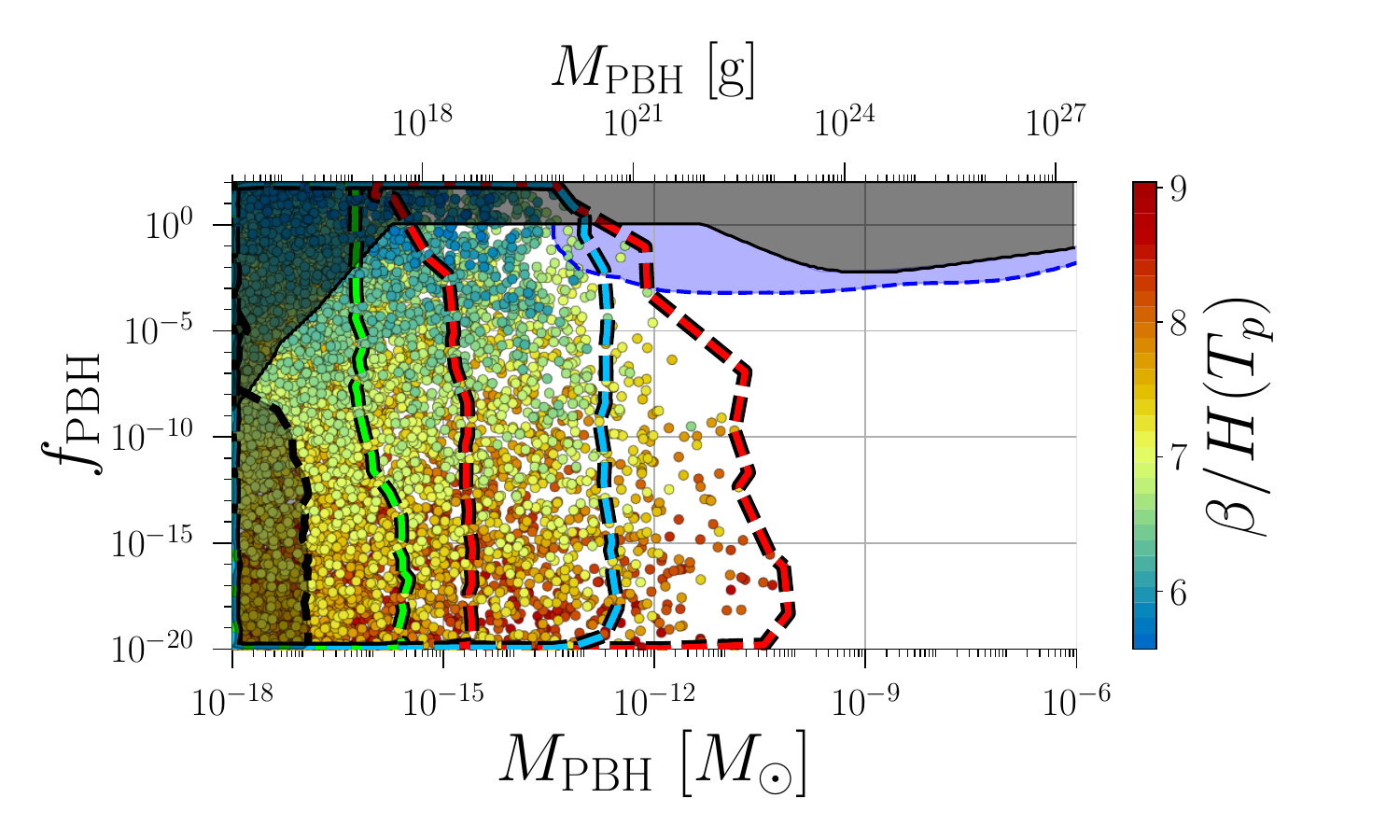}} \\
    \subfloat{\includegraphics[width=0.5\textwidth]{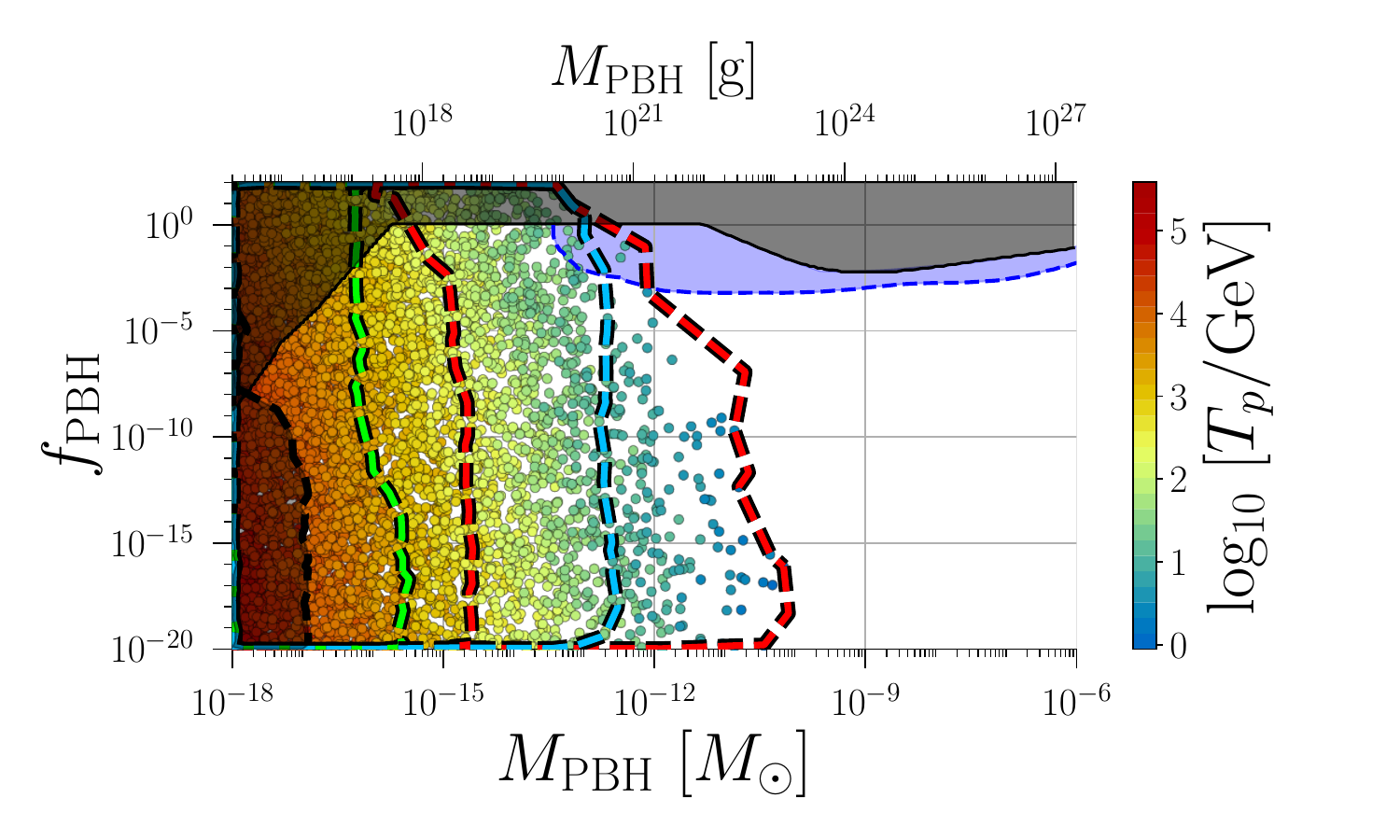}} 
    \subfloat{\includegraphics[width=0.5\textwidth]{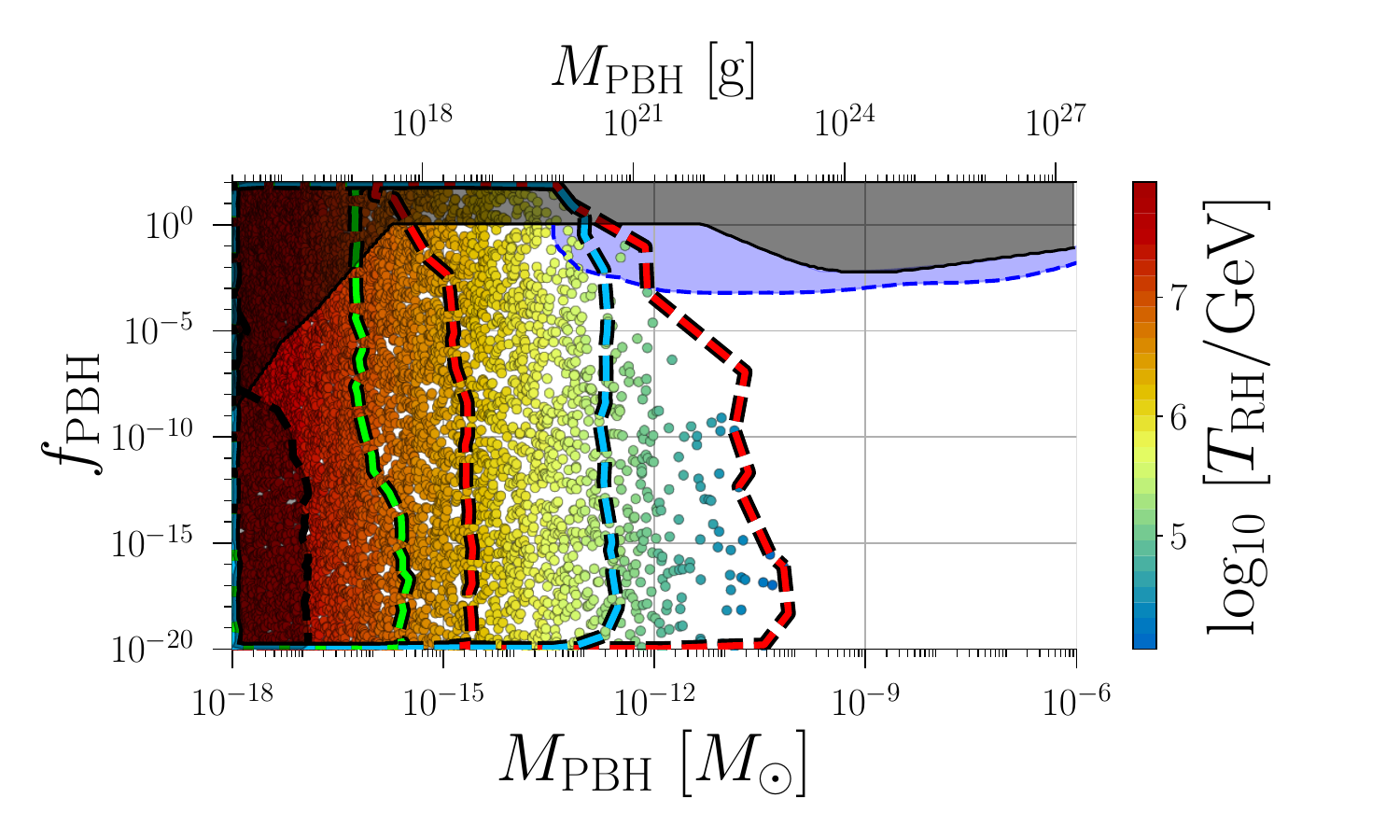}} \\
    \caption{\small Scatter plots of the PBH abundance, $f_\mathrm{PBH}$, as a function of $M_\mathrm{PBH}$. The color scales indicate the FOPT strength $\alpha$ (top-left panel), its inverse duration $\beta/H(T_p)$ (top-right panel), the percolation temperature $T_p$ (bottom-left), the reheating temperature $T_\mathrm{RH}$ (bottom-right panel). The shaded band enclosed by the black dashed contour is excluded by LVK data ($\mathrm{SNR_{LVK}} > 10$). The green, blue and red dashed contours enclose the regions within the reach of LIGO-O5, ET and LISA with $\mathrm{SNR} > 10$, respectively. The region above the solid black curve is excluded by combination of overabundant PBH production, $\gamma$-ray data and microlensing data. A dedicated microlensing survey of M31 by the Roman Space Telescope is sensitive to the region above the dashed blue curve~\cite{Drlica-Wagner:2022lbd}. 
    }
	\label{fig:PHBs_proj_plots_Thermo}
\end{figure*}
The PHB mass is displayed in grams on the top-horizontal axis and in solar masses on the bottom-horizontal axis.
The signal-to-noise ratio (SNR) is defined as
\begin{equation}\label{eq:SNR_calc}
    \mathrm{SNR} = \sqrt{\mathcal{T} \int df~ \frac{\Omega_{\mathrm{GW}}^2(f)}{\Omega_{\mathrm{Sens}}^2(f)}} \,,
\end{equation}
where $\Omega_{\mathrm{GW}}(f)$ is the predicted SGWB from the FOPT and cosmic strings, 
$\Omega_{\mathrm{Sens}}(f)$ is the experimental sensitivity, and $\mathcal{T} = 4~\mathrm{years}$ is the expected exposure time of future experiments. The SNR for Ligo-Virgo-Kagra (LVK) is obtained from current data~\cite{KAGRA:2021kbb}. The shaded band enclosed by the black dashed contour, 
with $M_\mathrm{PBH} \lesssim 1 \times 10^{-17} M_\odot$, is excluded by data from LVK's third observational run (O3) because the predicted SNR in this band exceeds 10.
The black solid contour represents a combined exclusion region derived from constraints on the overproduction of PBHs, the extragalatic $\gamma$-ray flux via Hawking radiation, and microlensing observations. The expected sensitivity for the Nancy Grace Roman Space Telescope is depicted by the blue shaded region~\cite{Drlica-Wagner:2022lbd}. The spread in points of a given color is due to variations in the critical threshold $\delta_c$.

Strong, sustained supercooling is a key requirement for PBH formation. We find $\alpha \gtrsim 10^9$ and $\beta/H(T_p) < 9$. Due to the exponential dependence of $f_\mathrm{PBH}$ on $\beta/H(T_p)$ in Eq.~\eqref{eq:abundance_PBH}, the DM relic abundance is saturated ($f_\mathrm{PBH} = 1$) in a narrow range $6 \lesssim \beta/H(T_p) \lesssim 7$ for PBH masses within $[10^{-16},10^{-11}]~M_\odot$. In this region, we also observe the strongest FOPTs, with $10^{15} \lesssim \alpha \lesssim 10^{18}$.

While $\beta/H(T_p)$ primarily determines $f_\mathrm{PBH}$, the reheating temperature sets the PBH mass via \cref{eq:pbhmasses}, according to which $M_\mathrm{PBH} \propto T_\mathrm{RH}^{-2}$. Consequently, lower values of $T_\mathrm{RH}$ result in heavier PBHs, as can be seen in the bottom-right panel. Recall that in the presence of strong supercooling $T_p \ll T_\mathrm{RH}$. For a fixed PBH mass and a correspondingly fixed reheating temperature, $\alpha T_p^4$ is a constant, which explains the observed color trends in the left panels. Note that larger PBH masses are obtained for stronger phase transitions. 

For reheating temperatures at the TeV scale, we find $M_\mathrm{PBH} \sim (10^{-12}-10^{-10})M_\odot$, while reheating temperatures near $10^8~\mathrm{GeV}$ yield lighter PBHs with $M_\mathrm{PBH} \sim 10^{-18}M_\odot$. The upper bound on PBH masses, $M_\mathrm{PBH} \lesssim 10^{-10}M_\odot$, arises from our requirement that phase transitions occur above the QCD scale (i.e., $T_p > 0.17~\mathrm{GeV}$). Note that all the points fall within the sensitivity reach of near-future GW experiments, as indicated by the green dashed contour (for LIGO-O5), the blue dashed contour (for ET) and the red dashed curve (for LISA). 

\subsubsection{Impact of the critical density threshold \texorpdfstring{$\delta_c$}{deltaC}}

\begin{figure*}[t]
	\centering
    \hspace*{-4em}
    \subfloat{\includegraphics[width=\textwidth]{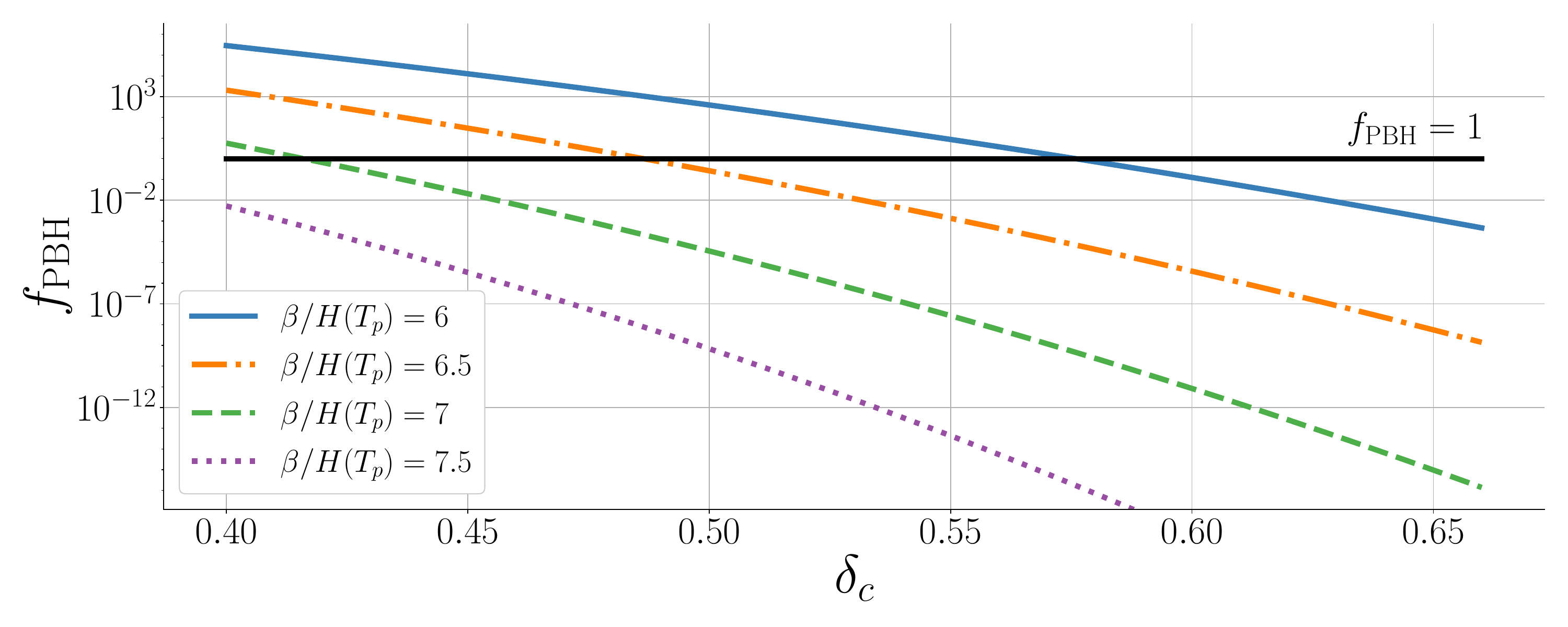}} \\
    \subfloat{\includegraphics[width=\textwidth]{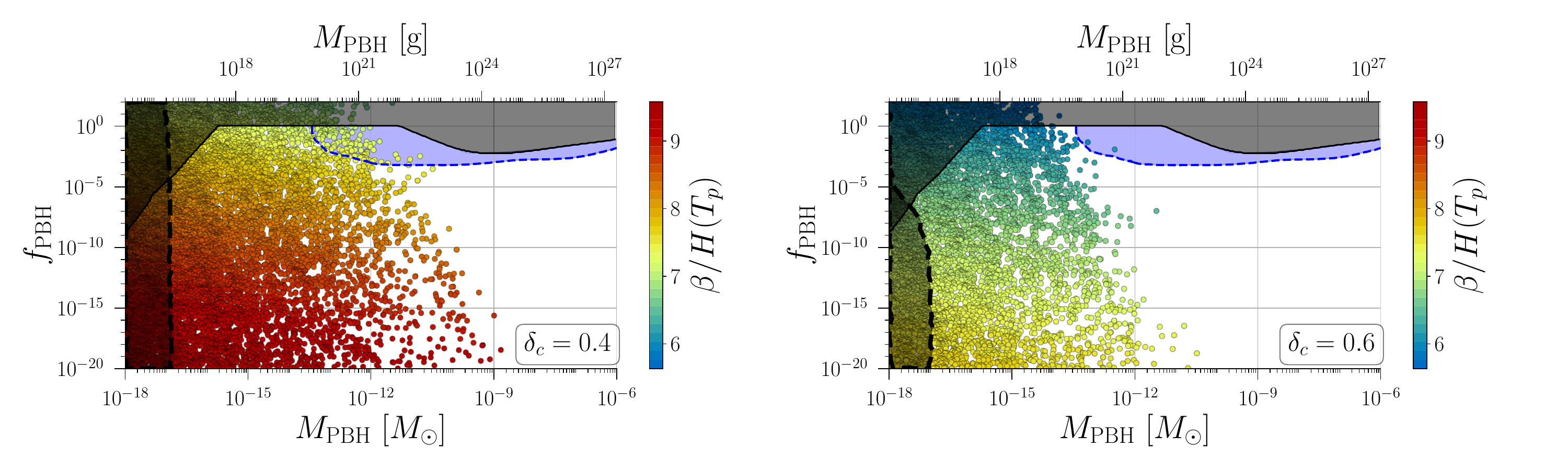}}
    \caption{Top panel: The DM abundance $f_\mathrm{PBH}$ as a function of the critical threshold $\delta_c$ for different values of the inverse time duration $\beta/H(T_p)$ and a fixed reheating temperature $T_\mathrm{RH} = 10^6~\mathrm{GeV}$. Bottom panels: Similar to Fig.~\ref{fig:PHBs_proj_plots_Thermo}, but the color scale represents the inverse duration $\beta/H(T_p)$ of the FOPT for $\delta_c = 0.4$ (left) and $\delta_c = 0.6$ (right). }
	\label{fig:PHBs_deltac}
\end{figure*}

The exponential dependence of $f_\mathrm{PBH}$ on the critical density threshold $\delta_c$ is highlighted in the top panel of \cref{fig:PHBs_deltac} for various values of $\beta/H(T_p)$, with the reheating temperature fixed at $T_\mathrm{RH} = 10^6~\mathrm{GeV}$. The value of $f_\mathrm{PBH}$ can vary by up to ten orders of magnitude within the range of $\delta_c$ obtained from simulations. This variation significantly affects the value of $\beta/H(T_p)$ required for all DM to be PBHs. Specifically, larger values of $\delta_c$ correspond to smaller values of $\beta/H(T_p)$. 

Although the reheating temperature influences $f_\mathrm{PBH}$ as in \cref{eq:abundance_PBH}, its precise value is irrelevant due to the strong correlation between $\beta/H(T_p)$ and $\delta_c$. This is confirmed in the bottom-right panel of \cref{fig:PHBs_proj_plots_Thermo}, which shows that varying $T_\mathrm{RH}$ has virtually no impact on $f_\mathrm{PBH}$.

\subsubsection{Impact of the model parameters on PBH formation}

In \cref{fig:PHBs_proj_plots_HEP}, we present the same projections as in \cref{fig:PHBs_proj_plots_Thermo} but focusing on the underlying model parameters.
\begin{figure*}[t]
	\centering
    \subfloat{\includegraphics[width=0.5\textwidth]{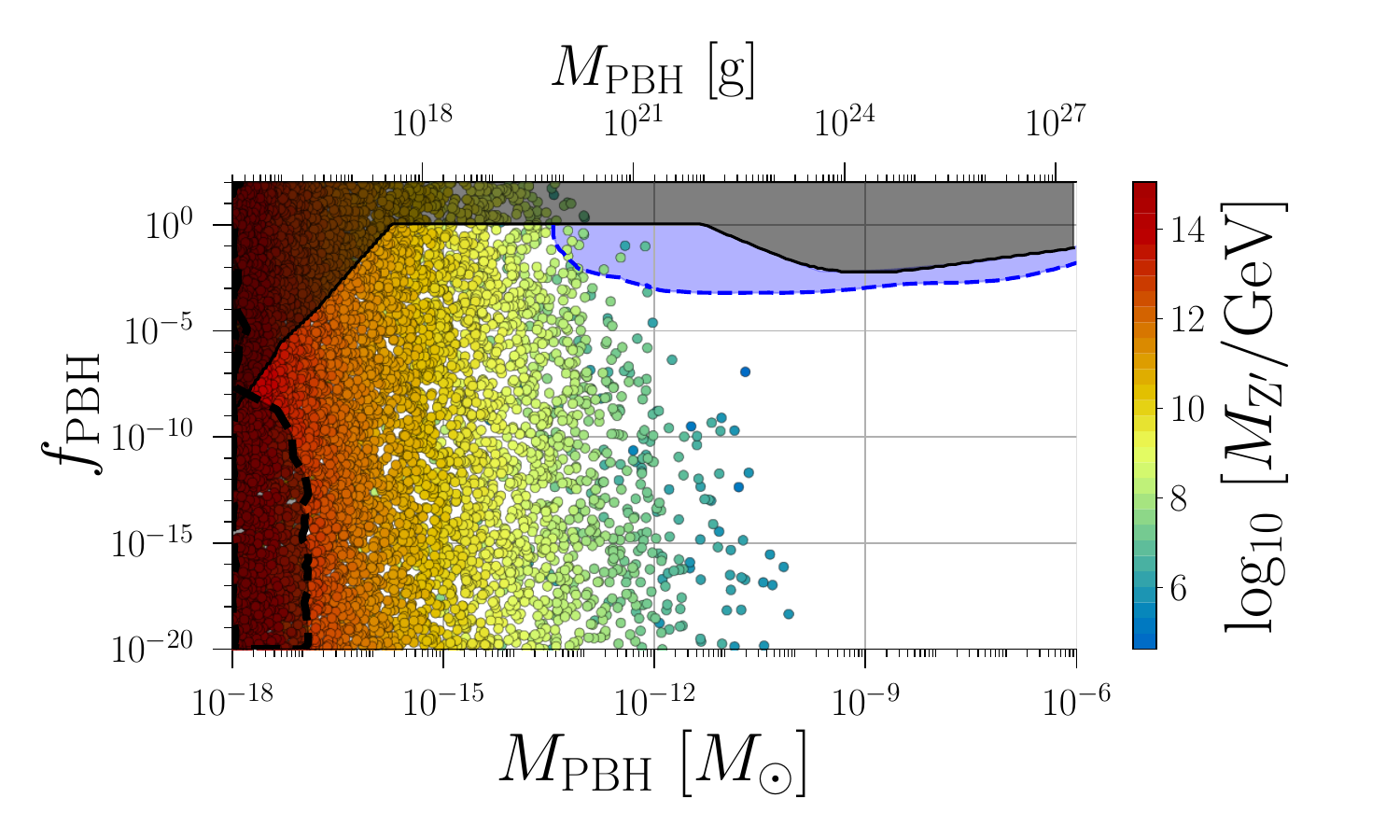}} 
    \subfloat{\includegraphics[width=0.5\textwidth]{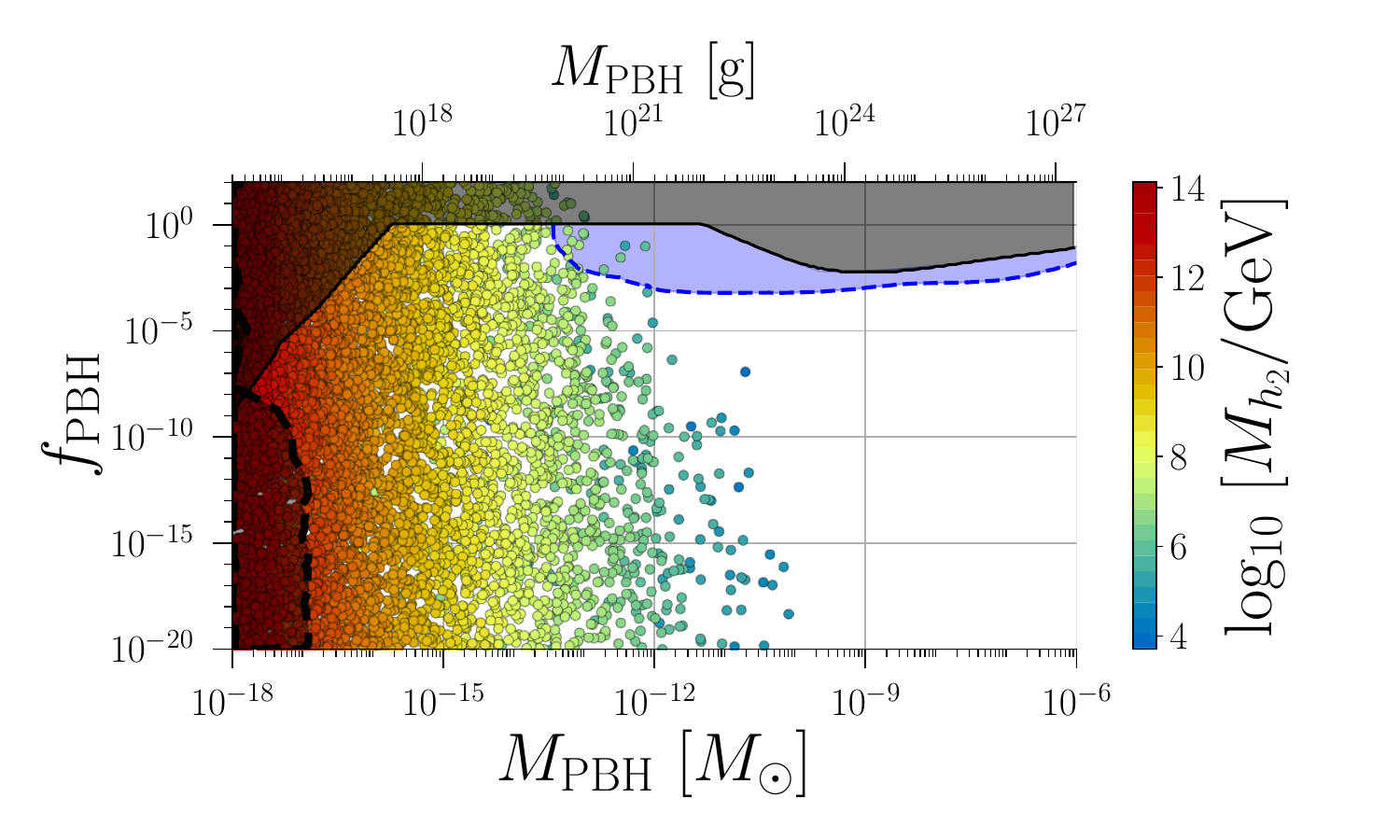}} \\
    \subfloat{\includegraphics[width=0.5\textwidth]{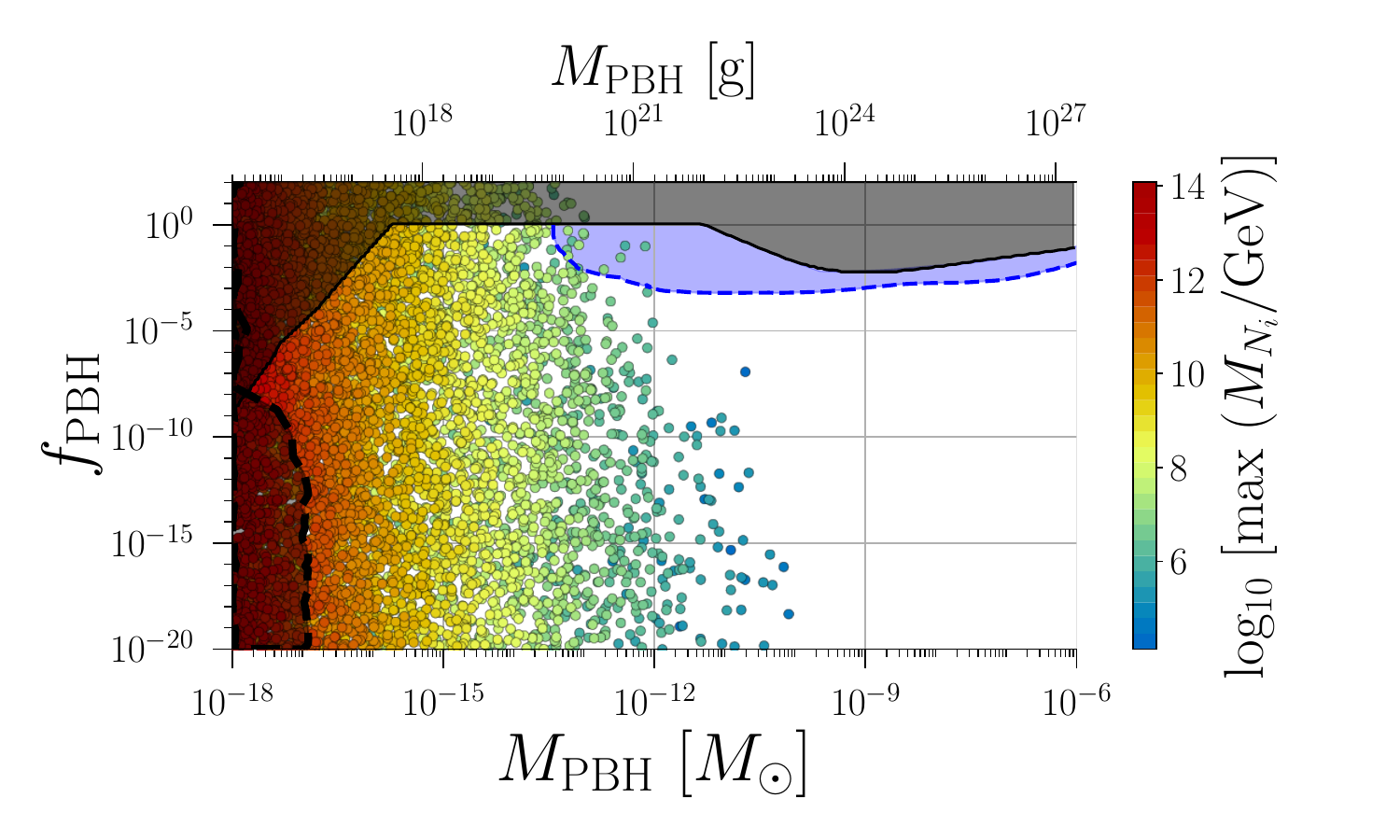}} 
    \subfloat{\includegraphics[width=0.5\textwidth]{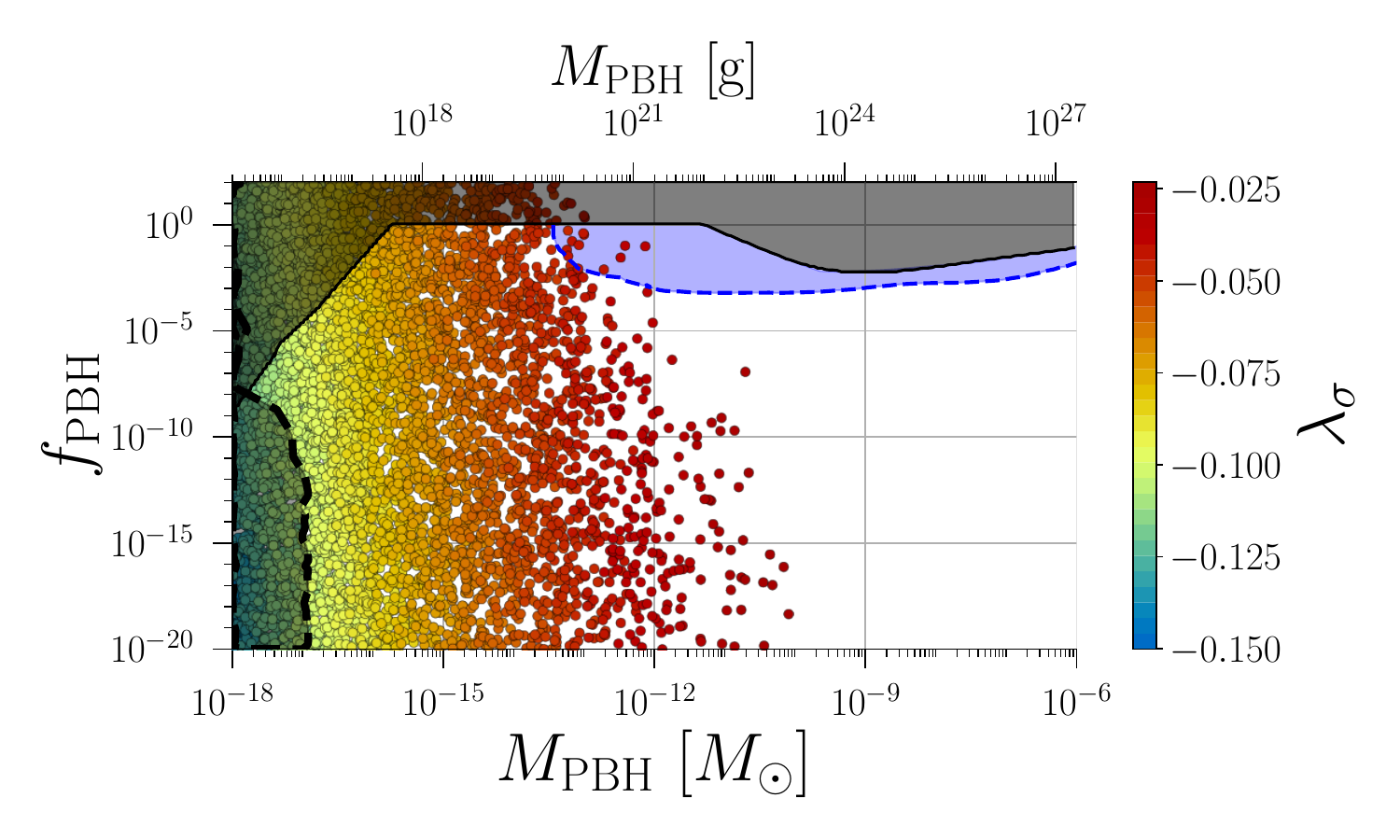}} \\
    \subfloat{\includegraphics[width=0.5\textwidth]{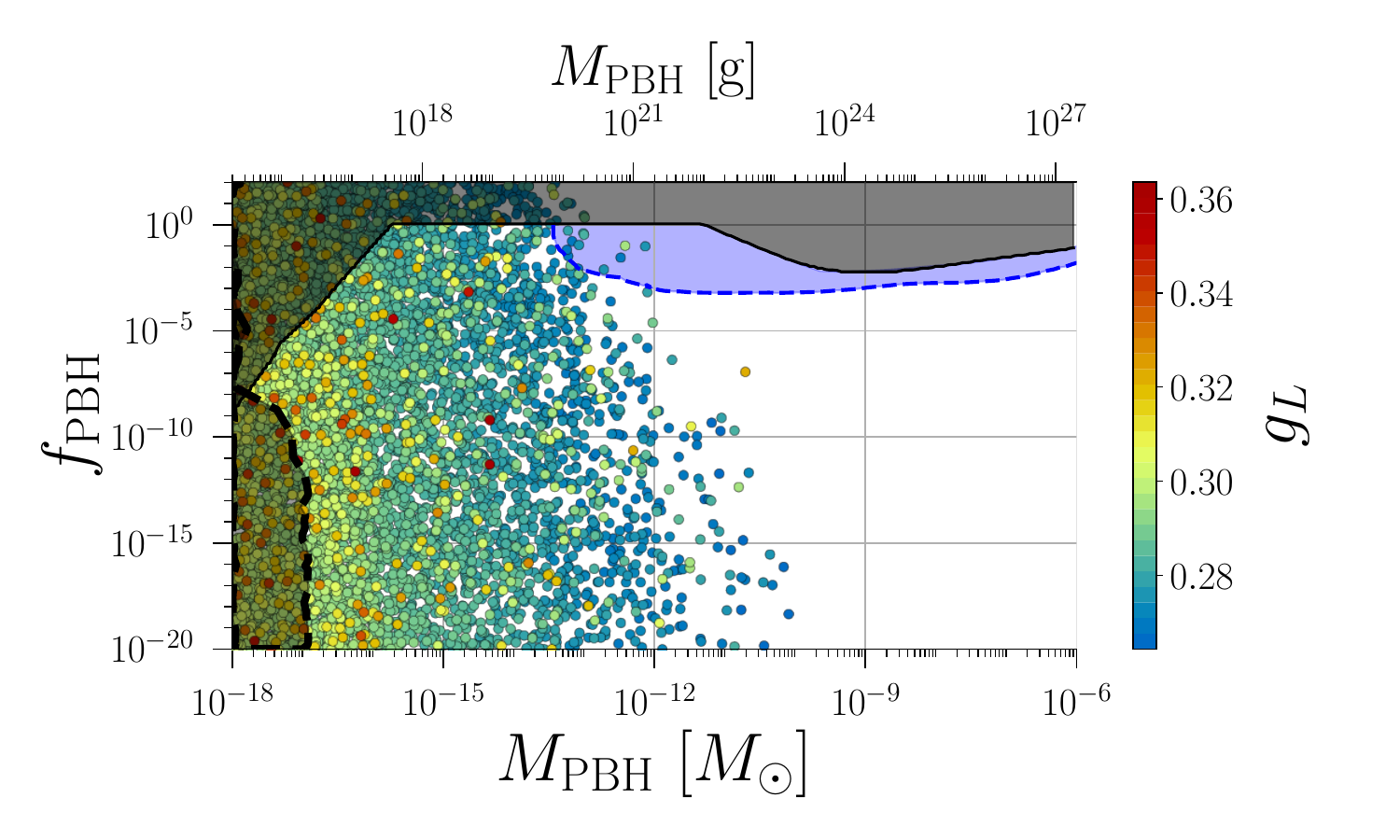}}  
    \subfloat{\includegraphics[width=0.5\textwidth]{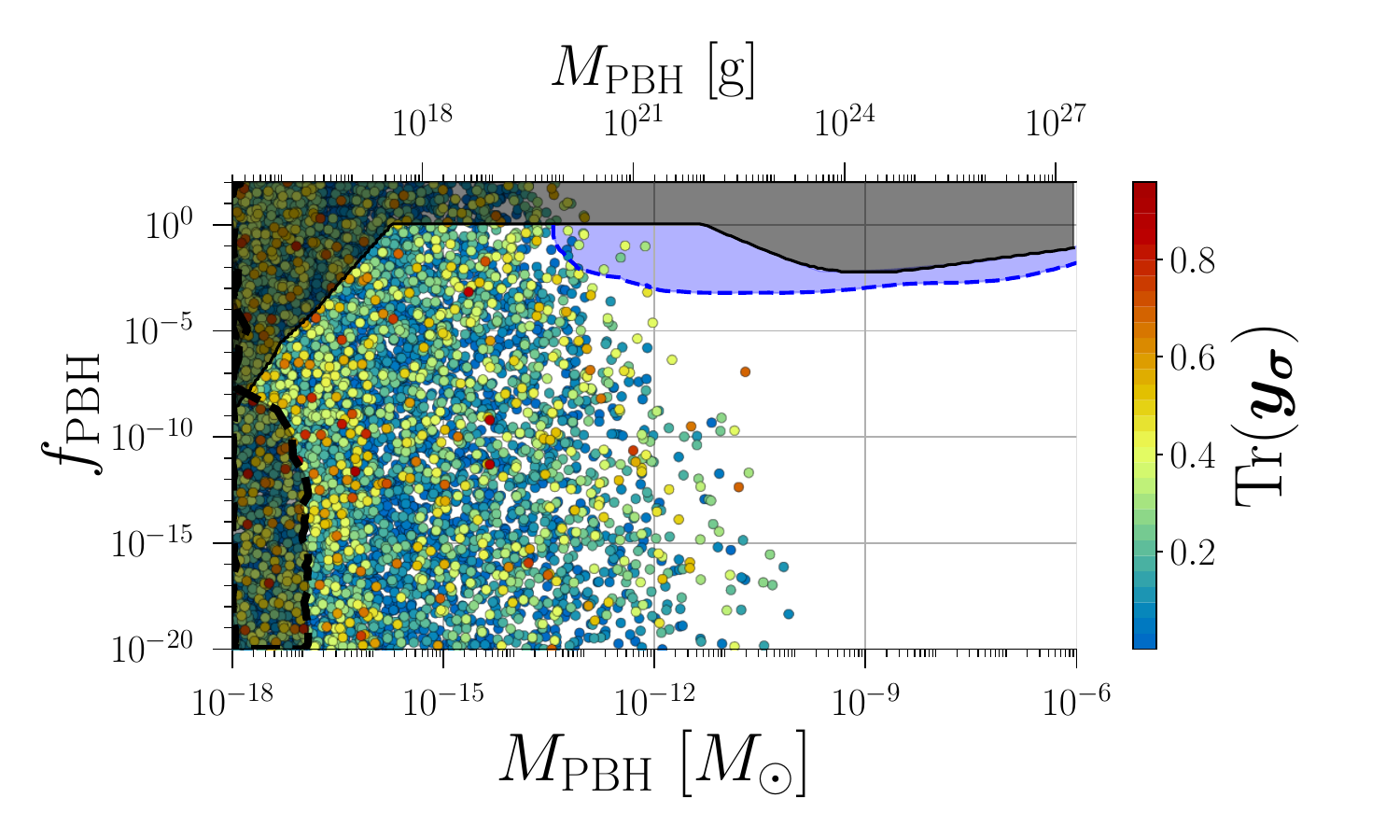}} 
    \caption{\small Similar to \cref{fig:PHBs_proj_plots_Thermo}, but the color scales represent the $\mathrm{Z^\prime}$ boson mass (top-left panel), the heavy Higgs $h_2$ mass (top-right panel), the largest right-handed neutrino mass (middle-left panel), the Majoron self-interaction coupling $\lambda_\sigma$ (middle-right panel), the gauge coupling $g_L$ (bottom-left panel) and the trace of the Yukawa matrix $\bm{y_\sigma}$ (bottom-right panel). }
	\label{fig:PHBs_proj_plots_HEP}
\end{figure*}
A strong correlation is observed between the PBH mass and the masses of the $\mathrm{Z^\prime}$, heavy Higgs boson and heavy neutrinos. The PBH mass scale is inversely related to the mass scale of these particles, $M_{\mathrm{Z^\prime}} \sim 10M_{h_2}$. At the upper end of PBH masses, $M_\mathrm{PBH} \sim 10^{-10}M_\odot$, the corresponding mass scale is $M_\mathrm{Z^\prime} \sim 10^4~\mathrm{GeV}$. At the lower end of the PBH mass spectrum, with $M_\mathrm{PBH} \sim 10^{-18} M_\odot$, the mass scale  approaches the GUT scale. This relationship arises because $M_\mathrm{Z^\prime} \sim T_{\rm RH}$. This is also reflected in the color gradient for $\lambda_\sigma$ (middle-right panel), since, within the parameter space relevant for PBH production, $M^2_{h_2} \sim -0.17 \lambda_\sigma v_\sigma^2$.  

The gauge coupling $g_L$ exhibits only a weak correlation with the PBH observables (bottom-left panel), although only a narrow range of $g_L \in [0.25,0.36]$ can produce the small values of $\beta/H(T_p)$ required for PBH formation~\cite{Goncalves:2024lrk}.

A key result of \cref{fig:PHBs_proj_plots_HEP} is the possibility of establishing a relationship between the model parameters and the conditions to obtain an observable SGWB and the entire DM relic abundance in the form of PBHs. We find $M_\mathrm{Z^\prime} \in [10^6,10^{12}]~\mathrm{GeV}$, with $g_L \in [0.25,0.30]$. This also implies a type-I seesaw scale within $[10^6,10^{10}]~\mathrm{GeV}$, which matches the mass range for the heavy Higgs boson $h_2$, along with $\lambda_\sigma \in [-0.075, -0.025]$. The neutrino Yukawa couplings characterized by $\mathrm{Tr}(\bm{y_\sigma})$ (bottom-right panel) remain largely unconstrained and can take values satisfying $10^{-6} \lesssim \mathrm{Tr}(\bm{y_\sigma}) \lesssim 1$. 
\subsubsection{Correlated signals: \texorpdfstring{$\gamma$}{gamma}--rays, microlensing, SGWB and magnetic fields}

In \cref{fig:PHBs_proj_plots_SNR}, we present the predicted SNR for the SGWB at LISA~\cite{LISA:2017pwj} (top-left panel), Einstein Telescope (ET)~\cite{Punturo:2010zz} (top-right panel), LIGO O5~\cite{LIGOScientific:2014pky} (bottom-left panel) and LVK~\cite{KAGRA:2021kbb} (bottom-right panel). 
\begin{figure*}[t]
	\centering
    \subfloat{\includegraphics[width=0.5\textwidth]{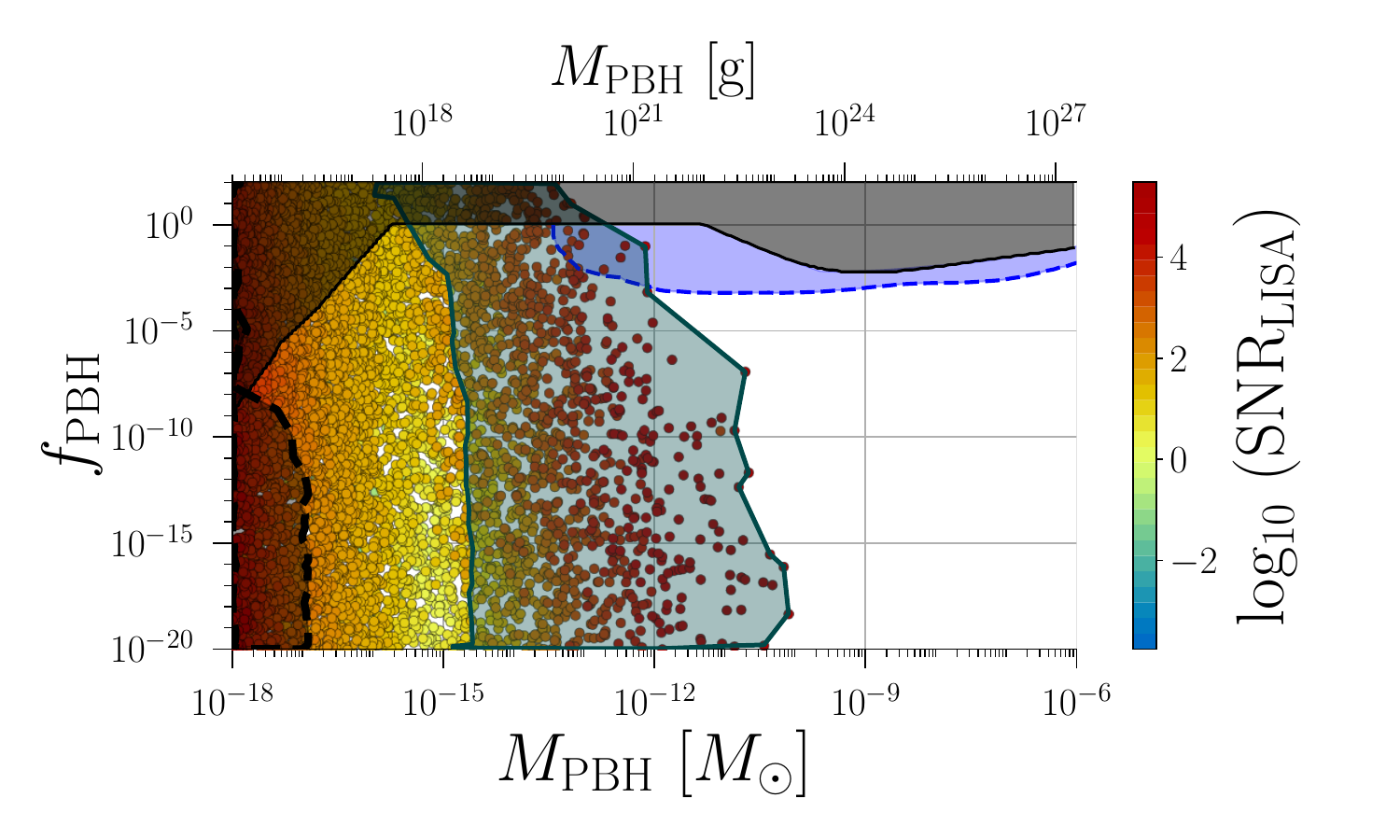}}
    \subfloat{\includegraphics[width=0.5\textwidth]{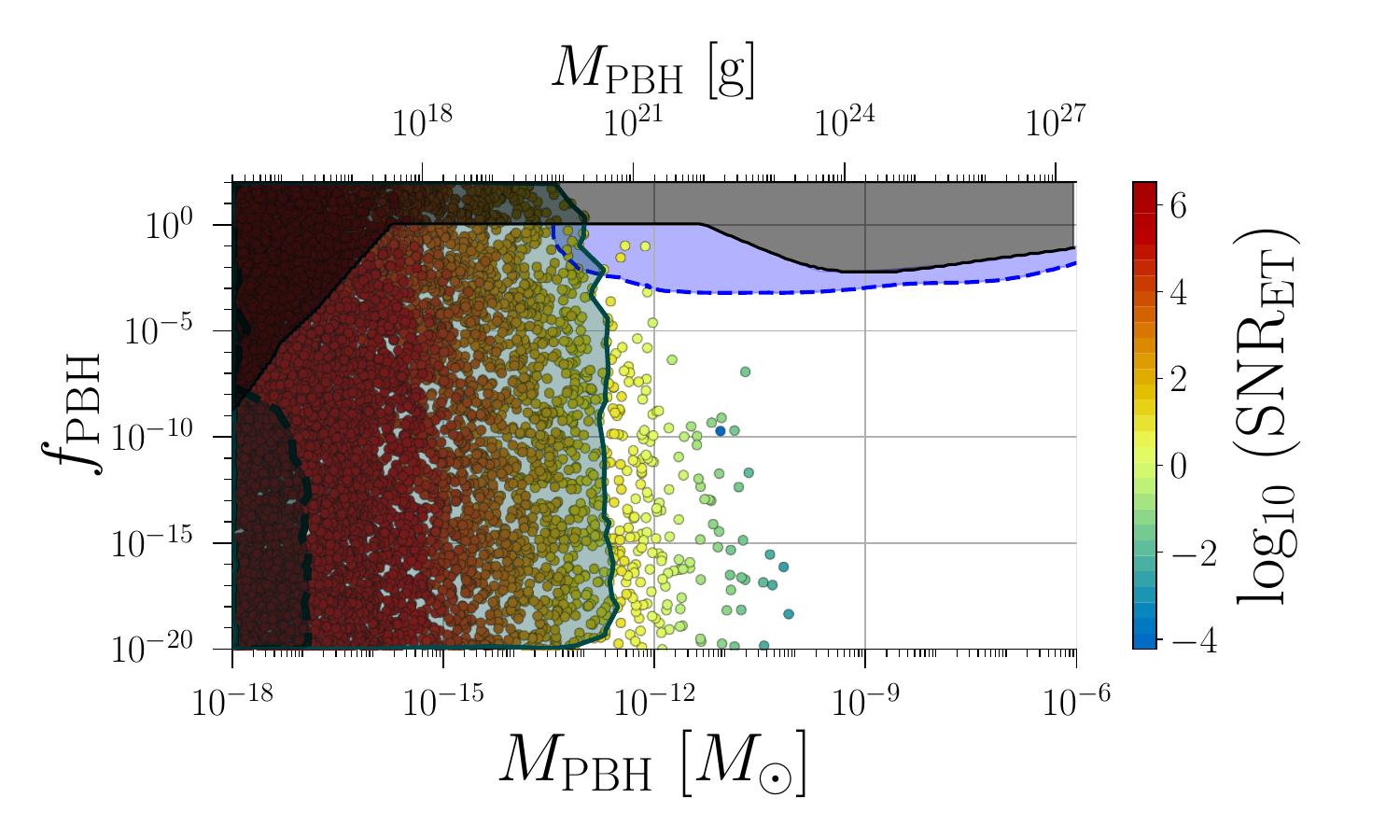}} \\
    \subfloat{\includegraphics[width=0.5\textwidth]{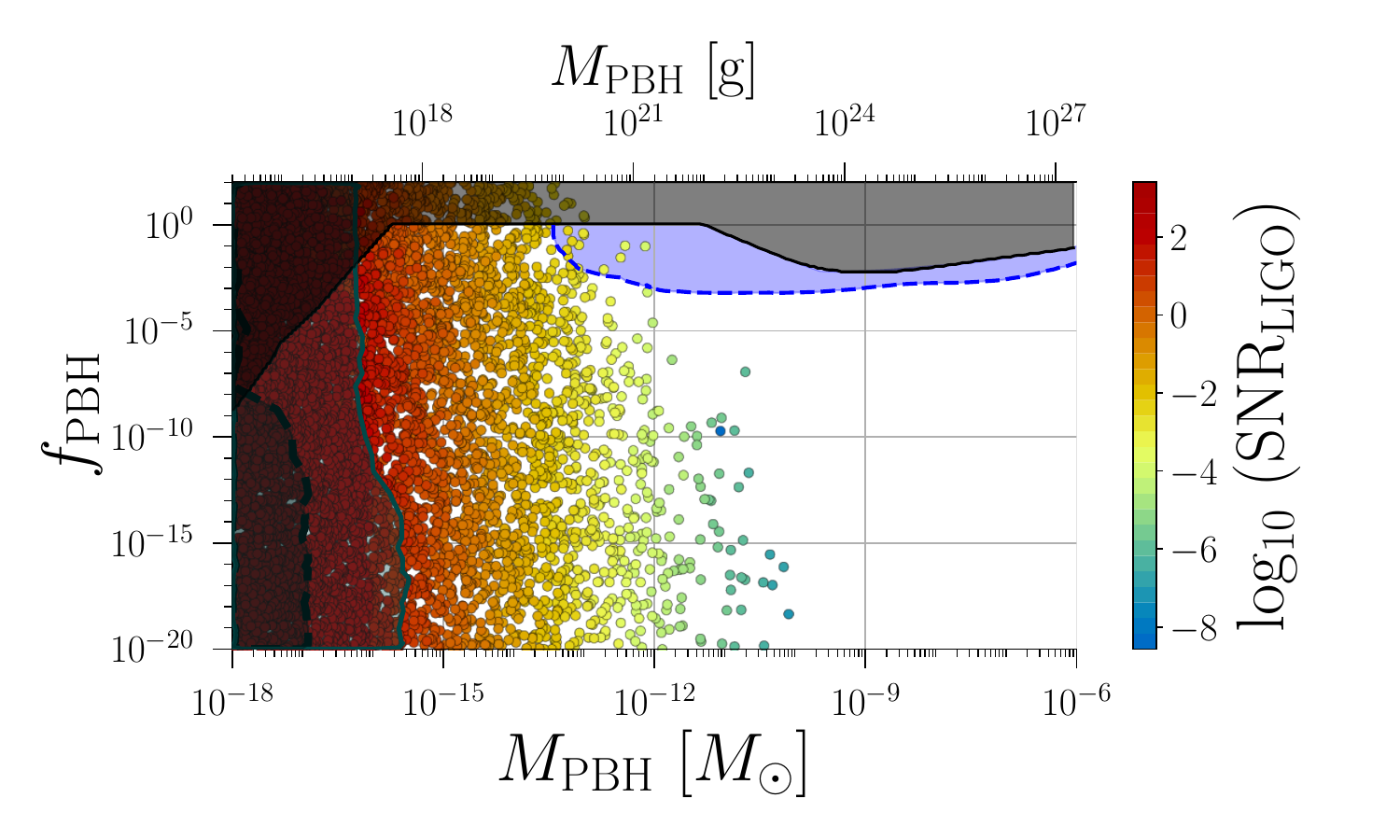}} 
    \subfloat{\includegraphics[width=0.5\textwidth]{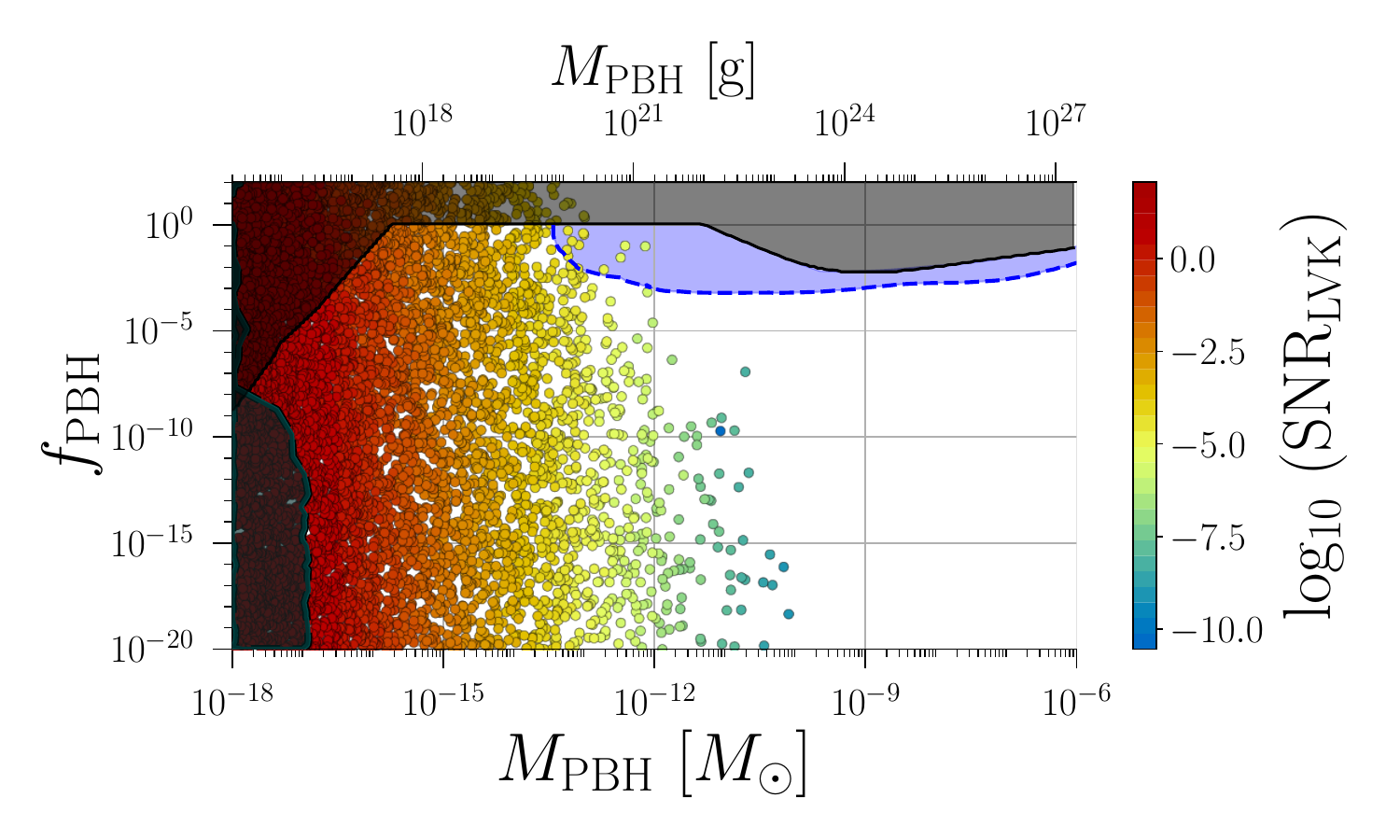}} 
    \caption{\small Similar to \cref{fig:PHBs_proj_plots_Thermo}, but here the color scales represent the SNR at LISA (top-left panel), ET (top-right panel), LIGO O5 (bottom-left panel) and LVK (bottom-right panel). The solid green contours enclose regions with $\mathrm{SNR}>10$. }
	\label{fig:PHBs_proj_plots_SNR}
\end{figure*}
 A signal is considered detectable if $\mathrm{SNR} > 10$. Experiments with sensitivity in the 1--100 Hz frequency range, such as LIGO and ET, primarily constrain the low-mass PBH region, while LIGO-O5 will be sensitive to masses below $\sim 10^{-17} M_\odot$ and ET can extend this sensitivity to PBH masses below $\sim 2\times 10^{-13} M_\odot$. In contrast, LISA is sensitive to PBH masses heavier than $1 \times 10^{-16} M_\odot$. Together, these experiments will cover the full range of PBH masses for $f_\mathrm{PBH}$ values down to $10^{-20}$, thus complementing $\gamma$-ray and microlensing observations. Current LVK data exclude PBH masses below $4 \times 10^{-18} M_\odot$.

Microlensing is a distinctive signal of compact objects. Notably, for PBHs with asteroid-scale masses between $10^{-17}~M_\odot$ and $10^{-12}~M_\odot$, the Schwarzschild radius becomes comparable to the optical wavelengths used in typical microlensing surveys. Then, wave optics effects become significant and lead to characteristic oscillatory patterns in the microlensing light curve. These oscillations serve as a signal for small lenses such as asteroid-sized PBHs, distinguishing them from larger astrophysical lenses.
A dedicated survey of the M31 galaxy by the Roman Space Telescope will be sensitive to a wide range of PBH masses, from $10^{-14}$ to $10^{-5}$ $M_\odot$~\cite{Drlica-Wagner:2022lbd}, thereby enhancing the ability to detect or rule out PBHs as a component of dark matter.

With this in mind, we compute the expected number of events at the Roman Space Telescope as 
\begin{equation}
    N_\mathrm{events} = T_\mathrm{obs} N_{\rm S} \int_{t_\mathrm{min}}^{t_\mathrm{max}} dt_{\rm E}\int dR_\mathrm{S} \int^0_1 dx \frac{d^2 \Gamma}{dx dt_{\rm E}} \frac{dn}{dR_\mathrm{S}}\,.
\end{equation}
Here, $x=D_L/D_S$ where $D_L$ and $D_S=778.5~\mathrm{kpc}$ are the distances from Earth to the lens and source (in M31), respectively, and $T_\mathrm{obs} = 6 \times 72~\mathrm{days}$ is the expected Roman observation time~\cite{2019ApJS2413P}, and $N_{\rm S} = 2.4 \times 10^{8}$ is the number of source stars in M31 \cite{2019ApJS2413P}. The duration for which the magnification remains above the detection threshold is $t_{\rm E}$ and is integrated from $t_\mathrm{min} = 15~\mathrm{minutes}$ (the proposed cadence \cite{2019ApJS2413P}) to $t_\mathrm{max} = 6\times 72~\mathrm{days}$ (total observation time). We take the stellar radius distribution $dn/dR_\mathrm{S}$ for the M31 galaxy from Ref.~\cite{Smyth:2019whb}. The differential event rate is given by
\begin{equation}
    \frac{d^2\Gamma}{dx dt_{\rm E}} = D_{\rm S} \frac{f_\mathrm{PBH}}{M_\mathrm{PBH}} \rho^\mathrm{DM}_\mathrm{M31}(r_\mathrm{M31})\frac{v^4_{\rm E}(x)}{v^2_\mathrm{M31}} e^{{-v^2_{\rm E}(x)}/v^2_\mathrm{M31}}\,,
    \label{dGamma}
\end{equation}
where $v_{\rm E}(x) = 2u_{1.34}(x)R_{\rm E}(x)/t_{\rm E}$, with $R_{\rm E}$ the Einstein radius and $u_{1.34} = 1$ the impact parameter for a point-like lens. The most probable source velocity in M31 is $v_\mathrm{M31} = 8.10\times 10^{-15}~\mathrm{kpc}/s$. For the PBH mass distribution $\rho^\mathrm{DM}_\mathrm{M31}$ we adopt the Navarro-Frenk-White DM profile:
\begin{equation}
\begin{aligned}
    \rho_\mathrm{M31}^\mathrm{DM}(r_\mathrm{M31}) &= \frac{\rho^\prime_\mathrm{M31}}{(r_\mathrm{M31}/r_\mathrm{M31}^\prime)(1+r_\mathrm{M31}/r_\mathrm{M31}^\prime)^2}\,, \\
    &r_\mathrm{M31} = \sqrt{R^2_\mathrm{sol} - 2xR_\mathrm{sol}D_s \cos\ell\cos b + x^2 D_S^2} \,,
\end{aligned}
\end{equation}
where the characteristic density is $\rho_\mathrm{M31} = 4.85\times 10^6~M_\odot/\mathrm{kpc}^3$, the scale radius is $r_\mathrm{M31} = 25~\mathrm{kpc}$, and $R_\mathrm{sol} = 8.5~\mathrm{kpc}$ is the distance from the center of the Milky Way to the Sun. The galactic coordinates of M31 are $(\ell,b)=(121.2\degree,-21.6\degree)$. 

In the left panel of \cref{fig:PHBs_proj_plots_SNRGamma} we show the parameter space with at least one microlensing event. 
\begin{figure*}[t]
	\centering
    \subfloat{\includegraphics[width=\textwidth]{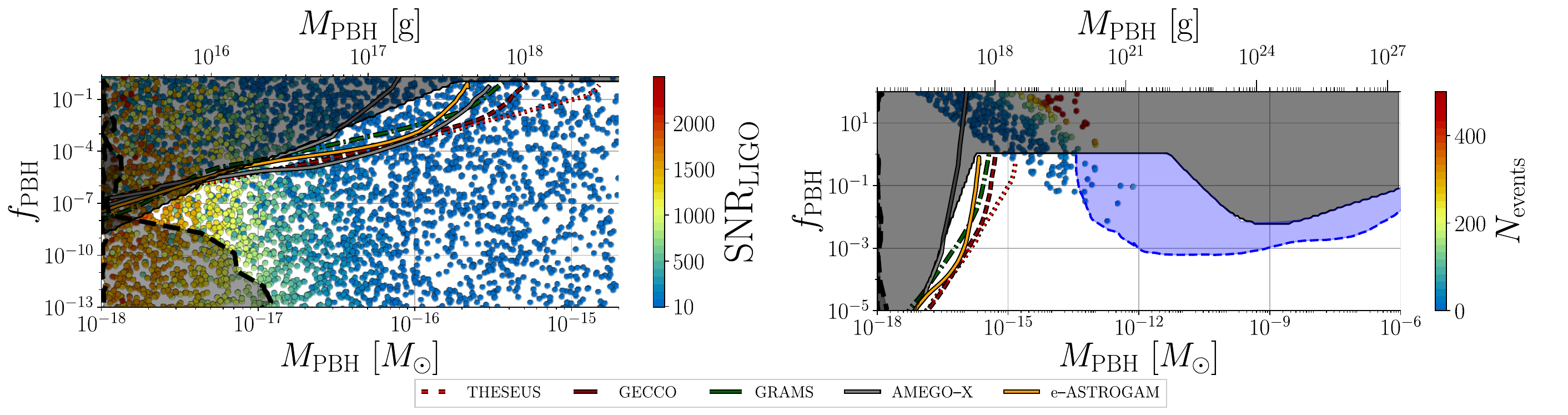}} 
    \caption{\small Left panel: Parameter space with at least one microlensing event. Right panel: The expected sensitivities of extragalactic $\gamma-$ray signals from Hawking radiation with SNR$>10$ for THESEUS (dotted red curve), GECCO (dashed brown curve), GRAMS (solid green curve), AMEGO$-$X (gray solid curve) and e$-$ASTROGAM (solid orange curve). The color scale in the right panel indicates the predicted SNR for LIGO O5. }\label{fig:PHBs_proj_plots_SNRGamma}
\end{figure*}
The Roman telescope can observe $\mathcal{O}(100)$ events for $M_\mathrm{PBH} \sim 10^{-12} M_\odot$ in a region consistent with current constraints.
From \cref{fig:PHBs_proj_plots_HEP} it is evident that the Roman telescope will target PeV mass scales, with $M_\mathrm{Z^\prime} \sim 10^5-10^8~\mathrm{GeV}$, $M_{h_2} \sim 10^4-10^7~\mathrm{GeV}$ and $\mathrm{max}(M_{N_i}) \equiv M_N \sim 10^6-10^7~\mathrm{GeV}$, $g_L \sim 0.28$, and $-0.025 \lesssim \lambda_\sigma \lesssim -0.05$. In this parameter space a correlated SGWB is detectable at both LISA and ET;  see \cref{fig:PHBs_proj_plots_SNR}.

The sensitivity of microlensing surveys falls for $M_\mathrm{PBH} < 10^{-15} M_\odot$, while $\gamma$-ray signals from Hawking evaporation become more sensitive. Several upcoming telescopes are designed to probe these masses, including THESEUS \cite{10.1117/12.2561012}, GECCO \cite{Orlando:2021get}, GRAMS \cite{Aramaki:2019bpi}, AMEGO-X \cite{Fleischhack:2021mhc} and e-ASTROGAM~\cite{e-ASTROGAM:2016bph}. To determine the regions with $\mathrm{SNR} > 10$, we utilized the \texttt{Isatis} script \cite{Auffinger:2022dic} from \texttt{BlackHawk}~\cite{Arbey:2019mbc, Arbey:2021mbl}, incorporating \texttt{Hazma}~\cite{Coogan:2019qpu} for low-energy hadronization.

In the right-panel of \cref{fig:PHBs_proj_plots_SNRGamma}, we present the $\mathrm{SNR} > 10$ sensitivity curves for these telescopes. We find that these facilities can probe a narrow region,  $M_\mathrm{PBH}\sim [10^{-17}, 10^{-15}]$~$M_\odot$ and $f_\mathrm{PBH} \sim 10^{-3}$, which coincides with the parameter space in which a SGWB is accessible at LIGO O5. From \cref{fig:PHBs_proj_plots_HEP}, we see that this corresponds to the regime of high-scale phase transitions, with $M_\mathrm{Z^\prime} \sim 10^{11}~\mathrm{GeV}$.

Figure~\ref{fig:primordial_Bfields} shows scatter plots in the $(M_\mathrm{PBH},f_\mathrm{PBH})$ plane, with the peak magnetic field strength (top panels) and peak coherence length (middle panels) indicated by the color scale.
\begin{figure*}[t]
	\centering
    \subfloat{\includegraphics[width=0.5\textwidth]{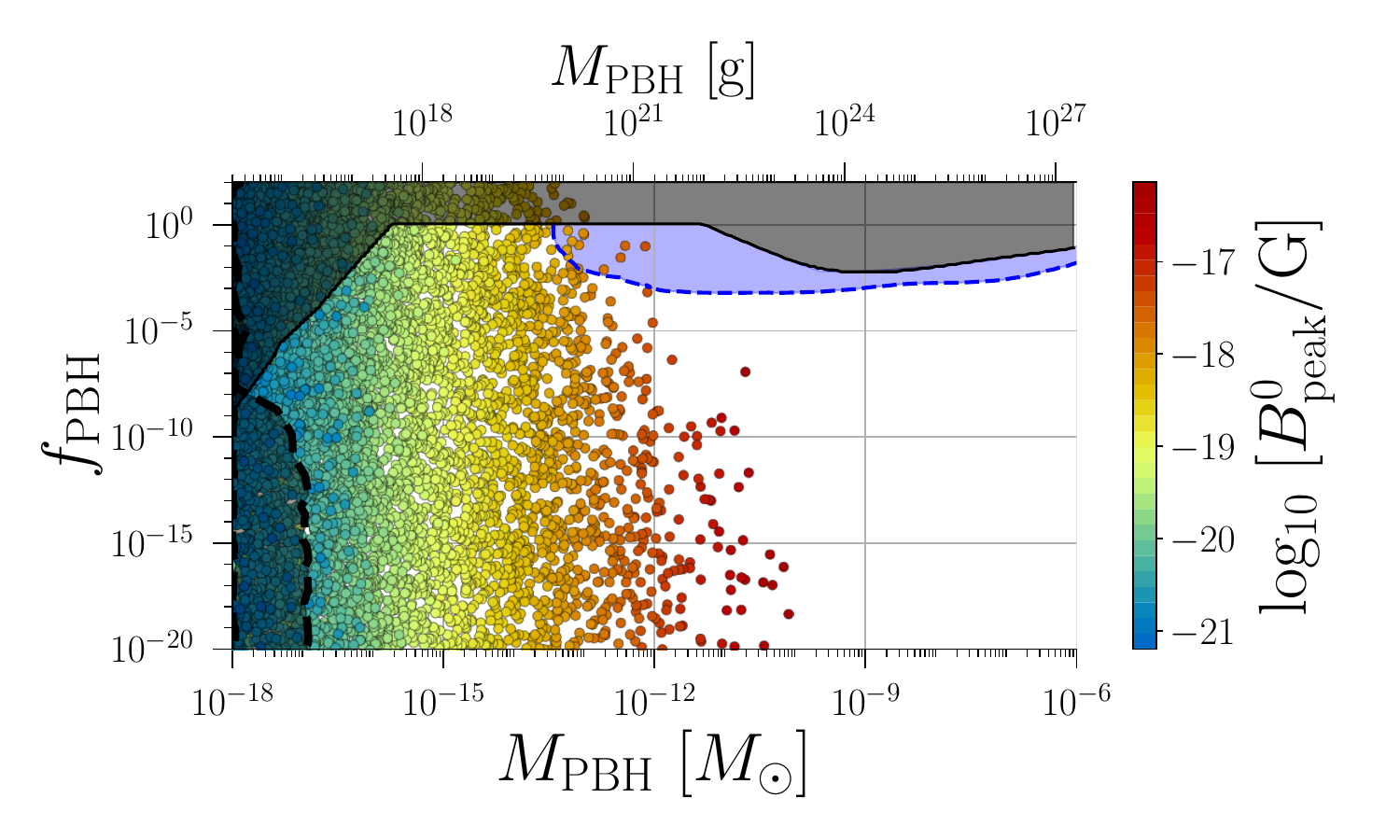}}
    \subfloat{\includegraphics[width=0.5\textwidth]{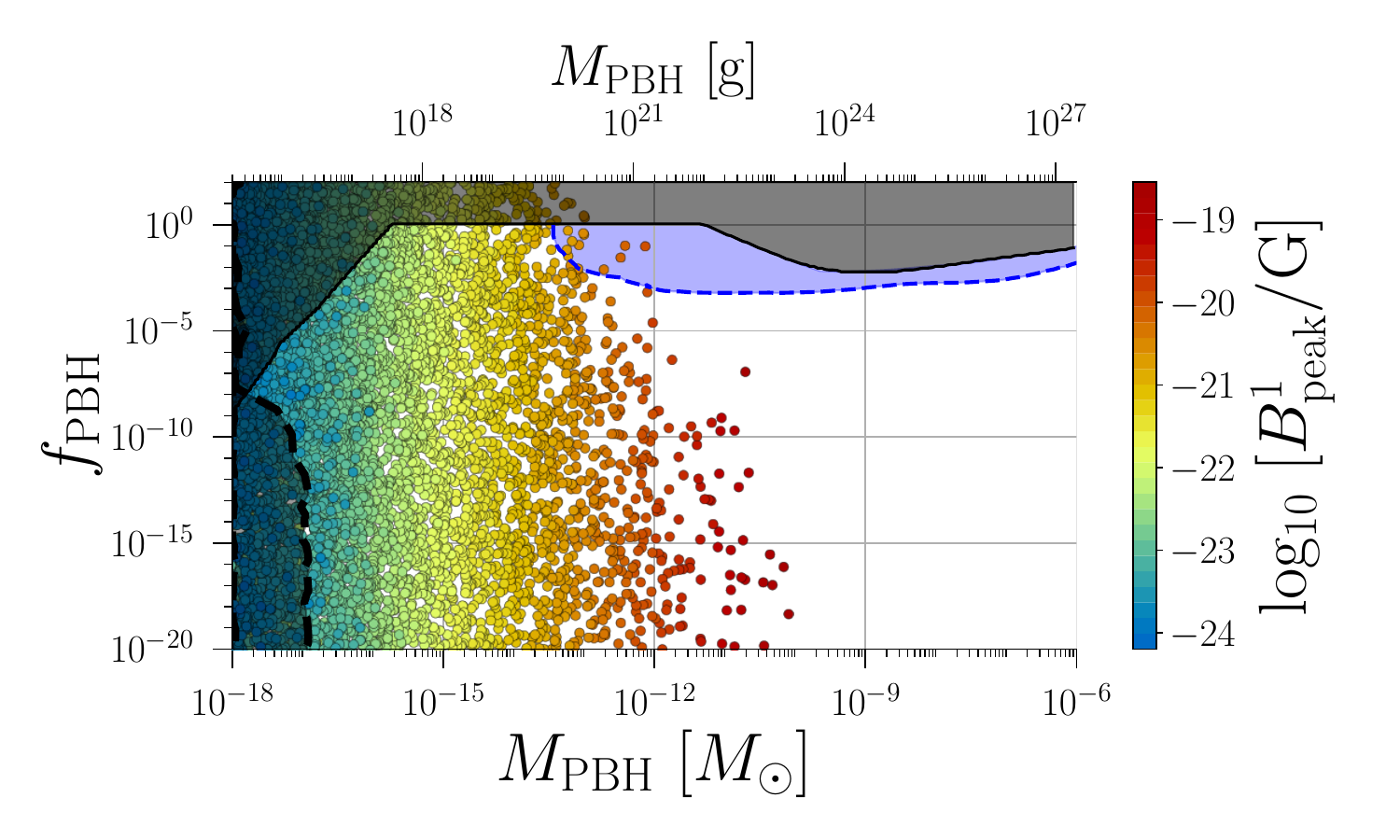}} \\
    \subfloat{\includegraphics[width=0.5\textwidth]{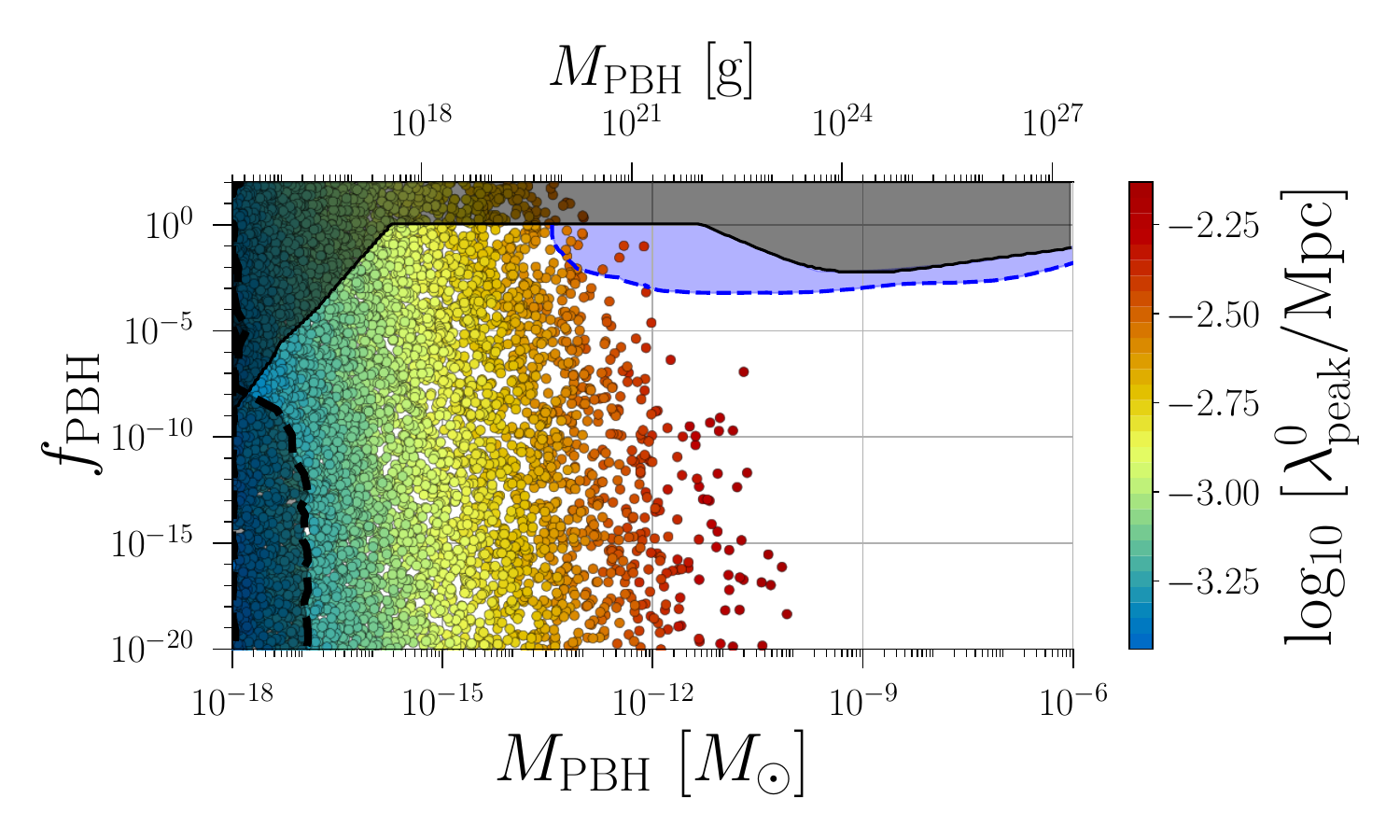}}
    \subfloat{\includegraphics[width=0.5\textwidth]{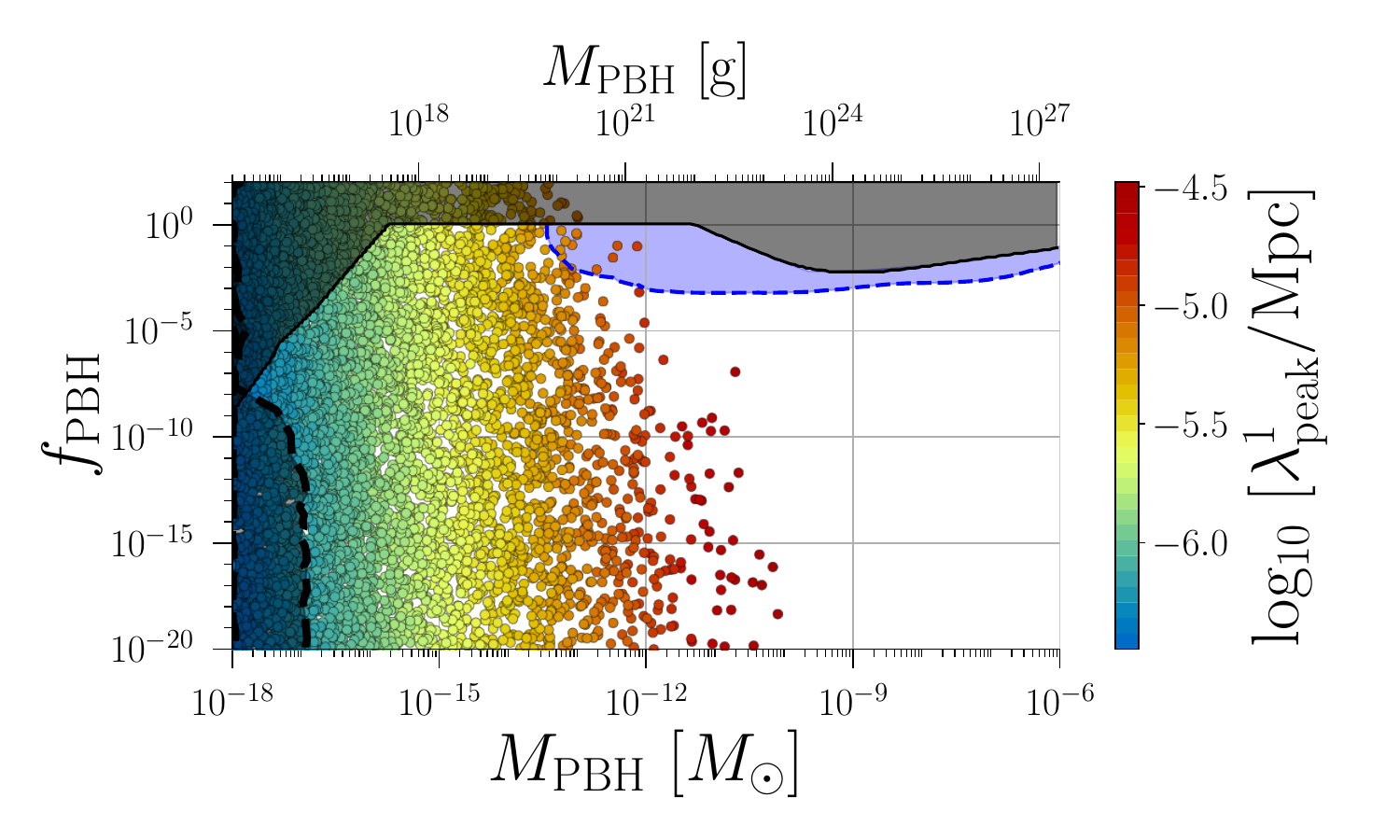}} \\
    \hspace*{3em} \subfloat{\includegraphics[width=0.95\textwidth]{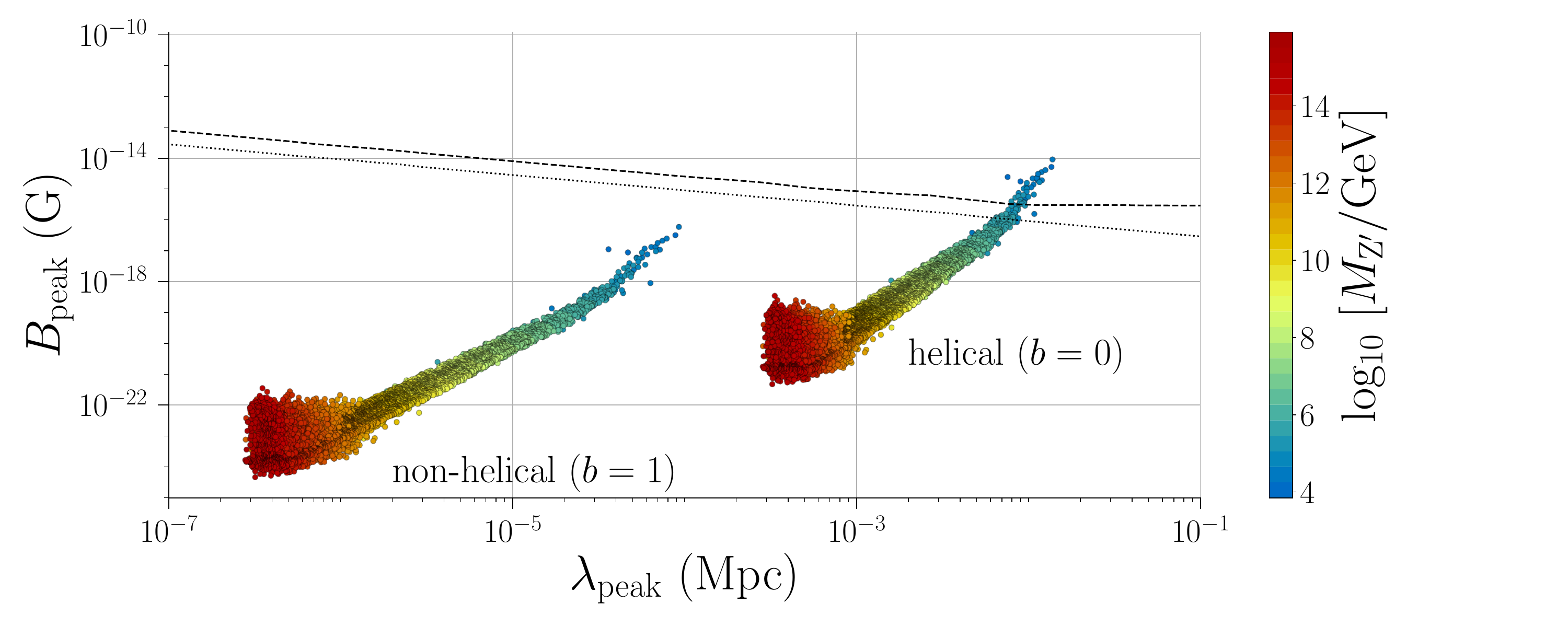}}
    \caption{Top and middle panels: Similar to \cref{fig:PHBs_proj_plots_Thermo}, but the color scale indicates the peak magnetic field strength $B_\mathrm{peak}$ and peak coherence length $\lambda_\mathrm{peak}$ for helical magnetic fields $b=0$ (left) and non-helical magnetic fields $b=1$ (right). Bottom panel: $B_\mathrm{peak}$ versus  
    $\lambda_\mathrm{peak}$ for $b=0$ and $b=1$. The black dashed line shows the lower limit from \cite{Fermi-LAT:2018jdy}, and the dotted black line shows the corresponding limit from \cite{MAGIC:2022piy}.
    } 
	\label{fig:primordial_Bfields}
\end{figure*}

\begin{figure*}[htb!]
	\centering
    \subfloat{\includegraphics[width=\textwidth]{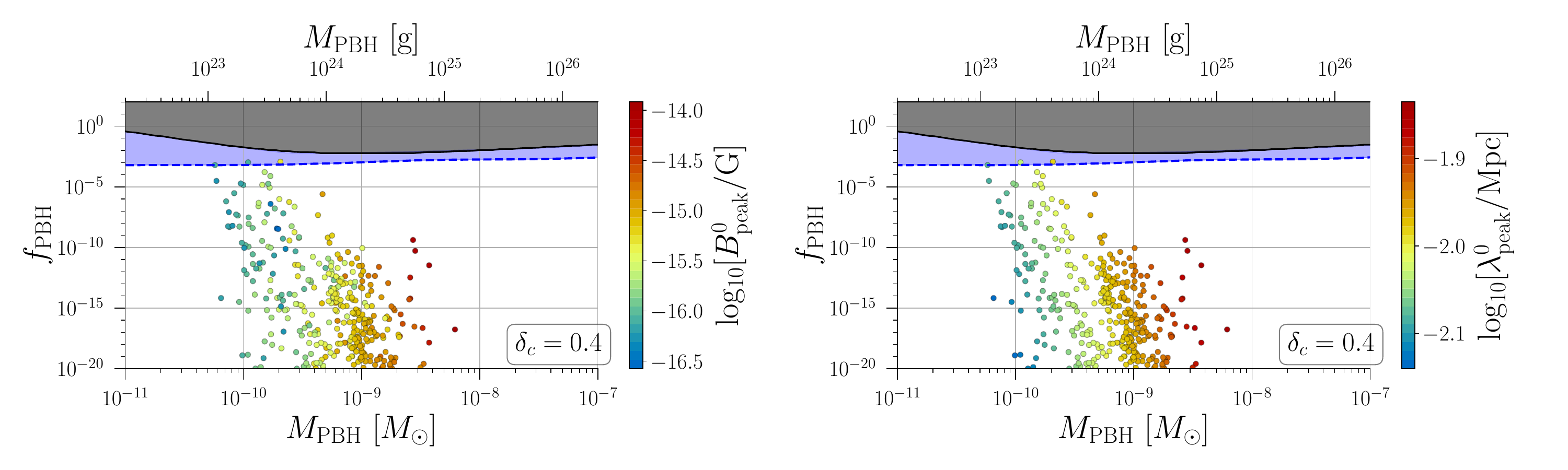}} \\
    \hspace*{3em} \subfloat{\includegraphics[width=0.95\textwidth]{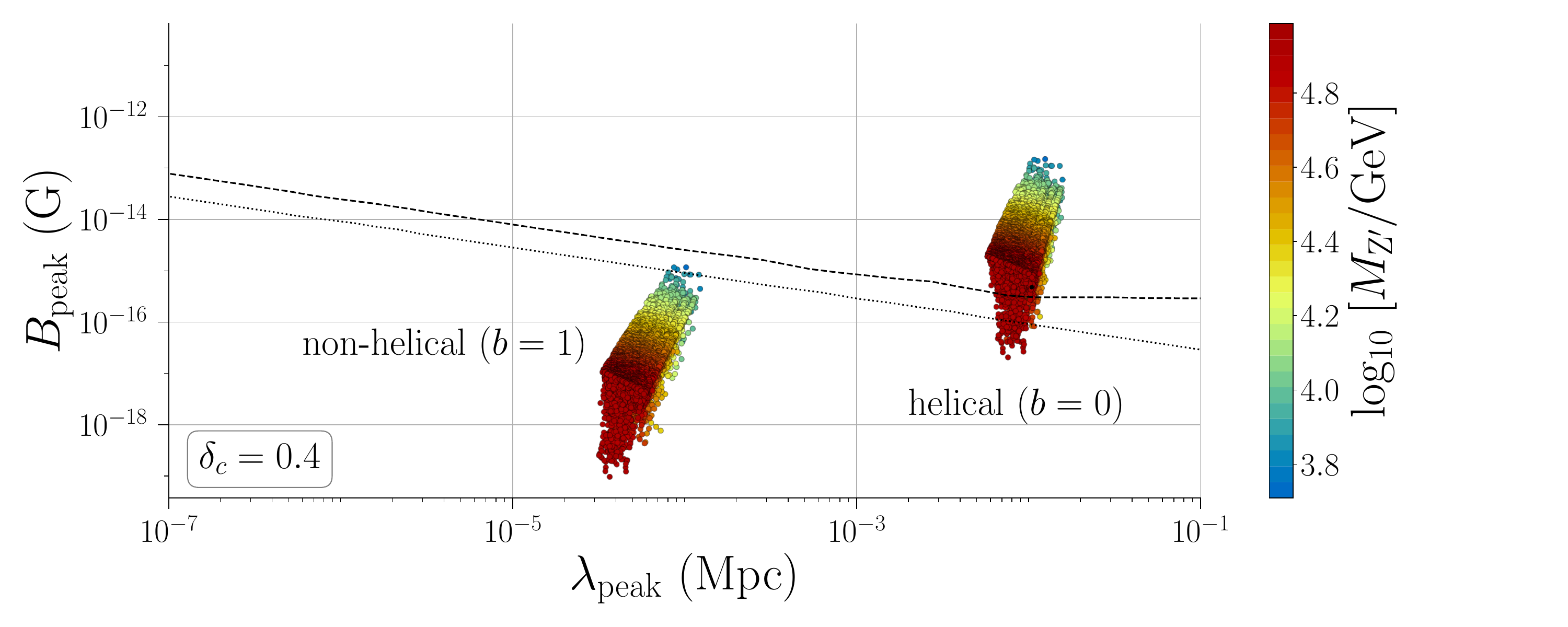}}
    \caption{Similar to Fig.~\ref{fig:primordial_Bfields} with $\delta_c = 0.40$. }
    
	\label{fig:primordial_Bfields_dedicated}
\end{figure*}

The left panels are for helical magnetic fields ($b=0$), and right panels are for non-helical magnetic fields ($b=1$). The bottom panel displays these points in the $(\lambda_\mathrm{peak},B_\mathrm{peak})$ plane, with the color scale indicating the $\mathrm{Z^\prime}$ mass. The black lines show lower limits from high-energy $\gamma$-ray observations of blazars.

For both $b=0$ and $b=1$, we observe a strong correlation with the PBH mass: lighter PBHs are associated with lower magnetic field strengths and shorter coherence lengths. Comparing with \cref{fig:PHBs_proj_plots_HEP,fig:PHBs_proj_plots_Thermo} we find that this trend is linked to both the dark sector masses and $\lambda_\sigma$. Specifically, $\mathcal{O}(10)$ TeV-scale phase transitions with $\lambda_\sigma \sim -0.025$ yield  $B_\mathrm{peak} \sim 10^{-16}~\mathrm{G}$ for $b=0$, and $B_\mathrm{peak} \sim 10^{-19}~\mathrm{G}$ for $b=1$. This behavior stems from the fact that the magnetic field strength scales as $B_\mathrm{peak} \propto T^{-p_b/2 - 2}_\mathrm{RH}$,  while the coherence length scales as $\lambda_\mathrm{peak} \propto T_\mathrm{RH}^{q_b-1}$ (see Eqs.~\eqref{eq:mag_fields}--\eqref{eq:redshift_magfields}), 
because the Hubble parameter $H \propto T^2$ during the radiation-dominated epoch. Since $q_b$ is always less than unity, higher reheating temperatures correspond to weaker magnetic fields and shorter coherence lengths. As expected, non-helical magnetic fields have lower amplitudes and shorter coherence lengths than helical magnetic fields. 

From the bottom panel of \cref{fig:primordial_Bfields} we observe that for $b=0$, $\mathrm{Z^\prime}$ masses in the range of approximately $[5, 100]~\mathrm{TeV}$ can generate magnetic fields that satisfy the lower limits from blazars, and are therefore observable.
For $b=1$, our general scan did not find points with magnetic fields above the blazar lower bounds.

 Since only a very narrow region of parameter space leads to  observable magnetic fields and PBH densities that almost saturate the DM abundance, our general scan did not find these solutions. Therefore, to properly highlight the interplay between PBH formation and magnetic field generation, we perform a focused scan in the ranges $g_L = [0.20,0.30]$ and $M_{h_2} = [150,10^4]~\mathrm{GeV}$, and show the results in Fig.~\ref{fig:primordial_Bfields_dedicated}. Note that the points from the focused scan are not shown in any other figure. For $b=0$, we find that the region with observable magnetic fields corresponds to PBH masses in the range $M_{\rm PBH} \sim  [10^{-10},10^{-8}]M_{\odot}$. In this region, current microlensing data strongly constrain $f_{\rm PBH}$ to values below $10^{-3}$, so observable magnetic fields are incompatible with PBHs that saturate the DM relic abundance. However, part of this parameter space lies within the projected sensitivity of Roman. Unlike in the general scan, for non-helical fields we find a very narrow region of  parameter space with $M_{\mathrm{Z}^\prime} \sim 6~\mathrm{TeV}$ which gives rise to visible magnetic fields.

\begin{figure*}[t]
	\centering
    \subfloat{\includegraphics[width=0.5\textwidth]{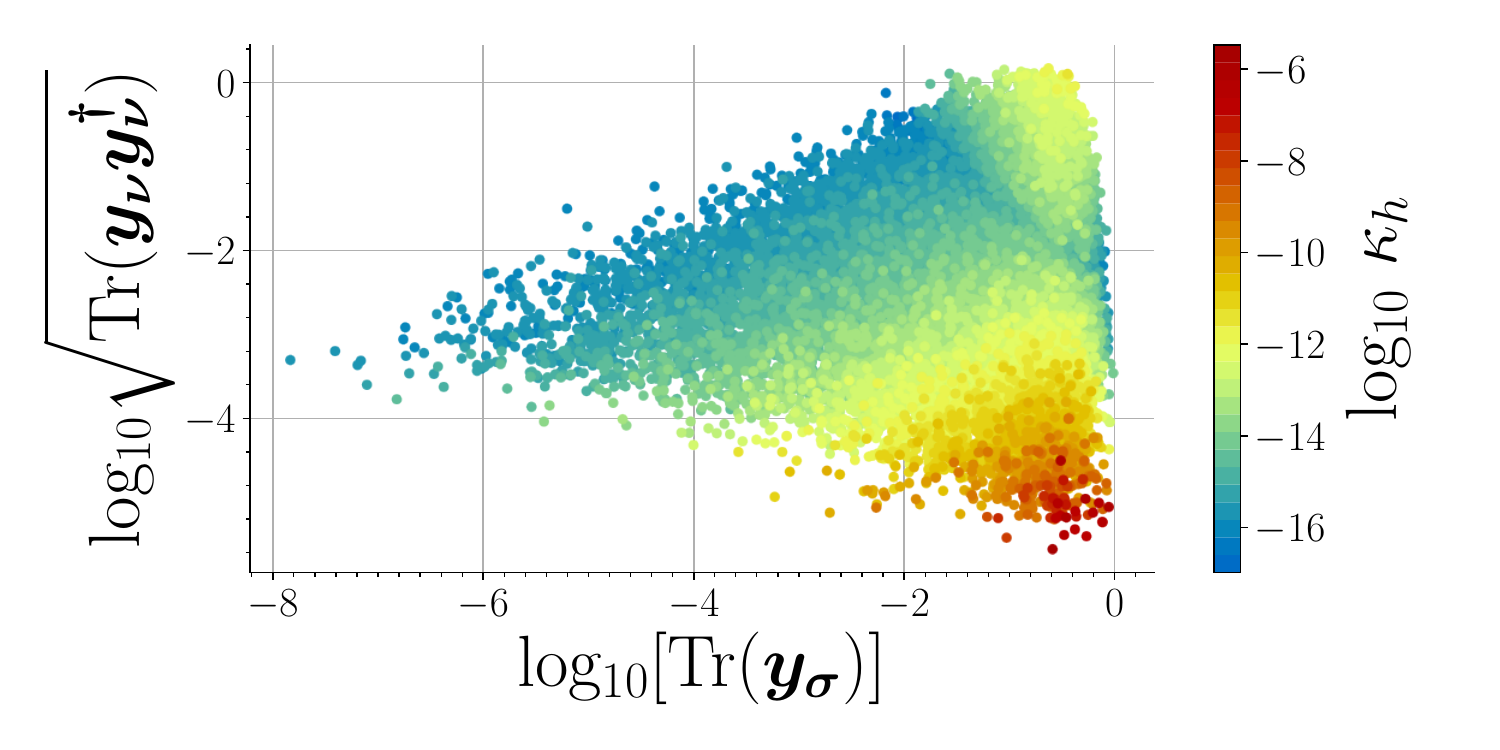}}
    \subfloat{\includegraphics[width=0.5\textwidth]{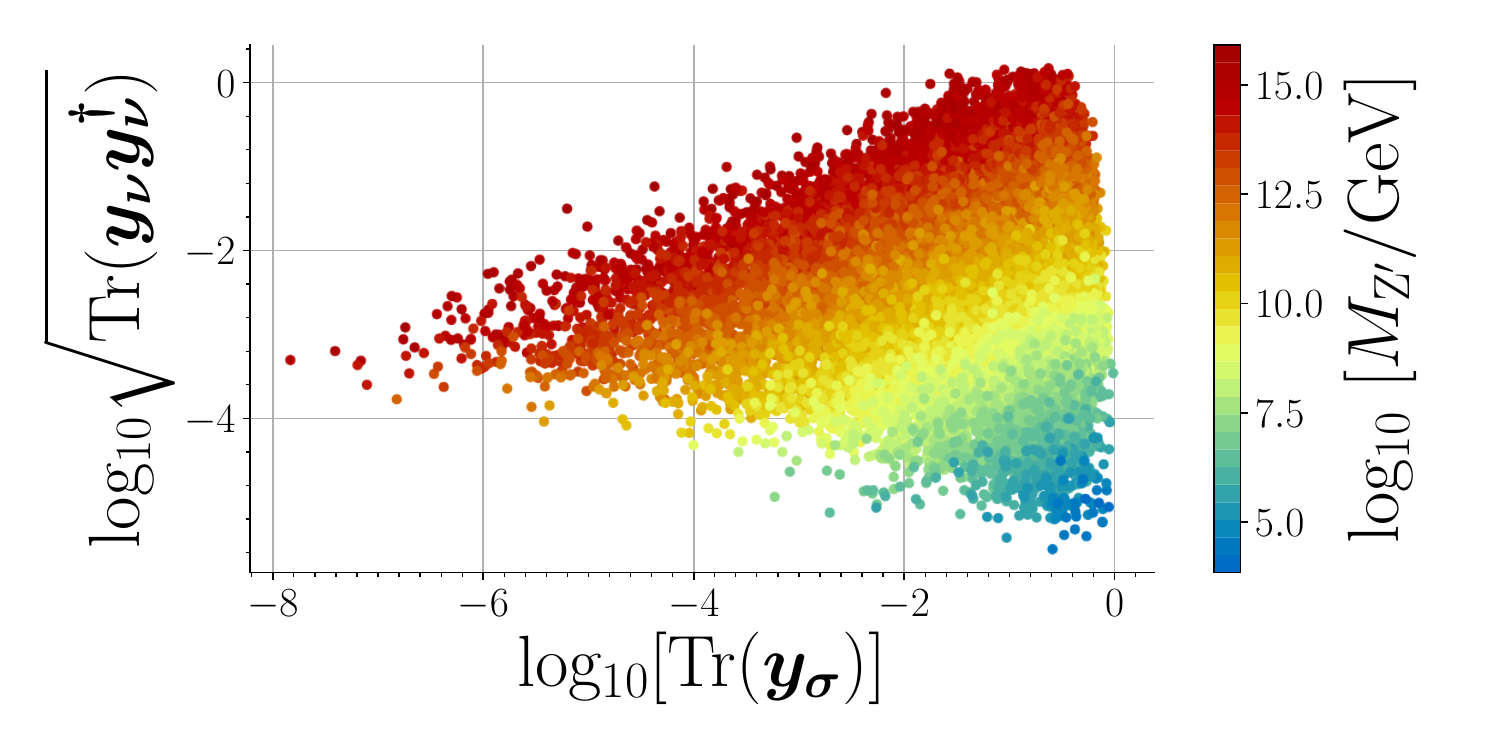}} \\
    \subfloat{\includegraphics[width=0.5\textwidth]{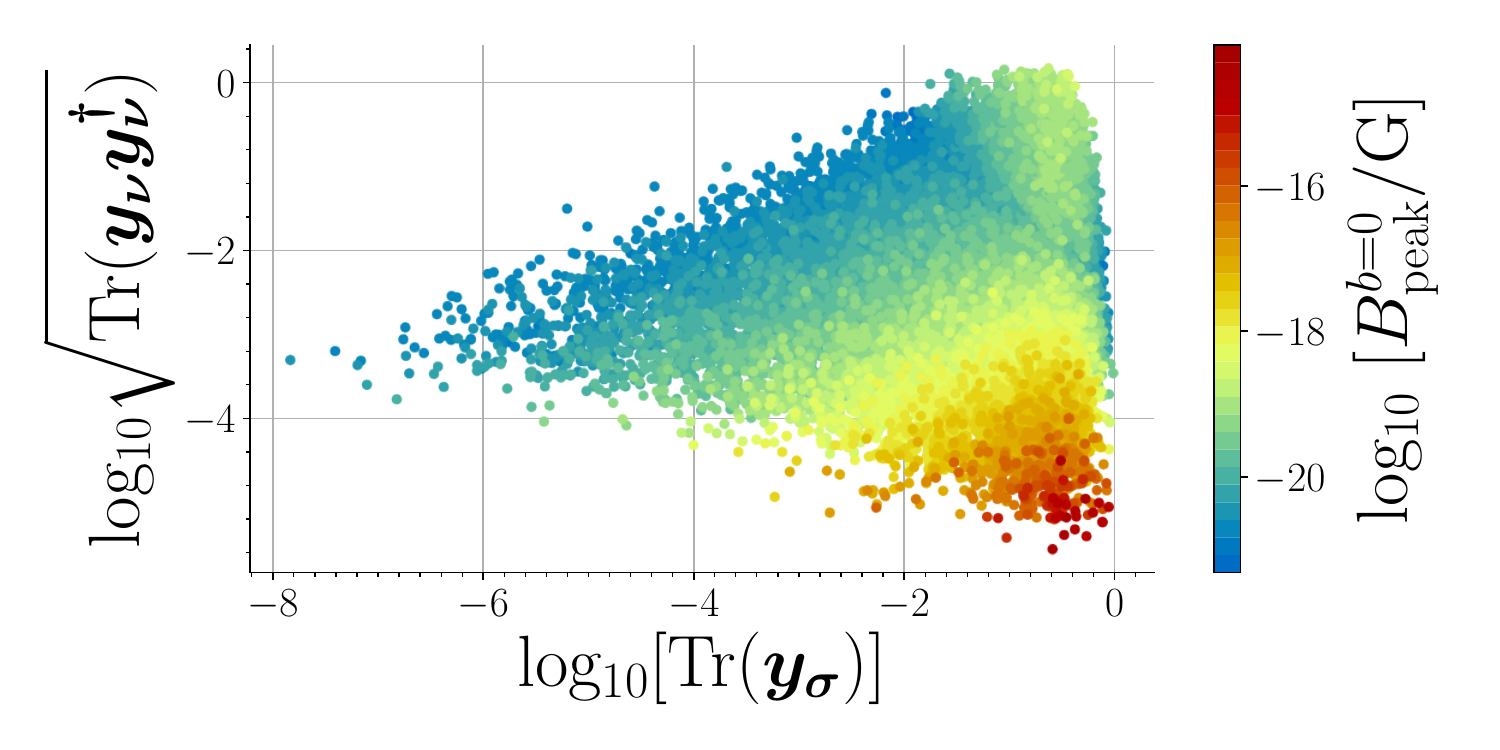}}
    \subfloat{\includegraphics[width=0.5\textwidth]{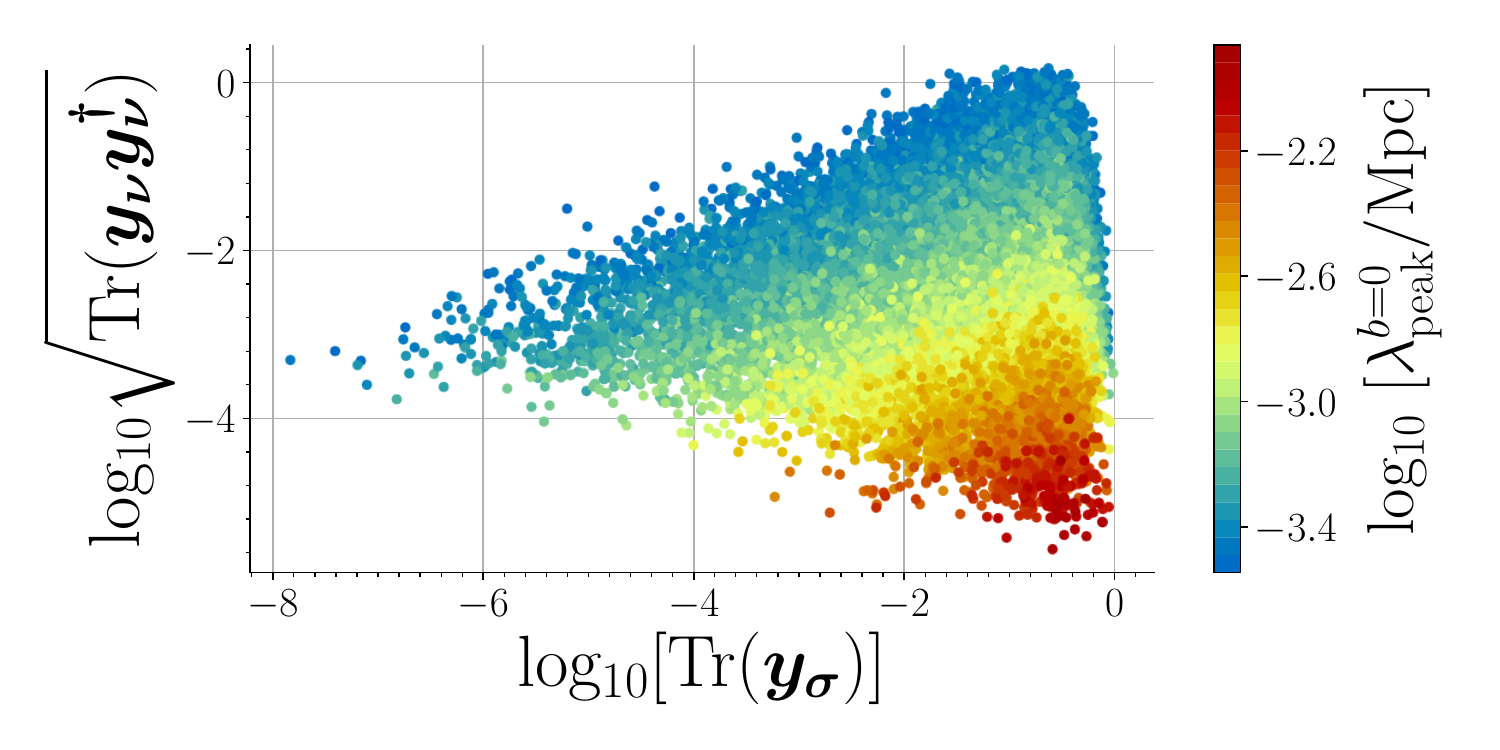}} \\
    \subfloat{\includegraphics[width=0.5\textwidth]{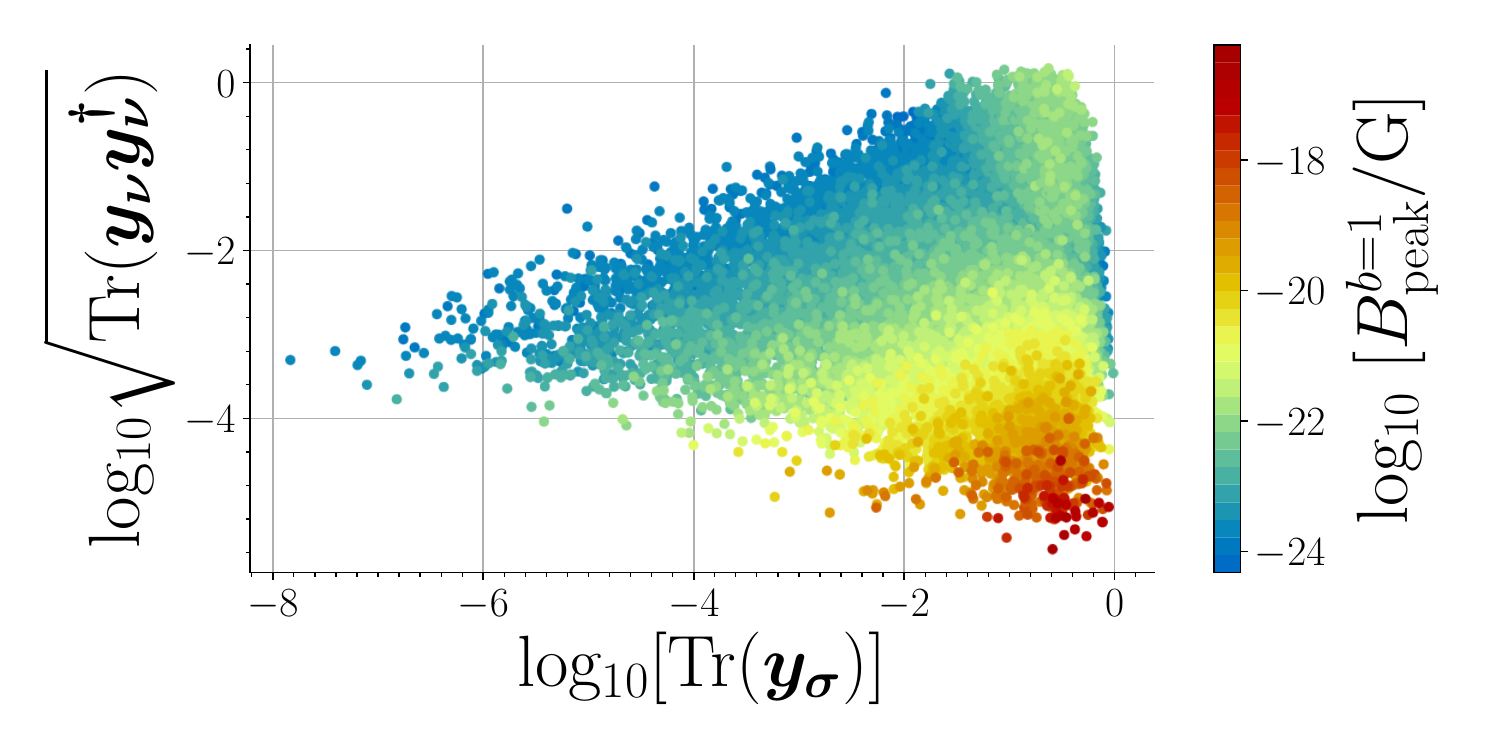}}
    \subfloat{\includegraphics[width=0.5\textwidth]{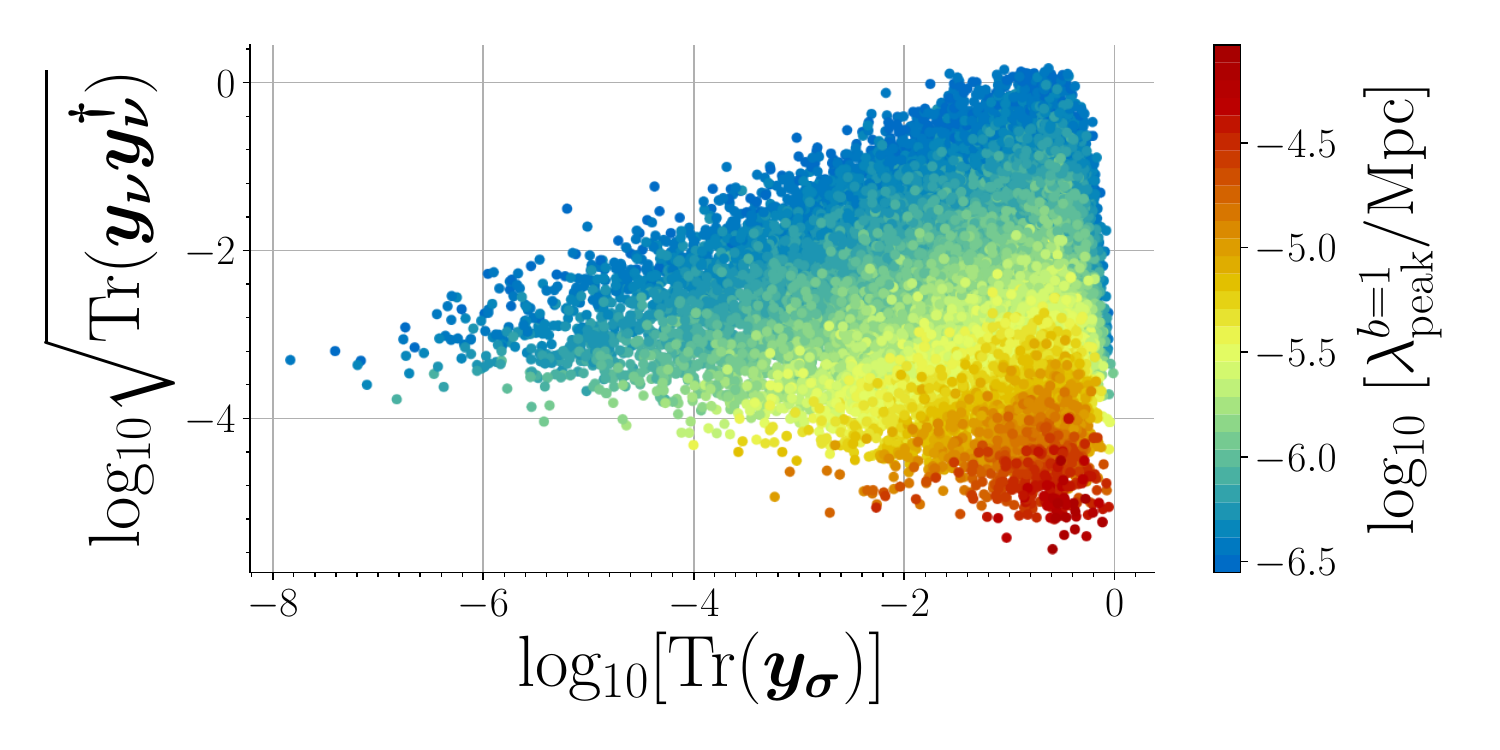}} 
    \caption{\small Scatter plots of Majorana Yukawa couplings $\mathrm{Tr}(\bm{y_\sigma})$ and the magnitude of the Dirac Yukawa couplings $\sqrt{\mathrm{Tr}(\bm{y_\nu} \bm{y_\nu}^\dagger)}$. The color scale indicates the suppression factor $\kappa_h$ (top-left panel), the mass of $\mathrm{Z^\prime}$ (top-right panel), the peak magnetic field strength (middle-left and bottom-left panels) and the correlation lengths (middle-right and bottom-right panels) for $b=0$ and $b=1$, respectively.}
	\label{fig:parameters_Bfields}
\end{figure*}

The role of the neutrino Yukawa couplings on the magnetic fields is illustrated in \cref{fig:parameters_Bfields}. We find that the suppression factor $\kappa_h$ strongly affects $B_{\rm peak}$, and that an inverse relationship exists between the $\mathrm{Z^\prime}$ mass and $\kappa_h$ because the decoupling between the dark and visible sectors $\sim v/v_\sigma$. As $\bm{y_\nu}$ is increased to $\mathcal{O}(0.1-1)$, the numerator in \cref{eq:kappa_h} increases while the denominator remains unchanged, so that $\kappa_h$ becomes larger. This consequently enhances $\kappa_h$ as evident by the color distribution in the top-right edge of the top-left panel. Nevertheless, sizable values of $\bm{y_\nu}$ are insufficient to generate observable magnetic fields, as the dark sector mass scale $M_\mathrm{Z^\prime} \sim 10^{15}~\mathrm{GeV}$ has a dominant effect. These enhanced couplings determine the bulge feature in the bottom panel of \cref{fig:primordial_Bfields}. We find that $\mathrm{Tr}(\bm{y_\sigma})$ is more weakly correlated with $B_\mathrm{peak}$ and $\lambda_\mathrm{peak}$ than is $\bm{y_\nu}$. For the latter, we find that lower values correspond to larger values of both $B_\mathrm{peak}$ and $\lambda_\mathrm{peak}$. This behavior is connected with the seesaw scale, since larger values of $\bm{y_\nu}$ are correlated with larger scales. For $b=0$, in the region where magnetic fields strengths are above the blazar bounds, we find $\sqrt{\mathrm{Tr}(\bm{y_\nu} \bm{y_\nu}^\dagger)} \sim [4\times 10^{-6}, 4\times 10^{-5}]$ which corresponds to dark sector scales $M_\mathrm{Z^\prime}\sim [5,100]~\mathrm{TeV}$. For $b=1$, from our focused scan we find a very small number of points with visible magnetic fields above the least conservative limit, with $M_{\rm Z^\prime} \sim 6~\mathrm{TeV}$ and $\sqrt{\mathrm{Tr}(\bm{y_\nu} \bm{y_\nu}^\dagger)} \sim 1.2\times 10^{-5}$.

\subsection{Scenarios with generic charge assignments}
\label{sec:generic_charges}

The previous results were obtained for the $\mathrm{U(1)_{B-L}}$ model. We now examine the case of generic $\mathrm{U(1)}^\prime$ charges. In Fig.~\ref{fig:PHBs_proj_genericcharges} we show projections in the $(M_\mathrm{PBH},f_\mathrm{PBH})$ plane with color scales for $g_L x_\mathcal{H}$ (left panel) and $g_L x_\sigma$ (right panel).
\begin{figure*}[t]
	\centering
    \subfloat{\includegraphics[width=0.5\textwidth]{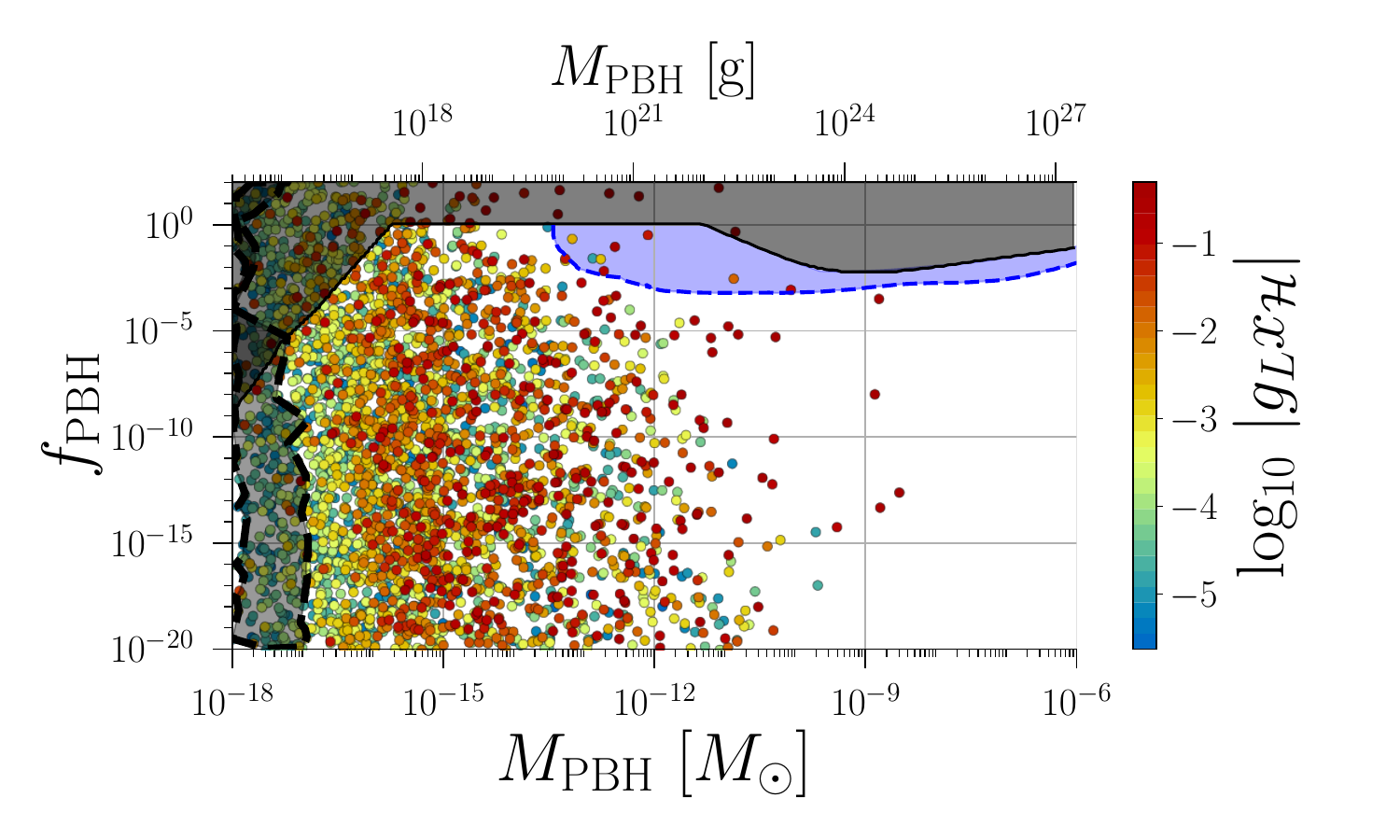}}
    \subfloat{\includegraphics[width=0.5\textwidth]{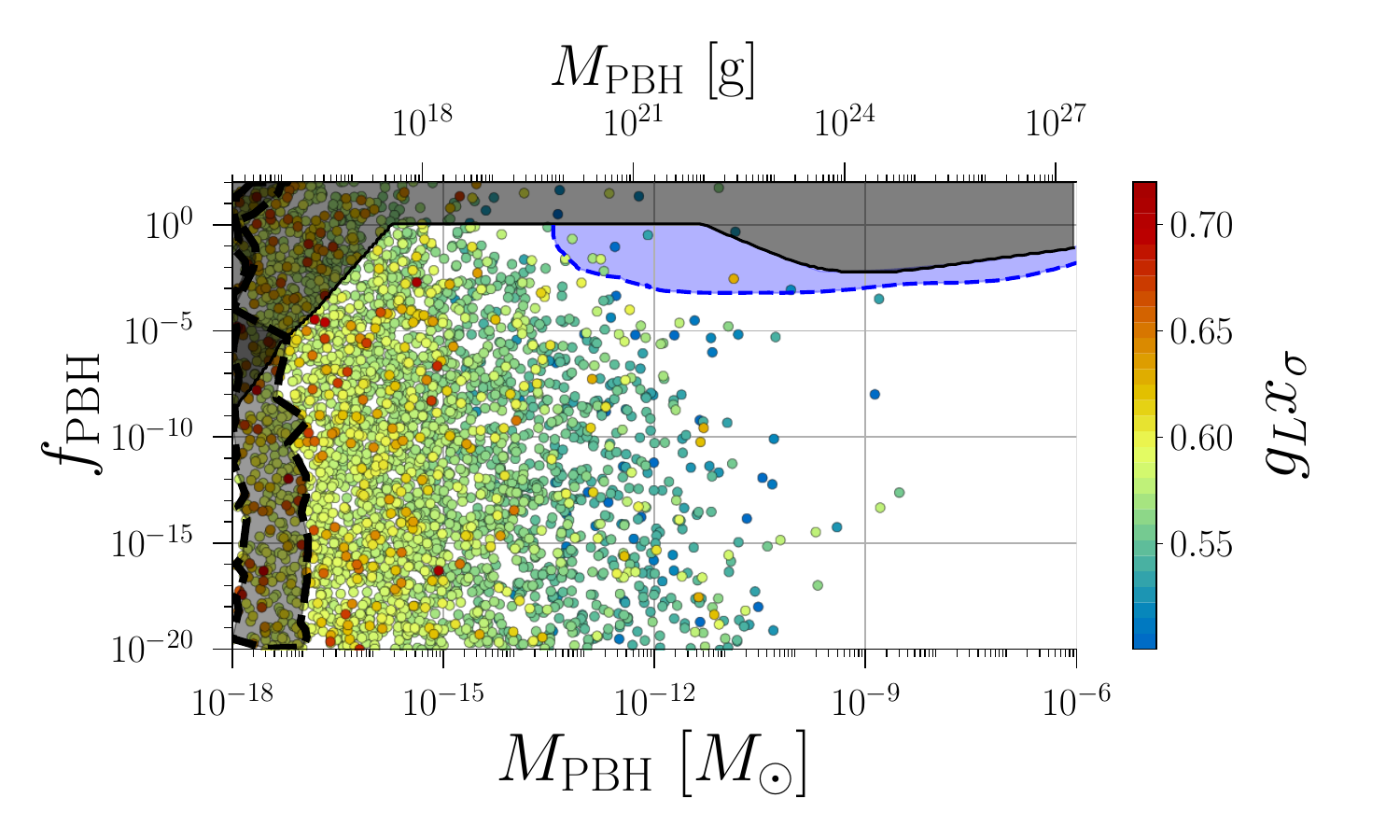}} 
    \caption{\small Similar to \cref{fig:PHBs_proj_plots_Thermo}, but  the color scale represents $g_L x_\mathcal{H}$ in the left panel and $g_L x_\sigma$ in the right panel.}
	\label{fig:PHBs_proj_genericcharges}
\end{figure*}
 We find that $x_\mathcal{H}$ introduces dispersion in the parameter space, resulting in a weak correlation with both $f_\mathrm{PBH}$ and $M_\mathrm{PBH}$. Similarly, the PBH parameters are weakly correlated with $g_L x_\sigma$, although as in the $\mathrm{U(1)_{B-L}}$ case, most points cluster in a narrow range $g_L x_\sigma = [0.50,0.70]$ to produce the small values of $\beta/H(T_p)$ required. Thermodynamic parameters, SGWB, and magnetic fields exhibit correlations consistent with those in the B$-$L model.

In \cref{fig:parameters_proj} we present scatter plots of various projections of the model's parameter space. The region enclosed by the black solid contour is excluded by LVK (with SNR $>10$) and the gray dotted contour defines the region where peak magnetic field strengths for $b=0$ exceed the lower bounds set by blazars.
\begin{figure*}[t]
	\centering
    \subfloat{\includegraphics[width=0.5\textwidth]{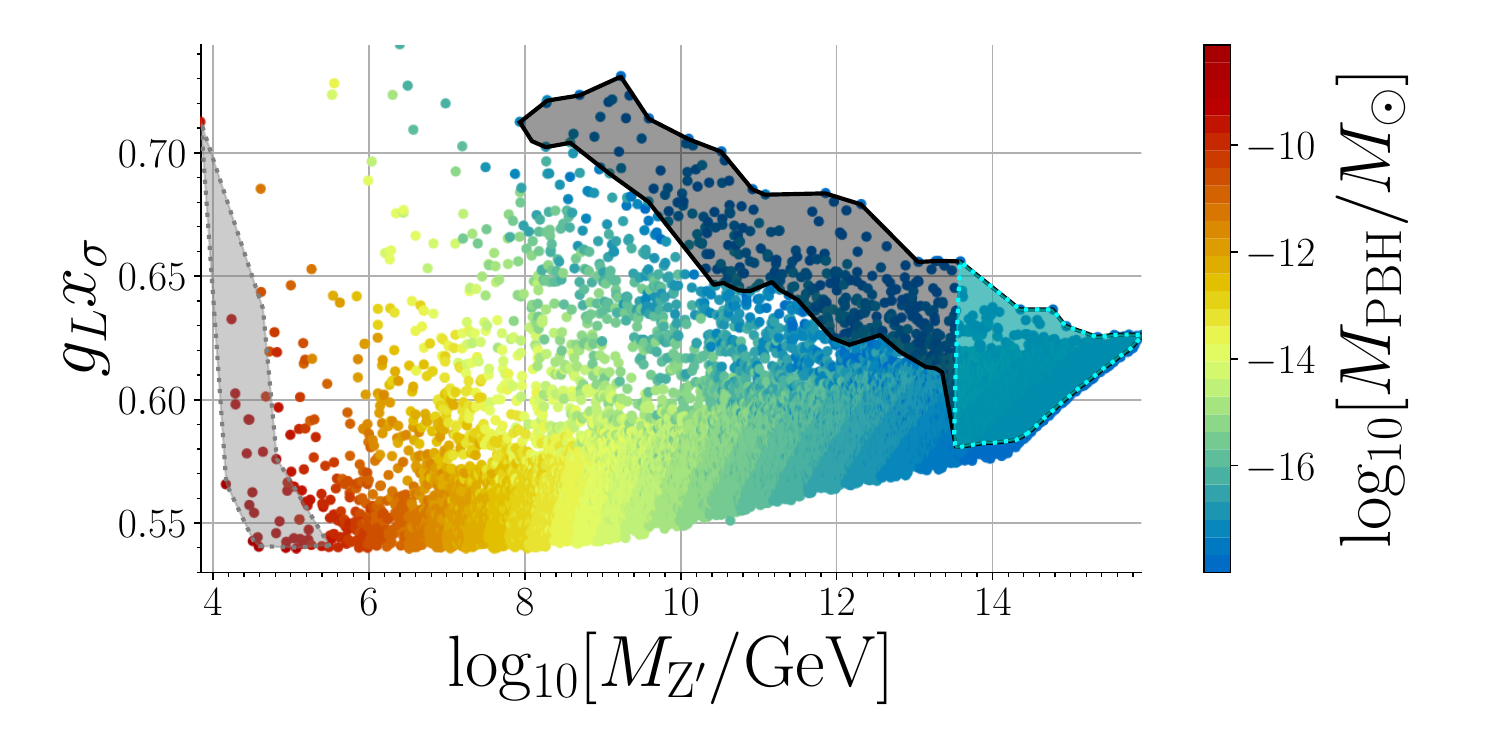}}
    \subfloat{\includegraphics[width=0.5\textwidth]{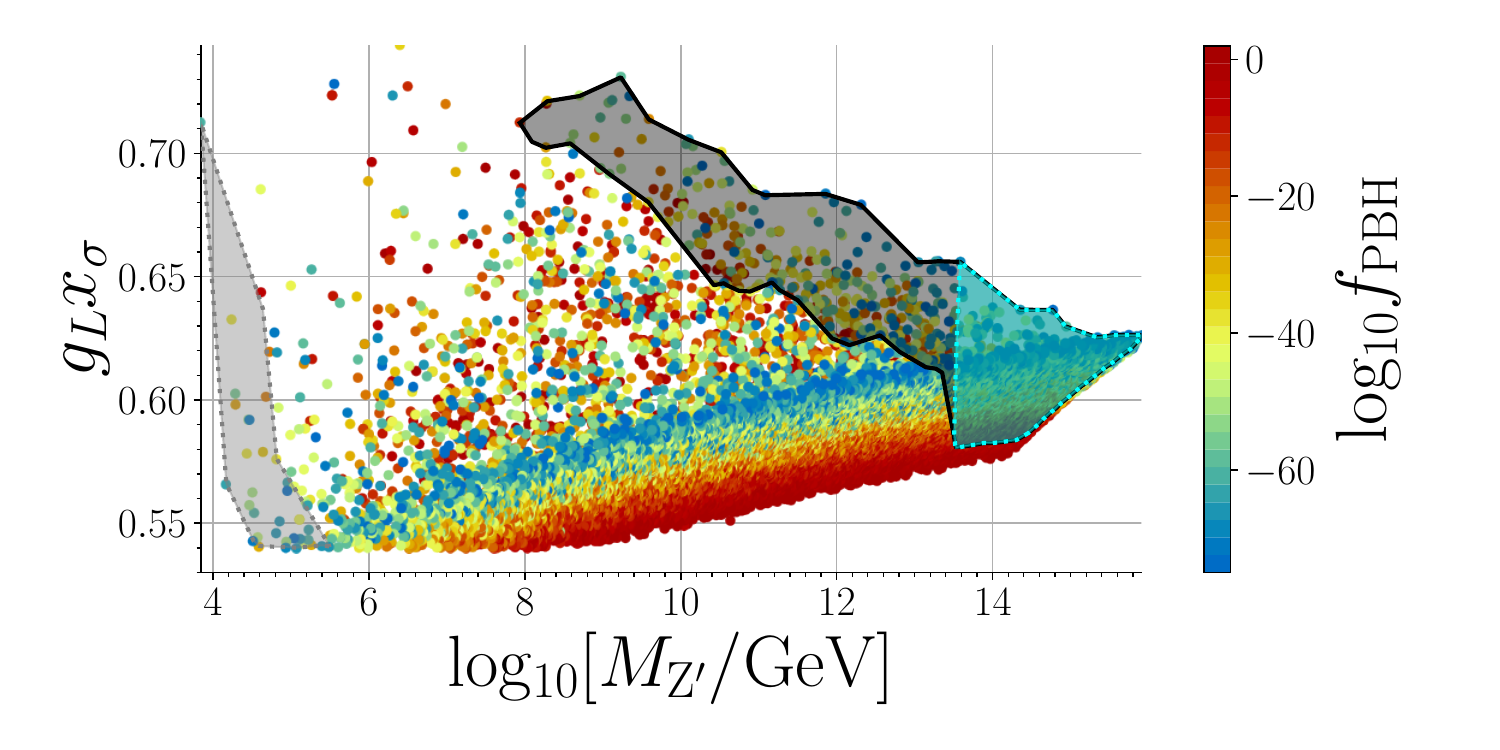}} \\
    \subfloat{\includegraphics[width=0.5\textwidth]{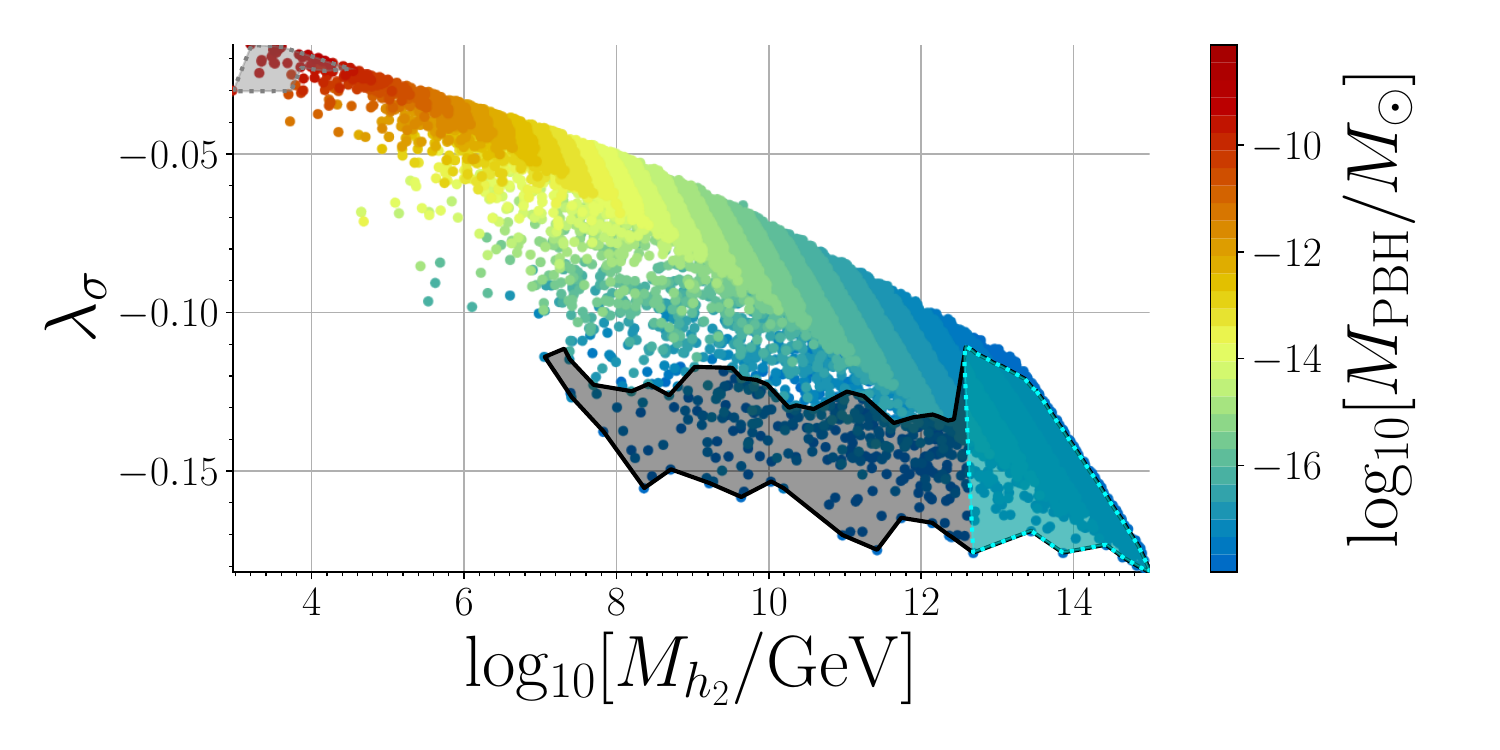}} 
    \subfloat{\includegraphics[width=0.5\textwidth]{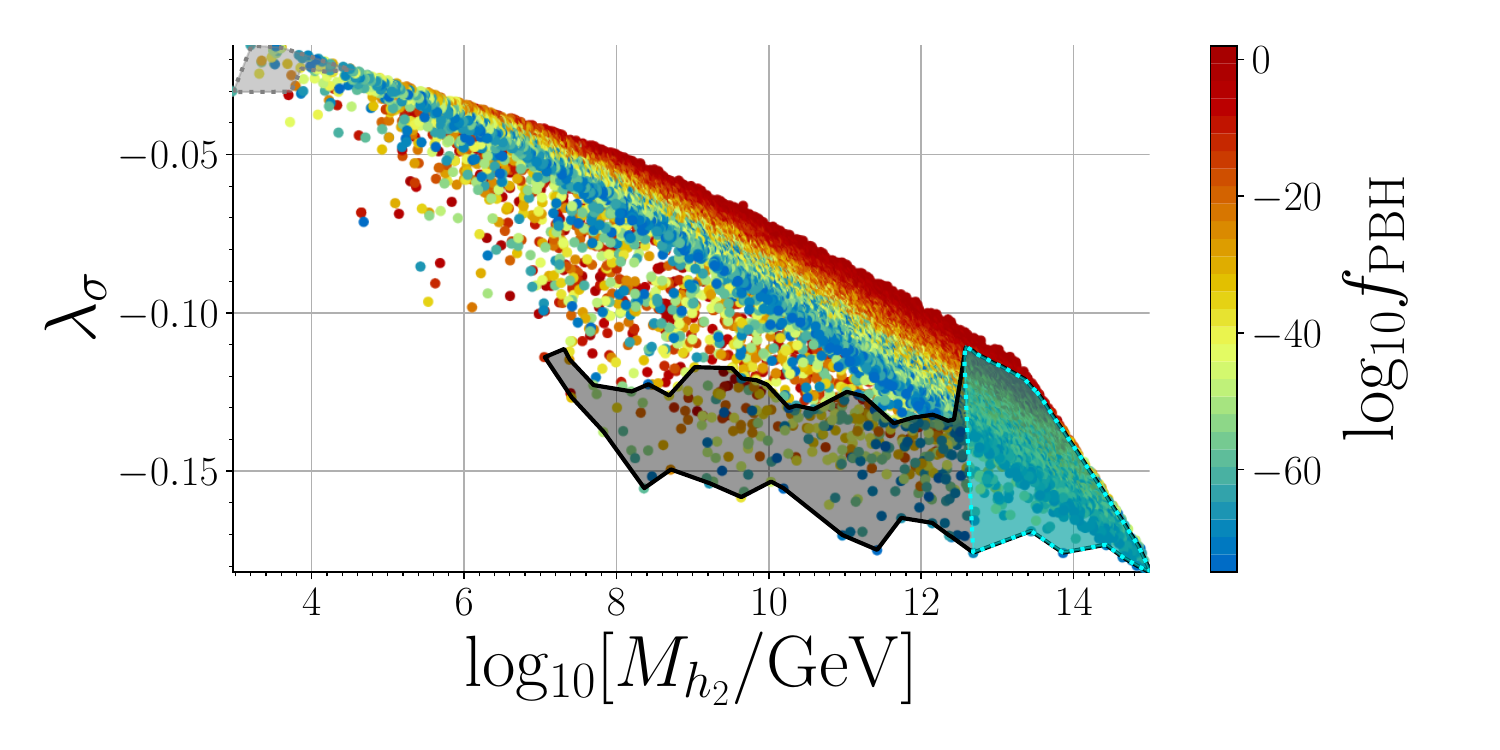}} \\
    \subfloat{\includegraphics[width=0.5\textwidth]{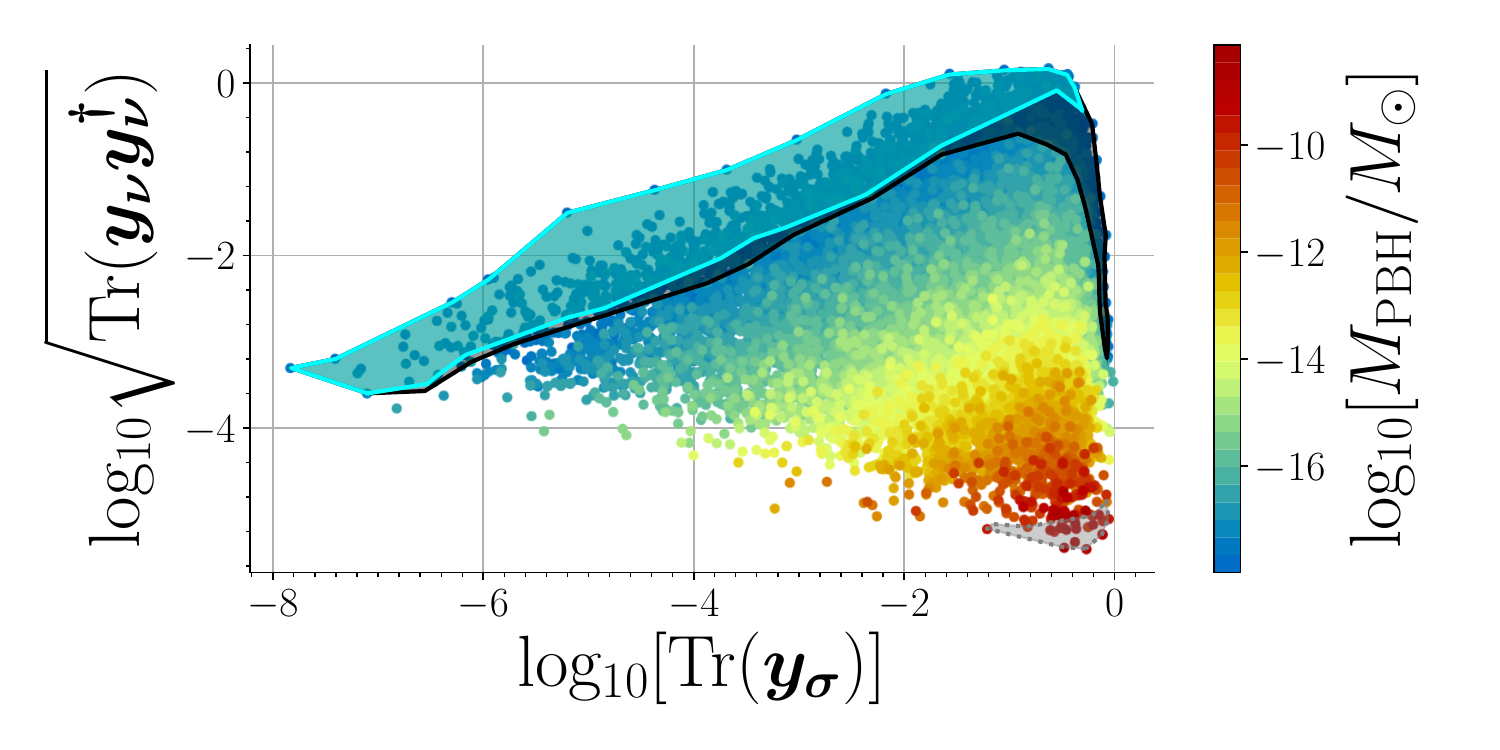}}
    \subfloat{\includegraphics[width=0.5\textwidth]{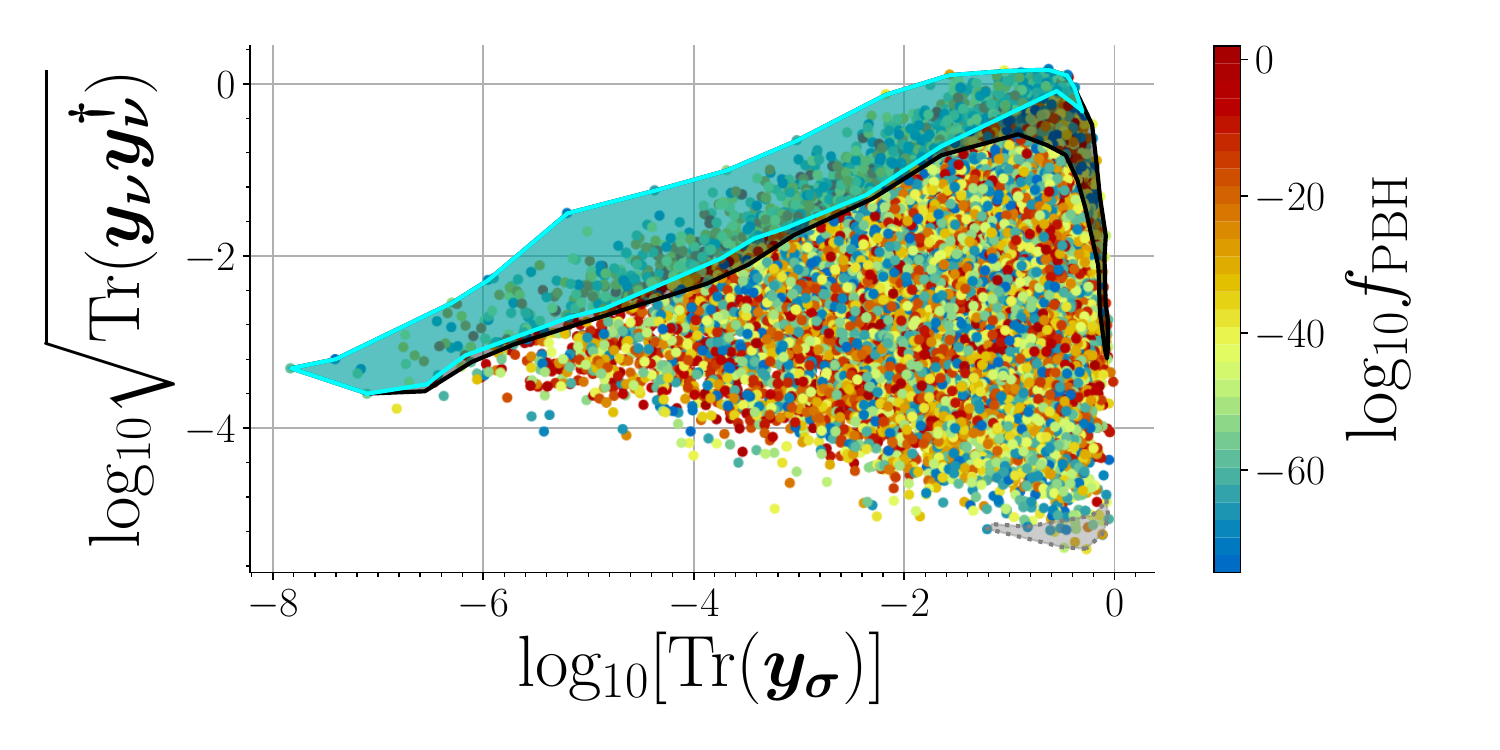}} \\
    \subfloat{\includegraphics[width=0.5\textwidth]{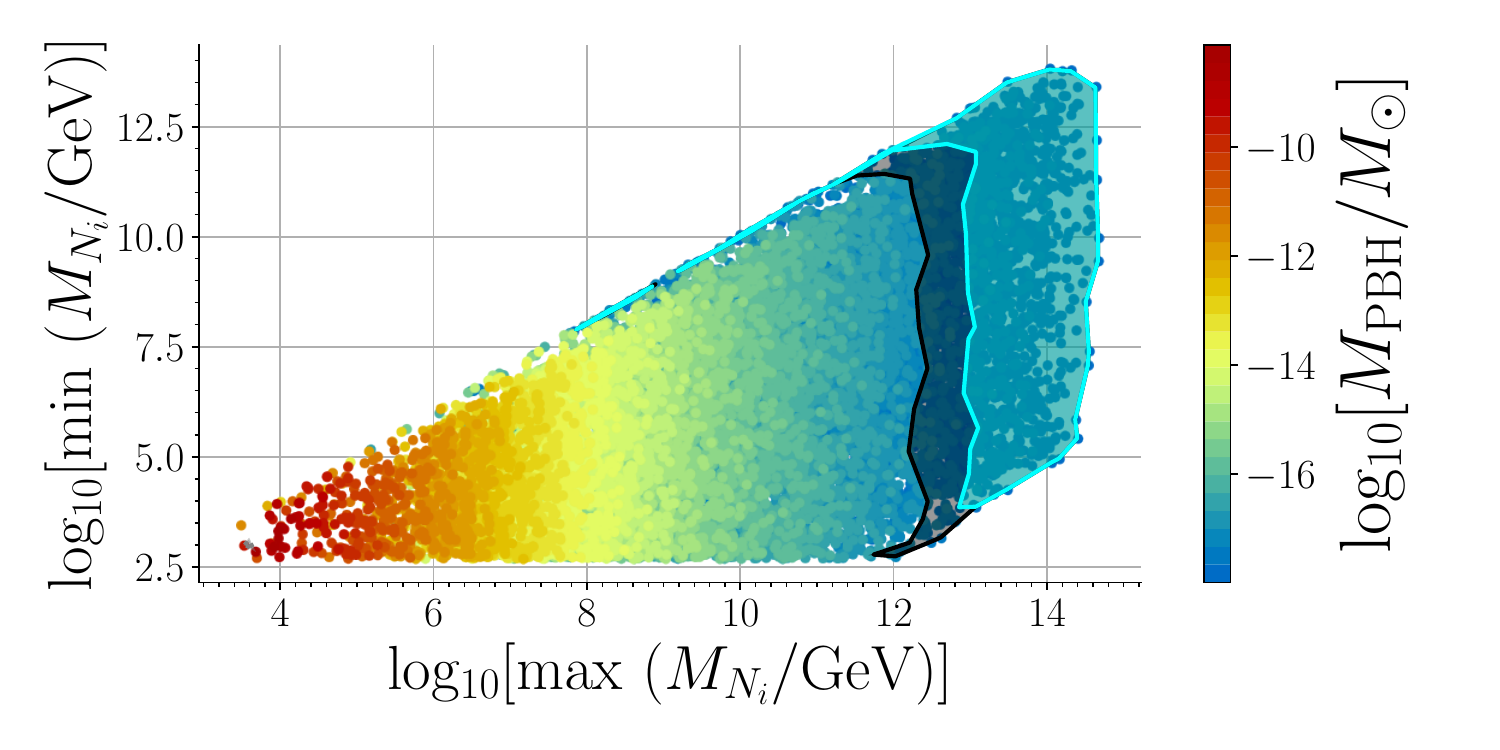}}
    \subfloat{\includegraphics[width=0.5\textwidth]{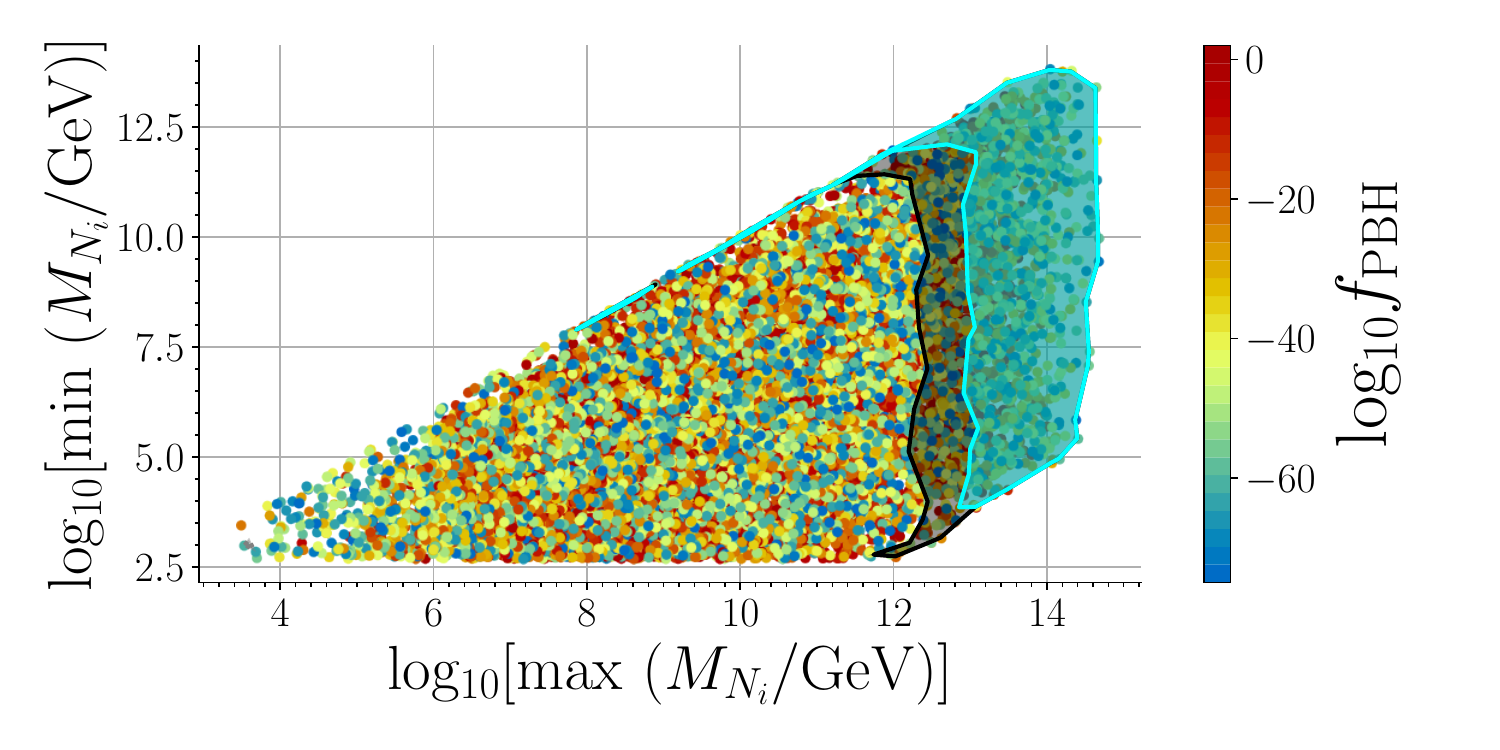}}
    \caption{\small Projections of the model's parameter space, with the PBH mass (left panels) and PBH abundance (right panels) indicated by the color scales. All points satisfy current constraints from microlensing and the extragalactic $\gamma$-ray background. The region enclosed by the black solid contour is excluded by LVK since $\mathrm{SNR}>10$. The region enclosed by the cyan solid contour is excluded by NANOGrav data. The gray dotted contour indicates the region where peak magnetic field strengths for $b=0$ exceed the blazar bounds.}
	\label{fig:parameters_proj}
\end{figure*}
In \cref{tab:BP} we present three benchmark points: BP-1 features microlensing events at the Roman telescope with PBHs saturating the DM relic abundance ($f_\mathrm{PBH} = 1$), BP-2 produces detectable $\gamma$-ray signals, and BP-3 generates detectable magnetic fields, microlensing events, $\gamma$-ray signals and GW signals, with a subdominant PBH abundance. BP-3 is a point from the focused scan in Fig.~\ref{fig:primordial_Bfields_dedicated}.
\begin{table}[t!]
\begin{adjustbox}{width=0.7\textwidth,center}
\begin{tabular}{|c|ccc|}
\hline
\hline
& {\bf BP-1} & {\bf BP-2} & {\bf BP-3}\\
\hline
\hline
$\lambda_\sigma$ & $-0.0439$ & $-0.0797$  & $-0.0229$\\
$\lambda_{\sigma h}$ & $-1.22\times 10^{-11}$ & $-1.77\times 10^{-18}$ & $-6.44\times 10^{-6}$\\
$g_L x_\sigma$ & $0.567$ & $0.575$ & $0.600$\\
$\mathrm{Tr}(\bm{y_\sigma)}$ & $0.392$ & $0.336$ & $0.713$\\
$\sqrt{\mathrm{Tr}(\bm{y_\nu} \bm{y_\nu}^\dagger})$ & $1.99\times 10^{-4}$ & $0.0128$ & $7.69\times 10^{-6}$\\
$M_\mathrm{Z^\prime}/\mathrm{GeV}$ & $2.17\times 10^{7}$ & $5.78\times 10^{10}$ & $3.16\times 10^4$\\
$M_{h_2}/\mathrm{GeV}$ & $2.38\times 10^6$ & $6.46\times 10^9$ & $3.61\times 10^{3}$\\
$M_N/\mathrm{GeV}$ & $7.58\times 10^6$ & $1.39\times 10^{10}$ & $1.32\times 10^4$\\
\hline
$T_p/\mathrm{GeV}$ & $19.0$ & $1.70\times 10^{3}$ & $0.502$\\
$T_\mathrm{RH}/\mathrm{GeV}$ & $3.05\times 10^5$ & $7.77\times 10^6$ & $6.31\times 10^{3}$\\
$T_c/\mathrm{GeV}$ & $1.13\times 10^6$ & $2.85\times 10^7$ & $2.62\times 10^4$\\
$\alpha$ & $6.61\times 10^{16}$ & $4.33\times 10^{14}$ & $2.50\times 10^{16}$\\
$\beta/H(T_p)$ & $6.68$ & $6.87$ & $7.26$\\
\hline
$\mathrm{SNR_\mathrm{LISA}}$ & $1.10\times 10^4$ & $8.59$ & $3.59\times 10^5$ \\
$\mathrm{SNR_\mathrm{ET}}$ & $25.7$ & $2.69\times 10^5$ & $4.22\times 10^{-4}$ \\
$\mathrm{SNR_\mathrm{LIGO}}$ & $0.00173$ & $13.1$ & $1.50\times 10^{-8}$ \\
$\mathrm{SNR_\mathrm{LVK}}$ & $5.62\times 10^{-6}$ & $0.110$ & $1.23\times 10^{-10}$ \\
\hline
$f_\mathrm{PBH}$ & $1.00$ & $0.0467$ & $0.00111$\\
$M_\mathrm{PBH}/M_\odot$ & $1.01\times 10^{-13}$ & $1.41\times 10^{-16}$ & $2.06\times 10^{-10}$\\
$N_\mathrm{events}$ & $109$ & $< 1$ & $226$ \\
$\delta_c$ & $0.448$ & $0.479$ & $0.400$ \\
$B^{b=0}_\mathrm{peak}/\mathrm{G}$ & $1.26\times 10^{-18}$ & $1.99\times 10^{-20}$ & $4.78\times 10^{-16}$ \\
$\lambda^{b=0}_\mathrm{peak}/\mathrm{Mpc}$ & $3.15\times 10^{-3}$ & $1.05\times 10^{-3}$ & $2.71\times 10^{-3}$ \\
$B^{b=1}_\mathrm{peak}/\mathrm{G}$ & $3.23\times 10^{-21}$ & $2.88\times 10^{-23}$ & $2.32\times 10^{-18}$ \\
$\lambda^{b=1}_\mathrm{peak}/\mathrm{Mpc}$ & $8.03\times 10^{-6}$ & $1.55\times 10^{-6}$ & $6.33\times 10^{-6}$ \\
\hline
\hline
\end{tabular}
\end{adjustbox}
\caption{\label{tab:BP}\small Model parameters, thermodynamic parameters and physical observables, for three benchmark points: BP-1, featuring microlensing events at the Roman telescope and $f_\mathrm{PBH} = 1$; BP-2, producing $\gamma$-ray signals with $\mathrm{SNR}>10$ in the proposed experiments considered; and BP-3, characterized by magnetic fields above the blazar limits, large number of microlensing events at the Roman telescope and detectable SGWB at interferometers. The model parameters are defined at $Q = M_\mathrm{Z^0}$.  }
\end{table}

In \cref{fig:parameters_proj}, the DM abundance shows a weak correlation with the model parameters.
However, along the lower edge of the regions in the right panels of the first two rows ($g_L x_\sigma \sim 0.55$ and $\lambda_\sigma \sim -0.063$) most points have $f_\mathrm{PBH}$ close to unity. The neutrino sector parameters are also weakly correlated with $f_\mathrm{PBH}$ as can be seen from the right panels of the last two rows. In contrast, all model parameters exhibit a strong correlation with the PBH mass. Specifically, larger values of the $\mathrm{Z^\prime}$, $h_2$ and $N_i$ masses correspond to lighter PBHs, as for the $\U{B-L}$ model. Also, higher values of $\mathrm{Tr}(\bm{y_\sigma})$ and lower values of $\sqrt{\mathrm{Tr}(\bm{y_\nu}\bm{y_\nu}^\dagger)}$ are associated with heavier PBHs.

For generic $\U{}^\prime$ models, we find that only a very narrow region of $g_L x_\sigma$ and $\lambda_\sigma$ can produce $\gamma$-ray signals at future experiments. This region, enclosed by the green dashed contour, features $\mathrm{Z'}$ masses in the interval $[10^{10}, 10^{13}]~\mathrm{GeV}$, and $h_2$ masses within $[10^9, 10^{12}]~\mathrm{GeV}$. The neutrino sector parameters in the last two rows are not constrained by $\gamma$-ray observations.

In the low $\mathrm{Z'}$-mass regime, between $5~\mathrm{TeV}$ and $100~\mathrm{TeV}$, it is possible to generate magnetic field strengths that exceed the current blazar bounds. This region, enclosed by the dashed gray contour, corresponds to a minimal hierarchy between the $h_1$ and $h_2$ masses, resulting in a larger efficiency factor $\kappa_h$. This also favors seesaw scales in the tens of TeV range and $\mathrm{Tr}(\bm{y_\sigma}) \lesssim 0.87$. 

\section{Summary}\label{sec:summary}

We investigated the production of primordial black holes and primordial magnetic fields from supercooled first-order phase transitions in a generic class of conformal $\mathrm{U(1)}^\prime$ models that incorporate neutrino masses via a type-I seesaw mechanism. 

These models exhibit strongly supercooled FOPTs, leading to vacuum-energy-dominated transitions that favor the formation of PBHs. Our analysis demonstrates that PBHs can form in a wide mass range, from $M_\mathrm{PBH} = 10^{-18} M_\odot$ to $10^{-9} M_\odot$ corresponding  to right-handed neutrino masses between $10^4$~GeV and $10^{14}~\mathrm{GeV}$, respectively, and $\mathrm{U(1)}^\prime$ gauge couplings in the range $g_L \sim 0.25$--$0.30$. 

We demonstrated that in the $\U{B-L}$ model, PBHs that make up the entire dark-matter abundance constrain the heavy neutrino masses to $10^4~\mathrm{GeV}- 10^{11}~\mathrm{GeV}$.  The predicted SGWB has frequencies detectable by LISA and the Einstein Telescope. Correlated microlensing events at the Roman Space Telescope correspond to $\mathrm{Z^\prime}$ boson masses of $10^6$~GeV and $10^{11}~\mathrm{GeV}$ and $M_\mathrm{PBH} \sim 10^{-12} M_\odot$. We also find that LISA will fully cover the regime in which PBHs account for all DM.

In addition to microlensing signals, PBHs with $M_\mathrm{PBH} \sim 10^{-16} M_\odot$ could emit detectable $\gamma$-ray signals from Hawking radiation at the planned telescopes THESEUS, GECCO, AMEGO-X, and e-ASTROGAM.

We also examined the implications of FOPTs in these models for the origin of large-scale primordial magnetic fields. After accounting for the inverse cascade in a radiation-dominated plasma and subsequent redshifting, from our general scan we find that coherence lengths can reach values of around $\lambda_\textrm{peak} \sim 10^{-4}$~Mpc to $10^{-2}\,\mathrm{Mpc}$ at peak field strengths $\sim 10^{-16}$~G to $10^{-14}$~G in the present Universe, depending on whether the initial magnetic field is maximally helical or non-helical, respectively. For helical fields, these coherence lengths and field strengths are in the ballpark required to exceed  IGMF lower bounds inferred from blazar observations. We also find a small region of parameter space in which PBHs saturate the DM relic abundance and magnetic fields can be observable at the Roman telescope. Magnetic fields exceeding current blazar-derived constraints are predicted for PBH masses near $10^{-9} M_\odot$. These magnetic fields arise preferably in scenarios with 3~TeV right-handed neutrinos, directly linking neutrino mass generation to magnetogenesis in conformal $\U{}^\prime$ models.

Our results highlight the rich interplay between PBH formation, gravitational wave astronomy, and multi-messenger probes of the early universe. We showed that viable PBH production is compatible with the generation of IGMFs and compelling neutrino phenomenology, yielding a region of parameter space wherein these phenomena originate from the same underlying dark-sector dynamics. Future gravitational wave detectors, microlensing surveys, and $\gamma$-ray observatories will be essential to test this scenario. Given the strong theoretical motivation for scale-invariant extensions of the Standard Model, further exploration of their implications, including for baryogenesis, neutrino physics, and the thermal history of the Universe, is an exciting direction for future research.

\section*{Acknowledgments}

J.G.~is directly funded by the Portuguese Foundation for Science and Technology (FCT - Funda\c{c}\~{a}o para a Ci\^{e}ncia e a Tecnologia) through the doctoral program grant with the reference 2021.04527.BD (\url{https://doi.org/10.54499/2021.04527.BD}).
D.M.~is supported in part by the U.S. Department of Energy under Grant No.DE-SC0010504.
A.P.M.~is supported by FCT through the project with reference 2024.05617.CERN (\url{https://doi.org/10.54499/2024.05617.CERN}).
R.P.~and J.G.~are supported in part by the Swedish Research Council grant, contract number 2016-05996. S.B. is supported by the STFC under grant ST/X000753/1. R.P.~also acknowledges support by the COST Action CA22130 (COMETA).
J.G.~and A.P.M.~are also supported by LIP and by the
Portuguese Foundation for Science and Technology (FCT), reference LA/P/0016/2020. 

\bibliographystyle{JHEP}
\bibliography{references}

@article{Oda:2015gna,
    author = "Oda, Satsuki and Okada, Nobuchika and Takahashi, Dai-suke",
    title = "{Classically conformal U(1)' extended standard model and Higgs vacuum stability}",
    eprint = "1504.06291",
    archivePrefix = "arXiv",
    primaryClass = "hep-ph",
    doi = "10.1103/PhysRevD.92.015026",
    journal = "Phys. Rev. D",
    volume = "92",
    number = "1",
    pages = "015026",
    year = "2015"
}

@article{Balaji:2024rvo,
    author = "Balaji, Shyam and Fairbairn, Malcolm and Olea-Romacho, Maria Olalla",
    title = "{Magnetogenesis with gravitational waves and primordial black hole dark matter}",
    eprint = "2402.05179",
    archivePrefix = "arXiv",
    primaryClass = "hep-ph",
    doi = "10.1103/PhysRevD.109.075048",
    journal = "Phys. Rev. D",
    volume = "109",
    number = "7",
    pages = "075048",
    year = "2024"
}

@article{Baldes:2023rqv,
    author = "Baldes, Iason and Olea-Romacho, Mar\'\i{}a Olalla",
    title = "{Primordial black holes as dark matter: interferometric tests of phase transition origin}",
    eprint = "2307.11639",
    archivePrefix = "arXiv",
    primaryClass = "hep-ph",
    doi = "10.1007/JHEP01(2024)133",
    journal = "JHEP",
    volume = "01",
    pages = "133",
    year = "2024"
}

@article{Salvio:2023ynn,
    author = "Salvio, Alberto",
    title = "{Supercooling in radiative symmetry breaking: theory extensions, gravitational wave detection and primordial black holes}",
    eprint = "2307.04694",
    archivePrefix = "arXiv",
    primaryClass = "hep-ph",
    doi = "10.1088/1475-7516/2023/12/046",
    journal = "JCAP",
    volume = "12",
    pages = "046",
    year = "2023"
}

@article{Stamou:2023vxu,
    author = "Stamou, Ioanna D.",
    title = "{Exploring critical overdensity thresholds in inflationary models of primordial black holes formation}",
    eprint = "2306.02758",
    archivePrefix = "arXiv",
    primaryClass = "astro-ph.CO",
    doi = "10.1103/PhysRevD.108.063515",
    journal = "Phys. Rev. D",
    volume = "108",
    number = "6",
    pages = "063515",
    year = "2023"
}

@article{Hawking:1982ga,
    author = "Hawking, S. W. and Moss, I. G. and Stewart, J. M.",
    title = "{Bubble Collisions in the Very Early Universe}",
    reportNumber = "Print-82-0180 (CAMBRIDGE)",
    doi = "10.1103/PhysRevD.26.2681",
    journal = "Phys. Rev. D",
    volume = "26",
    pages = "2681",
    year = "1982"
}

@article{Moss:1994iq,
    author = "Moss, I. G.",
    title = "{Singularity formation from colliding bubbles}",
    doi = "10.1103/PhysRevD.50.676",
    journal = "Phys. Rev. D",
    volume = "50",
    pages = "676--681",
    year = "1994"
}

@article{Kodama:1982sf,
    author = "Kodama, Hideo and Sasaki, Misao and Sato, Katsuhiko",
    title = "{Abundance of Primordial Holes Produced by Cosmological First Order Phase Transition}",
    reportNumber = "KUNS 642",
    doi = "10.1143/PTP.68.1979",
    journal = "Prog. Theor. Phys.",
    volume = "68",
    pages = "1979",
    year = "1982"
}

@article{Liu:2021svg,
    author = "Liu, Jing and Bian, Ligong and Cai, Rong-Gen and Guo, Zong-Kuan and Wang, Shao-Jiang",
    title = "{Primordial black hole production during first-order phase transitions}",
    eprint = "2106.05637",
    archivePrefix = "arXiv",
    primaryClass = "astro-ph.CO",
    doi = "10.1103/PhysRevD.105.L021303",
    journal = "Phys. Rev. D",
    volume = "105",
    number = "2",
    pages = "L021303",
    year = "2022"
}

@article{Hashino:2021qoq,
    author = "Hashino, Katsuya and Kanemura, Shinya and Takahashi, Tomo",
    title = "{Primordial black holes as a probe of strongly first-order electroweak phase transition}",
    eprint = "2111.13099",
    archivePrefix = "arXiv",
    primaryClass = "hep-ph",
    reportNumber = "OU-HET-1123",
    doi = "10.1016/j.physletb.2022.137261",
    journal = "Phys. Lett. B",
    volume = "833",
    pages = "137261",
    year = "2022"
}

@article{Lewicki:2023ioy,
    author = "Lewicki, Marek and Toczek, Piotr and Vaskonen, Ville",
    title = "{Primordial black holes from strong first-order phase transitions}",
    eprint = "2305.04924",
    archivePrefix = "arXiv",
    primaryClass = "astro-ph.CO",
    doi = "10.1007/JHEP09(2023)092",
    journal = "JHEP",
    volume = "09",
    pages = "092",
    year = "2023"
}

@article{Gouttenoire:2023naa,
    author = "Gouttenoire, Yann and Volansky, Tomer",
    title = "{Primordial black holes from supercooled phase transitions}",
    eprint = "2305.04942",
    archivePrefix = "arXiv",
    primaryClass = "hep-ph",
    doi = "10.1103/PhysRevD.110.043514",
    journal = "Phys. Rev. D",
    volume = "110",
    number = "4",
    pages = "043514",
    year = "2024"
}

@article{Musco:2020jjb,
    author = "Musco, Ilia and De Luca, Valerio and Franciolini, Gabriele and Riotto, Antonio",
    title = "{Threshold for primordial black holes. II. A simple analytic prescription}",
    eprint = "2011.03014",
    archivePrefix = "arXiv",
    primaryClass = "astro-ph.CO",
    doi = "10.1103/PhysRevD.103.063538",
    journal = "Phys. Rev. D",
    volume = "103",
    number = "6",
    pages = "063538",
    year = "2021"
}

@article{Escriva:2019phb,
    author = "Escriv\`a, Albert and Germani, Cristiano and Sheth, Ravi K.",
    title = "{Universal threshold for primordial black hole formation}",
    eprint = "1907.13311",
    archivePrefix = "arXiv",
    primaryClass = "gr-qc",
    reportNumber = "ICC-19-013",
    doi = "10.1103/PhysRevD.101.044022",
    journal = "Phys. Rev. D",
    volume = "101",
    number = "4",
    pages = "044022",
    year = "2020"
}

@article{Witten:1984rs,
    author = "Witten, Edward",
    title = "{Cosmic Separation of Phases}",
    reportNumber = "PRINT-84-0400 (IAS,PRINCETON)",
    doi = "10.1103/PhysRevD.30.272",
    journal = "Phys. Rev. D",
    volume = "30",
    pages = "272--285",
    year = "1984"
}

@article{Hogan:1986qda,
    author = "Hogan, C. J.",
    title = "{Gravitational radiation from cosmological phase transitions}",
    journal = "Mon. Not. Roy. Astron. Soc.",
    volume = "218",
    pages = "629--636",
    year = "1986"
}

@article{KAGRA:2021kbb,
    author = "Abbott, R. and others",
    collaboration = "KAGRA, Virgo, LIGO Scientific",
    title = "{Upper limits on the isotropic gravitational-wave background from Advanced LIGO and Advanced Virgo\textquoteright{}s third observing run}",
    eprint = "2101.12130",
    archivePrefix = "arXiv",
    primaryClass = "gr-qc",
    reportNumber = "LIGO-DCC-P2000314",
    doi = "10.1103/PhysRevD.104.022004",
    journal = "Phys. Rev. D",
    volume = "104",
    number = "2",
    pages = "022004",
    year = "2021"
}

@article{Kamionkowski:1993fg,
    author = "Kamionkowski, Marc and Kosowsky, Arthur and Turner, Michael S.",
    title = "{Gravitational radiation from first order phase transitions}",
    eprint = "astro-ph/9310044",
    archivePrefix = "arXiv",
    reportNumber = "IASSNS-HEP-93-44, FERMILAB-PUB-93-235-A",
    doi = "10.1103/PhysRevD.49.2837",
    journal = "Phys. Rev. D",
    volume = "49",
    pages = "2837--2851",
    year = "1994"
}

@article{Croon:2020cgk,
    author = "Croon, Djuna and Gould, Oliver and Schicho, Philipp and Tenkanen, Tuomas V. I. and White, Graham",
    title = "{Theoretical uncertainties for cosmological first-order phase transitions}",
    eprint = "2009.10080",
    archivePrefix = "arXiv",
    primaryClass = "hep-ph",
    reportNumber = "HIP-2020-26/TH",
    doi = "10.1007/JHEP04(2021)055",
    journal = "JHEP",
    volume = "04",
    pages = "055",
    year = "2021"
}

@article{Marzo:2018nov,
    author = "Marzo, Carlo and Marzola, Luca and Vaskonen, Ville",
    title = "{Phase transition and vacuum stability in the classically conformal B\textendash{}L model}",
    eprint = "1811.11169",
    archivePrefix = "arXiv",
    primaryClass = "hep-ph",
    reportNumber = "KCL-PH-TH/2018-68",
    doi = "10.1140/epjc/s10052-019-7076-x",
    journal = "Eur. Phys. J. C",
    volume = "79",
    number = "7",
    pages = "601",
    year = "2019"
}

@article{Carr:1975qj,
    author = "Carr, Bernard J.",
    title = "{The Primordial black hole mass spectrum}",
    doi = "10.1086/153853",
    journal = "Astrophys. J.",
    volume = "201",
    pages = "1--19",
    year = "1975"
}

@article{Esteban:2024eli,
    author = "Esteban, Ivan and Gonzalez-Garcia, M. C. and Maltoni, Michele and Martinez-Soler, Ivan and Pinheiro, Jo\~ao Paulo and Schwetz, Thomas",
    title = "{NuFit-6.0: updated global analysis of three-flavor neutrino oscillations}",
    eprint = "2410.05380",
    archivePrefix = "arXiv",
    primaryClass = "hep-ph",
    reportNumber = "IFT-UAM/CSIC-24-140, YITP-SB-2024-24, IPPP/24/64, IPPP/24/64, IFT-UAM/CSIC-24-140, YITP-SB-2024-24",
    doi = "10.1007/JHEP12(2024)216",
    journal = "JHEP",
    volume = "12",
    pages = "216",
    year = "2024"
}

@article{Chikashige:1980qk,
    author = "Chikashige, Y. and Mohapatra, Rabindra N. and Peccei, R. D.",
    title = "{Spontaneously Broken Lepton Number and Cosmological Constraints on the Neutrino Mass Spectrum}",
    reportNumber = "MPI-PAE/PTh 40/80",
    doi = "10.1103/PhysRevLett.45.1926",
    journal = "Phys. Rev. Lett.",
    volume = "45",
    pages = "1926",
    year = "1980"
}

@article{Chikashige:1980ui,
    author = "Chikashige, Y. and Mohapatra, Rabindra N. and Peccei, R. D.",
    title = "{Are There Real Goldstone Bosons Associated with Broken Lepton Number?}",
    reportNumber = "MPI-PAE-PTH-36-80",
    doi = "10.1016/0370-2693(81)90011-3",
    journal = "Phys. Lett. B",
    volume = "98",
    pages = "265--268",
    year = "1981"
}

@article{Gelmini:1980re,
    author = "Gelmini, G. B. and Roncadelli, M.",
    title = "{Left-Handed Neutrino Mass Scale and Spontaneously Broken Lepton Number}",
    reportNumber = "MPI-PAE-PTH-50-80",
    doi = "10.1016/0370-2693(81)90559-1",
    journal = "Phys. Lett. B",
    volume = "99",
    pages = "411--415",
    year = "1981"
}

@article{Cordero-Carrion:2019qtu,
    author = "Cordero-Carri\'on, Isabel and Hirsch, Martin and Vicente, Avelino",
    title = "{General parametrization of Majorana neutrino mass models}",
    eprint = "1912.08858",
    archivePrefix = "arXiv",
    primaryClass = "hep-ph",
    reportNumber = "IFIC/19-59",
    doi = "10.1103/PhysRevD.101.075032",
    journal = "Phys. Rev. D",
    volume = "101",
    number = "7",
    pages = "075032",
    year = "2020"
}

@article{Coleman:1973jx,
    author = "Coleman, Sidney R. and Weinberg, Erick J.",
    title = "{Radiative Corrections as the Origin of Spontaneous Symmetry Breaking}",
    doi = "10.1103/PhysRevD.7.1888",
    journal = "Phys. Rev. D",
    volume = "7",
    pages = "1888--1910",
    year = "1973"
}

@article{Goncalves:2024lrk,
    author = "Gon\c{c}alves, João and Marfatia, Danny and Morais, Ant\'onio P. and Pasechnik, Roman",
    title = "{Gravitational waves from supercooled phase transitions in conformal Majoron models of neutrino mass}",
    eprint = "2412.02645",
    archivePrefix = "arXiv",
    primaryClass = "hep-ph",
    doi = "10.1007/JHEP02(2025)110",
    journal = "JHEP",
    volume = "02",
    pages = "110",
    year = "2025"
}

@article{Ellis:2020nnr,
    author = "Ellis, John and Lewicki, Marek and Vaskonen, Ville",
    title = "{Updated predictions for gravitational waves produced in a strongly supercooled phase transition}",
    eprint = "2007.15586",
    archivePrefix = "arXiv",
    primaryClass = "astro-ph.CO",
    reportNumber = "KCL-PH-TH/2020-40, CERN-TH-2020-129",
    doi = "10.1088/1475-7516/2020/11/020",
    journal = "JCAP",
    volume = "11",
    pages = "020",
    year = "2020"
}

@article{ATLAS:2019erb,
    author = "Aad, Georges and others",
    collaboration = "ATLAS",
    title = "{Search for high-mass dilepton resonances using 139 fb$^{-1}$ of $pp$ collision data collected at $\sqrt{s}=$13 TeV with the ATLAS detector}",
    eprint = "1903.06248",
    archivePrefix = "arXiv",
    primaryClass = "hep-ex",
    reportNumber = "CERN-EP-2019-030",
    doi = "10.1016/j.physletb.2019.07.016",
    journal = "Phys. Lett. B",
    volume = "796",
    pages = "68--87",
    year = "2019"
}

@article{ATLAS:2017eiz,
    author = "Aaboud, Morad and others",
    collaboration = "ATLAS",
    title = "{Search for additional heavy neutral Higgs and gauge bosons in the ditau final state produced in 36 fb$^{-1}$ of pp collisions at $ \sqrt{s}=13 $ TeV with the ATLAS detector}",
    eprint = "1709.07242",
    archivePrefix = "arXiv",
    primaryClass = "hep-ex",
    reportNumber = "CERN-EP-2017-199",
    doi = "10.1007/JHEP01(2018)055",
    journal = "JHEP",
    volume = "01",
    pages = "055",
    year = "2018"
}

@article{ATLAS:2018tfk,
    author = "Aaboud, Morad and others",
    collaboration = "ATLAS",
    title = "{Search for resonances in the mass distribution of jet pairs with one or two jets identified as $b$-jets in proton-proton collisions at $\sqrt{s}=13$ TeV with the ATLAS detector}",
    eprint = "1805.09299",
    archivePrefix = "arXiv",
    primaryClass = "hep-ex",
    reportNumber = "CERN-EP-2018-075",
    doi = "10.1103/PhysRevD.98.032016",
    journal = "Phys. Rev. D",
    volume = "98",
    pages = "032016",
    year = "2018"
}

@article{ATLAS:2020lks,
    author = "Aad, Georges and others",
    collaboration = "ATLAS",
    title = "{Search for $ t\overline{t} $ resonances in fully hadronic final states in $pp$ collisions at $ \sqrt{s} $ = 13 TeV with the ATLAS detector}",
    eprint = "2005.05138",
    archivePrefix = "arXiv",
    primaryClass = "hep-ex",
    reportNumber = "CERN-EP-2020-055",
    doi = "10.1007/JHEP10(2020)061",
    journal = "JHEP",
    volume = "10",
    pages = "061",
    year = "2020"
}

@article{Drlica-Wagner:2022lbd,
    author = "Drlica-Wagner, Alex and others",
    title = "{Report of the Topical Group on Cosmic Probes of Dark Matter for Snowmass 2021}",
    eprint = "2209.08215",
    archivePrefix = "arXiv",
    primaryClass = "hep-ph",
    reportNumber = "FERMILAB-FN-1211-PPD",
    month = "9",
    year = "2022"
}

@article{Gould:2021oba,
    author = "Gould, Oliver and Tenkanen, Tuomas V. I.",
    title = "{On the perturbative expansion at high temperature and implications for cosmological phase transitions}",
    eprint = "2104.04399",
    archivePrefix = "arXiv",
    primaryClass = "hep-ph",
    reportNumber = "NORDITA 2021-010",
    doi = "10.1007/JHEP06(2021)069",
    journal = "JHEP",
    volume = "06",
    pages = "069",
    year = "2021"
}

@article{Caprini:2024hue,
    author = "Caprini, Chiara and Jinno, Ryusuke and Lewicki, Marek and Madge, Eric and Merchand, Marco and Nardini, Germano and Pieroni, Mauro and Roper Pol, Alberto and Vaskonen, Ville",
    collaboration = "LISA Cosmology Working Group",
    title = "{Gravitational waves from first-order phase transitions in LISA: reconstruction pipeline and physics interpretation}",
    eprint = "2403.03723",
    archivePrefix = "arXiv",
    primaryClass = "astro-ph.CO",
    reportNumber = "LISA-COSWG-24-01, CERN-TH-2024-029",
    month = "3",
    year = "2024"
}

@article{LISA:2017pwj,
    author = "Amaro-Seoane, Pau and others",
    collaboration = "LISA",
    title = "{Laser Interferometer Space Antenna}",
    eprint = "1702.00786",
    archivePrefix = "arXiv",
    primaryClass = "astro-ph.IM",
    month = "2",
    year = "2017"
}

@article{LIGOScientific:2014pky,
    author = "Aasi, J. and others",
    collaboration = "LIGO Scientific",
    title = "{Advanced LIGO}",
    eprint = "1411.4547",
    archivePrefix = "arXiv",
    primaryClass = "gr-qc",
    doi = "10.1088/0264-9381/32/7/074001",
    journal = "Class. Quant. Grav.",
    volume = "32",
    pages = "074001",
    year = "2015"
}

@article{Punturo:2010zz,
    author = "Punturo, M. and others",
    editor = "Ricci, Fulvio",
    title = "{The Einstein Telescope: A third-generation gravitational wave observatory}",
    doi = "10.1088/0264-9381/27/19/194002",
    journal = "Class. Quant. Grav.",
    volume = "27",
    pages = "194002",
    year = "2010"
}

@article{2019ApJS2413P,
    author = {{Penny}, Matthew T. and {Gaudi}, B. Scott and {Kerins}, Eamonn and {Rattenbury}, Nicholas J. and {Mao}, Shude and {Robin}, Annie C. and {Calchi Novati}, Sebastiano},
    title = "{Predictions of the WFIRST Microlensing Survey. I. Bound Planet Detection Rates}",
    journal = {APJS},
    year = 2019,
    month = mar,
    volume = {241},
    number = {1},
    eid = {3},
    pages = {3},
    doi = {10.3847/1538-4365/aafb69},
    archivePrefix = {arXiv},
    eprint = {1808.02490},
    primaryClass = {astro-ph.EP},
    adsurl = {https://ui.adsabs.harvard.edu/abs/2019ApJS..241....3P}
}

@article{Smyth:2019whb,
    author = "Smyth, Nolan and Profumo, Stefano and English, Samuel and Jeltema, Tesla and McKinnon, Kevin and Guhathakurta, Puragra",
    title = "{Updated Constraints on Asteroid-Mass Primordial Black Holes as Dark Matter}",
    eprint = "1910.01285",
    archivePrefix = "arXiv",
    primaryClass = "astro-ph.CO",
    doi = "10.1103/PhysRevD.101.063005",
    journal = "Phys. Rev. D",
    volume = "101",
    number = "6",
    pages = "063005",
    year = "2020"
}

@article{Aramaki:2019bpi,
    author = "Aramaki, Tsuguo and Hansson Adrian, Per and Karagiorgi, Georgia and Odaka, Hirokazu",
    title = "{Dual MeV Gamma-Ray and Dark Matter Observatory - GRAMS Project}",
    eprint = "1901.03430",
    archivePrefix = "arXiv",
    primaryClass = "astro-ph.HE",
    doi = "10.1016/j.astropartphys.2019.07.002",
    journal = "Astropart. Phys.",
    volume = "114",
    pages = "107--114",
    year = "2020"
}

@article{Orlando:2021get,
    author = "Orlando, Elena and others",
    title = "{Exploring the MeV sky with a combined coded mask and Compton telescope: the Galactic Explorer with a Coded aperture mask Compton telescope (GECCO)}",
    eprint = "2112.07190",
    archivePrefix = "arXiv",
    primaryClass = "astro-ph.HE",
    doi = "10.1088/1475-7516/2022/07/036",
    journal = "JCAP",
    volume = "07",
    number = "07",
    pages = "036",
    year = "2022"
}

@article{e-ASTROGAM:2016bph,
    author = "De Angelis, A. and others",
    collaboration = "e-ASTROGAM",
    title = "{The e-ASTROGAM mission}",
    eprint = "1611.02232",
    archivePrefix = "arXiv",
    primaryClass = "astro-ph.HE",
    doi = "10.1007/s10686-017-9533-6",
    journal = "Exper. Astron.",
    volume = "44",
    number = "1",
    pages = "25--82",
    year = "2017"
}

@article{Arbey:2019mbc,
    author = "Arbey, Alexandre and Auffinger, J\'er\'emy",
    title = "{BlackHawk: A public code for calculating the Hawking evaporation spectra of any black hole distribution}",
    eprint = "1905.04268",
    archivePrefix = "arXiv",
    primaryClass = "gr-qc",
    reportNumber = "CERN-TH-2019-067",
    doi = "10.1140/epjc/s10052-019-7161-1",
    journal = "Eur. Phys. J. C",
    volume = "79",
    number = "8",
    pages = "693",
    year = "2019"
}

@article{Arbey:2021mbl,
    author = "Arbey, Alexandre and Auffinger, J\'er\'emy",
    title = "{Physics Beyond the Standard Model with BlackHawk v2.0}",
    eprint = "2108.02737",
    archivePrefix = "arXiv",
    primaryClass = "gr-qc",
    reportNumber = "CERN-TH-2021-117",
    doi = "10.1140/epjc/s10052-021-09702-8",
    journal = "Eur. Phys. J. C",
    volume = "81",
    pages = "910",
    year = "2021"
}

@article{Coogan:2019qpu,
    author = "Coogan, Adam and Morrison, Logan and Profumo, Stefano",
    title = "{Hazma: A Python Toolkit for Studying Indirect Detection of Sub-GeV Dark Matter}",
    eprint = "1907.11846",
    archivePrefix = "arXiv",
    primaryClass = "hep-ph",
    doi = "10.1088/1475-7516/2020/01/056",
    journal = "JCAP",
    volume = "01",
    pages = "056",
    year = "2020"
}

@article{Auffinger:2022dic,
    author = "Auffinger, J\'er\'emy",
    title = "{Limits on primordial black holes detectability with Isatis: a BlackHawk tool}",
    eprint = "2201.01265",
    archivePrefix = "arXiv",
    primaryClass = "astro-ph.HE",
    doi = "10.1140/epjc/s10052-022-10199-y",
    journal = "Eur. Phys. J. C",
    volume = "82",
    number = "4",
    pages = "384",
    year = "2022"
}

@article{Goncalves:2025uwh,
    author = "Gon{\c{c}}alves, Jo{\~a}o and Marfatia, Danny and Morais, Ant{\'o}nio P. and Pasechnik, Roman",
    title = "{Supercooled phase transitions in conformal dark sectors explain NANOGrav data}",
    eprint = "2501.11619",
    archivePrefix = "arXiv",
    primaryClass = "hep-ph",
    doi = "10.1016/j.physletb.2025.139829",
    journal = "Phys. Lett. B",
    volume = "869",
    pages = "139829",
    year = "2025"
}

@inproceedings{10.1117/12.2561012,
author = {C. Labanti and others},
title = {{The X/Gamma-ray Imaging Spectrometer (XGIS) on-board THESEUS: design, main characteristics, and concept of operation}},
volume = {11444},
booktitle = {Space Telescopes and Instrumentation 2020: Ultraviolet to Gamma Ray},
editor = {Jan-Willem A. den Herder and Shouleh Nikzad and Kazuhiro Nakazawa},
organization = {International Society for Optics and Photonics},
publisher = {SPIE},
pages = {114442K},
year = {2020},
doi = {10.1117/12.2561012},
URL = {https://doi.org/10.1117/12.2561012}
}

@article{Planck:2018vyg,
    author = "Aghanim, N. and others",
    collaboration = "Planck",
    title = "{Planck 2018 results. VI. Cosmological parameters}",
    eprint = "1807.06209",
    archivePrefix = "arXiv",
    primaryClass = "astro-ph.CO",
    doi = "10.1051/0004-6361/201833910",
    journal = "Astron. Astrophys.",
    volume = "641",
    pages = "A6",
    year = "2020",
    note = "[Erratum: Astron.Astrophys. 652, C4 (2021)]"
}

@article{Fermi-LAT:2018jdy,
    author = "Ackermann, M. and others",
    collaboration = "Fermi-LAT",
    title = "{The Search for Spatial Extension in High-latitude Sources Detected by the $Fermi$ Large Area Telescope}",
    eprint = "1804.08035",
    archivePrefix = "arXiv",
    primaryClass = "astro-ph.HE",
    doi = "10.3847/1538-4365/aacdf7",
    journal = "Astrophys. J. Suppl.",
    volume = "237",
    number = "2",
    pages = "32",
    year = "2018"
}

@article{MAGIC:2022piy,
    author = "Acciari, V. A. and others",
    collaboration = "MAGIC",
    title = "{A lower bound on intergalactic magnetic fields from time variability of 1ES 0229+200 from MAGIC and Fermi/LAT observations}",
    eprint = "2210.03321",
    archivePrefix = "arXiv",
    primaryClass = "astro-ph.HE",
    doi = "10.1051/0004-6361/202244126",
    journal = "Astron. Astrophys.",
    volume = "670",
    pages = "A145",
    year = "2023"
}

@article{Ellis:2019tjf,
    author = "Ellis, John and Fairbairn, Malcolm and Lewicki, Marek and Vaskonen, Ville and Wickens, Alastair",
    title = "{Intergalactic Magnetic Fields from First-Order Phase Transitions}",
    eprint = "1907.04315",
    archivePrefix = "arXiv",
    primaryClass = "astro-ph.CO",
    reportNumber = "KCL-PH-TH/2019-60, CERN-TH-2019-104",
    doi = "10.1088/1475-7516/2019/09/019",
    journal = "JCAP",
    volume = "09",
    pages = "019",
    year = "2019"
}

@article{Tevzadze:2012kk,
    author = "Tevzadze, Alexander G. and Kisslinger, Leonard and Brandenburg, Axel and Kahniashvili, Tina",
    title = "{Magnetic Fields from QCD Phase Transitions}",
    eprint = "1207.0751",
    archivePrefix = "arXiv",
    primaryClass = "astro-ph.CO",
    reportNumber = "NORDITA-2012-52",
    doi = "10.1088/0004-637X/759/1/54",
    journal = "Astrophys. J.",
    volume = "759",
    pages = "54",
    year = "2012"
}

@article{ATLAS:2023oaq,
    author = "Aad, Georges and others",
    collaboration = "ATLAS",
    title = "{Combined Measurement of the Higgs Boson Mass from the H\textrightarrow{}\ensuremath{\gamma}\ensuremath{\gamma} and H\textrightarrow{}ZZ*\textrightarrow{}4\ensuremath{\ell} Decay Channels with the ATLAS Detector Using s=7, 8, and 13~TeV pp Collision Data}",
    eprint = "2308.04775",
    archivePrefix = "arXiv",
    primaryClass = "hep-ex",
    reportNumber = "CERN-EP-2023-156",
    doi = "10.1103/PhysRevLett.131.251802",
    journal = "Phys. Rev. Lett.",
    volume = "131",
    number = "25",
    pages = "251802",
    year = "2023"
}

@article{AlvesBatista:2021sln,
    author = "Alves Batista, Rafael and Saveliev, Andrey",
    title = "{The Gamma-ray Window to Intergalactic Magnetism}",
    eprint = "2105.12020",
    archivePrefix = "arXiv",
    primaryClass = "astro-ph.HE",
    doi = "10.3390/universe7070223",
    journal = "Universe",
    volume = "7",
    number = "7",
    pages = "223",
    year = "2021"
}

@article{Vachaspati:1991nm,
    author = "Vachaspati, T.",
    title = "{Magnetic fields from cosmological phase transitions}",
    doi = "10.1016/0370-2693(91)90051-Q",
    journal = "Phys. Lett. B",
    volume = "265",
    pages = "258--261",
    year = "1991"
}

@article{Sigl:1996dm,
    author = "Sigl, Guenter and Olinto, Angela V. and Jedamzik, Karsten",
    title = "{Primordial magnetic fields from cosmological first order phase transitions}",
    eprint = "astro-ph/9610201",
    archivePrefix = "arXiv",
    doi = "10.1103/PhysRevD.55.4582",
    journal = "Phys. Rev. D",
    volume = "55",
    pages = "4582--4590",
    year = "1997"
}

@article{Copi:2008he,
    author = "Copi, Craig J. and Ferrer, Francesc and Vachaspati, Tanmay and Achucarro, Ana",
    title = "{Helical Magnetic Fields from Sphaleron Decay and Baryogenesis}",
    eprint = "0801.3653",
    archivePrefix = "arXiv",
    primaryClass = "astro-ph",
    doi = "10.1103/PhysRevLett.101.171302",
    journal = "Phys. Rev. Lett.",
    volume = "101",
    pages = "171302",
    year = "2008"
}

@article{Chu:2011tx,
    author = "Chu, Yi-Zen and Dent, James B. and Vachaspati, Tanmay",
    title = "{Magnetic Helicity in Sphaleron Debris}",
    eprint = "1105.3744",
    archivePrefix = "arXiv",
    primaryClass = "hep-th",
    doi = "10.1103/PhysRevD.83.123530",
    journal = "Phys. Rev. D",
    volume = "83",
    pages = "123530",
    year = "2011"
}

@article{Vachaspati:2001nb,
    author = "Vachaspati, Tanmay",
    title = "{Estimate of the primordial magnetic field helicity}",
    eprint = "astro-ph/0101261",
    archivePrefix = "arXiv",
    doi = "10.1103/PhysRevLett.87.251302",
    journal = "Phys. Rev. Lett.",
    volume = "87",
    pages = "251302",
    year = "2001"
}

@article{PhysRev.82.863,
         title = {Cosmic Radiation and Cosmic Magnetic Fields. II. Origin of Cosmic Magnetic Fields},
         author = {Biermann, Ludwig and Schl\"uter, Arnulf},
         journal = {Phys. Rev.},
         volume = {82},
         issue = {6},
         pages = {863--868},
         numpages = {0},
         year = {1951},
         month = {Jun},
         publisher = {American Physical Society},
         doi = {10.1103/PhysRev.82.863},
         url = {https://link.aps.org/doi/10.1103/PhysRev.82.863}
}

@article{Brandenburg:1996fc,
    author = "Brandenburg, Axel and Enqvist, Kari and Olesen, Poul",
    title = "{Large scale magnetic fields from hydromagnetic turbulence in the very early universe}",
    eprint = "astro-ph/9602031",
    archivePrefix = "arXiv",
    reportNumber = "NORDITA-96-6-A",
    doi = "10.1103/PhysRevD.54.1291",
    journal = "Phys. Rev. D",
    volume = "54",
    pages = "1291--1300",
    year = "1996"
}

@article{Christensson:2000sp,
    author = "Christensson, Mattias and Hindmarsh, Mark and Brandenburg, Axel",
    title = "{Inverse cascade in decaying 3-D magnetohydrodynamic turbulence}",
    eprint = "astro-ph/0011321",
    archivePrefix = "arXiv",
    doi = "10.1103/PhysRevE.64.056405",
    journal = "Phys. Rev. E",
    volume = "64",
    pages = "056405",
    year = "2001"
}

@article{Kahniashvili:2010gp,
    author = "Kahniashvili, Tina and Brandenburg, Axel and Tevzadze, Alexander G. and Ratra, Bharat",
    title = "{Numerical simulations of the decay of primordial magnetic turbulence}",
    eprint = "1004.3084",
    archivePrefix = "arXiv",
    primaryClass = "astro-ph.CO",
    reportNumber = "NORDITA-PREPRINT-NORDITA-2010-20",
    doi = "10.1103/PhysRevD.81.123002",
    journal = "Phys. Rev. D",
    volume = "81",
    pages = "123002",
    year = "2010"
}

@article{Brandenburg:2017neh,
    author = "Brandenburg, Axel and Kahniashvili, Tina and Mandal, Sayan and Roper Pol, Alberto and Tevzadze, Alexander G. and Vachaspati, Tanmay",
    title = "{Evolution of hydromagnetic turbulence from the electroweak phase transition}",
    eprint = "1711.03804",
    archivePrefix = "arXiv",
    primaryClass = "astro-ph.CO",
    reportNumber = "NORDITA-2017-116",
    doi = "10.1103/PhysRevD.96.123528",
    journal = "Phys. Rev. D",
    volume = "96",
    number = "12",
    pages = "123528",
    year = "2017"
}

@article{HESS:2023zwb,
    author = "Aharonian, F. and others",
    collaboration = "H.E.S.S., Fermi-LAT",
    title = "{Constraints on the Intergalactic Magnetic Field Using Fermi-LAT and H.E.S.S. Blazar Observations}",
    eprint = "2306.05132",
    archivePrefix = "arXiv",
    primaryClass = "astro-ph.HE",
    doi = "10.3847/2041-8213/acd777",
    journal = "Astrophys. J. Lett.",
    volume = "950",
    number = "2",
    pages = "L16",
    year = "2023"
}

@article{Neronov:2010gir,
    author = "Neronov, A. and Vovk, I.",
    title = "{Evidence for strong extragalactic magnetic fields from Fermi observations of TeV blazars}",
    eprint = "1006.3504",
    archivePrefix = "arXiv",
    primaryClass = "astro-ph.HE",
    doi = "10.1126/science.1184192",
    journal = "Science",
    volume = "328",
    pages = "73--75",
    year = "2010"
}

@article{Banerjee:2004df,
    author = "Banerjee, Robi and Jedamzik, Karsten",
    title = "{The Evolution of cosmic magnetic fields: From the very early universe, to recombination, to the present}",
    eprint = "astro-ph/0410032",
    archivePrefix = "arXiv",
    doi = "10.1103/PhysRevD.70.123003",
    journal = "Phys. Rev. D",
    volume = "70",
    pages = "123003",
    year = "2004"
}

@article{Brandenburg:2017rnt,
    author = "Brandenburg, Axel and Kahniashvili, Tina and Mandal, Sayan and Roper Pol, Alberto and Tevzadze, Alexander G. and Vachaspati, Tanmay",
    title = "{The dynamo effect in decaying helical turbulence}",
    eprint = "1710.01628",
    archivePrefix = "arXiv",
    primaryClass = "physics.flu-dyn",
    reportNumber = "NORDITA-2017-099",
    doi = "10.1103/PhysRevFluids.4.024608",
    journal = "Phys. Rev. Fluids.",
    volume = "4",
    pages = "024608",
    year = "2019"
}

@article{Armua:2022rvx,
    author = "Armua, Andres and Berera, Arjun and Figueroa, Jaime Calderon",
    title = "{Parameter study of decaying magnetohydrodynamic turbulence}",
    eprint = "2212.02418",
    archivePrefix = "arXiv",
    primaryClass = "physics.plasm-ph",
    doi = "10.1103/PhysRevE.107.055206",
    journal = "Phys. Rev. E",
    volume = "107",
    number = "5",
    pages = "055206",
    year = "2023"
}

@article{RoperPol:2023bqa,
    author = "Roper Pol, A. and Neronov, A. and Caprini, C. and Boyer, T. and Semikoz, D.",
    title = "{LISA and $\gamma$-ray telescopes as multi-messenger probes of a first-order cosmological phase transition}",
    eprint = "2307.10744",
    archivePrefix = "arXiv",
    primaryClass = "astro-ph.CO",
    month = "7",
    year = "2023"
}

@article{Kahniashvili:2009qi,
    author = "Kahniashvili, Tina and Tevzadze, Alexander G. and Ratra, Bharat",
    title = "{Phase Transition Generated Cosmological Magnetic Field at Large Scales}",
    eprint = "0907.0197",
    archivePrefix = "arXiv",
    primaryClass = "astro-ph.CO",
    doi = "10.1088/0004-637X/726/2/78",
    journal = "Astrophys. J.",
    volume = "726",
    pages = "78",
    year = "2011"
}

@article{Durrer:2013pga,
    author = "Durrer, Ruth and Neronov, Andrii",
    title = "{Cosmological Magnetic Fields: Their Generation, Evolution and Observation}",
    eprint = "1303.7121",
    archivePrefix = "arXiv",
    primaryClass = "astro-ph.CO",
    doi = "10.1007/s00159-013-0062-7",
    journal = "Astron. Astrophys. Rev.",
    volume = "21",
    pages = "62",
    year = "2013"
}

@article{Caprini:2019egz,
    author = "Caprini, Chiara and others",
    title = "{Detecting gravitational waves from cosmological phase transitions with LISA: an update}",
    eprint = "1910.13125",
    archivePrefix = "arXiv",
    primaryClass = "astro-ph.CO",
    reportNumber = "DESY-19-159, IPPP/19/27, HIP-2019-14/TH, MITP/19-066, IFT-UAM/CSIC-19-139",
    doi = "10.1088/1475-7516/2020/03/024",
    journal = "JCAP",
    volume = "03",
    pages = "024",
    year = "2020"
}

@article{Fleischhack:2021mhc,
    author = "Fleischhack, Henrike",
    title = "{AMEGO-X: MeV gamma-ray Astronomy in the Multi-messenger Era}",
    eprint = "2108.02860",
    archivePrefix = "arXiv",
    primaryClass = "astro-ph.IM",
    doi = "10.22323/1.395.0649",
    journal = "PoS",
    volume = "ICRC2021",
    pages = "649",
    year = "2021"
}

@article{CamargoNevesdaCunha:2022mvg,
    author = "Camargo Neves da Cunha, Disrael and Ringeval, Christophe and Bouchet, Fran{\c{c}}ois R.",
    title = "{Stochastic gravitational waves from long cosmic strings}",
    eprint = "2205.04349",
    archivePrefix = "arXiv",
    primaryClass = "astro-ph.CO",
    doi = "10.1088/1475-7516/2022/09/078",
    journal = "JCAP",
    volume = "09",
    pages = "078",
    year = "2022"
}

@article{Blanco-Pillado:2024aca,
    author = "Blanco-Pillado, Jose J. and Cui, Yanou and Kuroyanagi, Sachiko and Lewicki, Marek and Nardini, Germano and Pieroni, Mauro and Rybak, Ivan Yu. and Sousa, Lara and Wachter, Jeremy M.",
    collaboration = "LISA Cosmology Working Group",
    title = "{Gravitational waves from cosmic strings in LISA: reconstruction pipeline and physics interpretation}",
    eprint = "2405.03740",
    archivePrefix = "arXiv",
    primaryClass = "astro-ph.CO",
    reportNumber = "LISA-COSWG-24-02, CERN-TH-2024-085",
    doi = "10.1088/1475-7516/2025/05/006",
    journal = "JCAP",
    volume = "05",
    pages = "006",
    year = "2025"
}

@article{Kibble:1976sj,
    author = "Kibble, T. W. B.",
    title = "{Topology of Cosmic Domains and Strings}",
    reportNumber = "ICTP/75/5",
    doi = "10.1088/0305-4470/9/8/029",
    journal = "J. Phys. A",
    volume = "9",
    pages = "1387--1398",
    year = "1976"
}

@article{Marfatia:2023fvh,
    author = "Marfatia, Danny and Zhou, Ye-Ling",
    title = "{Gravitational waves from cosmic superstrings and gauge strings}",
    eprint = "2312.10455",
    archivePrefix = "arXiv",
    primaryClass = "hep-ph",
    doi = "10.1007/JHEP07(2024)204",
    journal = "JHEP",
    volume = "07",
    pages = "204",
    year = "2024"
}

@article{Planck:2015fie,
    author = "Ade, P. A. R. and others",
    collaboration = "Planck",
    title = "{Planck 2015 results. XIII. Cosmological parameters}",
    eprint = "1502.01589",
    archivePrefix = "arXiv",
    primaryClass = "astro-ph.CO",
    doi = "10.1051/0004-6361/201525830",
    journal = "Astron. Astrophys.",
    volume = "594",
    pages = "A13",
    year = "2016"
}

@article{NANOGrav:2023hvm,
    author = "Afzal, Adeela and others",
    collaboration = "NANOGrav",
    title = "{The NANOGrav 15 yr Data Set: Search for Signals from New Physics}",
    eprint = "2306.16219",
    archivePrefix = "arXiv",
    primaryClass = "astro-ph.HE",
    reportNumber = "FERMILAB-PUB-23-589-T",
    doi = "10.3847/2041-8213/acdc91",
    journal = "Astrophys. J. Lett.",
    volume = "951",
    number = "1",
    pages = "L11",
    year = "2023",
    note = "[Erratum: Astrophys.J.Lett. 971, L27 (2024), Erratum: Astrophys.J. 971, L27 (2024)]"
}

@article{Franciolini:2025ztf,
    author = "Franciolini, Gabriele and Gouttenoire, Yann and Jinno, Ryusuke",
    title = "{Curvature Perturbations from First-Order Phase Transitions: Implications to Black Holes and Gravitational Waves}",
    eprint = "2503.01962",
    archivePrefix = "arXiv",
    primaryClass = "hep-ph",
    month = "3",
    year = "2025"
}

@article{Hashino:2025fse,
    author = "Hashino, Katsuya and Kanemura, Shinya and Takahashi, Tomo and Tanaka, Masanori and Yoo, Chul-Moon",
    title = "{Super-critical primordial black hole formation via delayed first-order electroweak phase transition}",
    eprint = "2501.11040",
    archivePrefix = "arXiv",
    primaryClass = "hep-ph",
    reportNumber = "OU-HET-1260",
    month = "1",
    year = "2025"
}

@article{Flores:2024lng,
    author = "Flores, Marcos M. and Kusenko, Alexander and Sasaki, Misao",
    title = "{Revisiting formation of primordial black holes in a supercooled first-order phase transition}",
    eprint = "2402.13341",
    archivePrefix = "arXiv",
    primaryClass = "hep-ph",
    reportNumber = "IPMU23-0053, YITP-23-169",
    doi = "10.1103/PhysRevD.110.015005",
    journal = "Phys. Rev. D",
    volume = "110",
    number = "1",
    pages = "015005",
    year = "2024"
}
\end{document}